\documentclass[aps,twocolumn,superscriptaddress,altaffilletter,lengthcheck,
tightenlines,showpacs,showkeys]{revtex4}
\usepackage{multirow}
\usepackage{amsmath,amssymb}
\usepackage[pdftex]{graphicx}
\usepackage[english]{babel}
\usepackage{hyperref}
\usepackage{color}
\usepackage[dvipdf]{epsfig}
\usepackage{color}

\begin{document}
\title{Scattering and absorption of  a scalar field  impinging on a charged black hole in the Einstein-Maxwell-dilaton theory}
\author{Mart\'{\i}n G. Richarte}\email{martin@df.uba.ar (corresponding author)}
\affiliation{Departamento de F\'isica - Universidade Federal do Esp\'irito Santo,   29075-910  Vit\'oria, ES, Brazil}
\affiliation{PPGCosmo, CCE - Universidade Federal do Esp\'irito Santo,  29075-910 Vit\'oria, ES, Brazil}
\affiliation{Departamento de F\'isica, Facultad de Ciencias Exactas y Naturales,
Universidad de Buenos Aires, Ciudad Universitaria 1428, Pabell\'on I,  Buenos Aires, Argentina}
\author{\'Ebano L. Martins}\email{ebano.lm@gmail.com}
\affiliation{Departamento de F\'isica - Universidade Federal do Esp\'irito Santo,   29075-910  Vit\'oria, ES, Brazil}
\author{J\'ulio C. Fabris}\email{julio.fabris@cosmo-ufes.org}
\affiliation{PPGCosmo, CCE - Universidade Federal do Esp\'irito Santo,  29075-910 Vit\'oria, ES, Brazil}
\affiliation{N\'ucleo Cosmo-ufes\&Departamento de F\'isica - Universidade Federal do Esp\'irito Santo,   29075-910  Vit\'oria, ES, Brazil}
\affiliation{National Research Nuclear University MEPhI, Kashirskoe sh. 31, Moscow 115409, Russia}
\bibliographystyle{plain}

\begin{abstract}
This study revisits the absorption and scattering process by which a massless scalar field impinges on a charged dilatonic black hole. First, we review the classical analysis to obtain the deflection angle and the differential scattering cross-section in terms of the mass, electric charge and dilatonic charge. Then, using the partial wave method, we determine the total absorption cross-section numerically in terms of the decoupling parameter called $M\omega$, finding that the amplitude of the dilatonic black hole is lower than the Reissner-Nordstr\"om one for mild frequencies. In the high-frequency limit, the absorption cross-section exhibits two different complex behaviors;  the fine structure and the hyperfine structure. For the differential scattering cross-section, smaller values of $M\omega$ lead to more significant amplitudes; the opposite scenario is obtained by increasing the charge-to-mass ratio. To fully grasp the main properties of the charged dilatonic black hole, we consider a different framework where the compact object is impinged by a charged massive scalar field.  The superradiant effect is lessened for intermediate frequency concerning the Reissner-Nordstr\"om case. However, this effect does not necessarily imply the existence of any dynamical instability. In order to trigger the superradiant instability, unstable modes must remain trapped outside the event horizon with a mechanism based on the reflecting-mirror boundary conditions. In this way, a charged scalar field plus a charged black hole configure a charged black hole bomb. We provide an analytic formula (lower bound) for the values of the charge field, which can trigger this superradiant instability. We extend this minimal setup by considering the dilaton perturbations while freezing the other degree of freedom. The new perturbation scheme enhances the superradiance scattering and reduces the lower bound of the charge-to-mass ratio to develop a superradiant instability.

\end{abstract}
\date{\today}
\maketitle


\section{Introduction}

 General Relativity (GR)  admits  the existence of several kinds of compact objects such as stars and black holes. The latter ones are the final stage of a  star  that has collapsed in on itself. They have such a strong gravitational field that it pulls in everything within their surrounding environments: gas, interstellar dust, and even light \cite{book0}, \cite{book1}, \cite{book3}. These astrophysical objects have been detected by different telescopes/satellites and other ground-based devices \cite{book2}. The first indirect detection of a black hole in astrophysics  goes back  to 1964 when a powerful source of X-ray was detected in Earth coming from a distant constellation called Cygnus \cite{cx}. Since then, the shreds of evidence supporting the existence of black holes in the Universe have been increasing at a fantastic pace, especially for the last 30 years\cite{book2}. In particular, the first detection by the LIGO interferometer of  a gravitational wave signal produced by the merging of two black holes in 2015 is considered as a strong confirmation of GR \cite{let1}. In 2019, the first image of the black hole  surrounding  at the center of a massive galaxy in the nearby Virgo galaxy cluster,  Messier 87*, was obtained by the EHT collaboration \cite{let2}. Both collaborations are  sources of new findings about black hole properties (mass, spin) and its environments (accretion disk).

Among the most straightforward generalizations of GR, there is a class of frameworks in which, besides the Hilbert-Einstein term, the additional degree of freedoms are minimally coupled to gravity. One of the most popular ones is the so-called Einstein-Maxwell-dilaton (EMd) gravity, where a scalar field (dilaton) is coupled to the kinetic term of a gauge $U(1)$ field. A self-interacting potential could be included in addition to the kinetic term associated with the dilaton field. 
Historically speaking, the EMd gravity  emerges  as the bosonic sector of $SU(4)$ version of $N=4$ supergravity theory,  but it also appears in the low-energy limit of the heterotic string as an effective field theory \cite{di1}, \cite{di2}. 
The recent interest in the black hole solutions within the EMd gravity relies on the fact that the dilaton field adds a new charge to the theory besides the mass and charge parameters. So the question is  whether the dilaton field  is radiated away from the black hole or if this global  charge  persists, evading  the no-hair conjecture that states that the final state of a black is only characterized by its mass, charge, and angular momentum \cite{book0}-\cite{book1}. 
This point is still debatable, provided it depends on how the dilaton field is coupled to the gauge field along with the extremality character of the black hole \cite{nh1}. The violation of the no-hair conjecture has been explored in different setups. In the case of EM gravity plus a self-interacting scalar field, not coupled to the gauge field,  a comparison between the shadow produced by a black hole with hair and the shadow of M87*  obtained by the Event Horizon Telescope led to solid constraints on the breaking of the no-hair conjecture.\cite{nh2}. Moreover,  the possibility of observing in Nature hairy black holes in the EMd framework was confronted with the waveform of gravitational waves generated by a binary system \cite{nh3}, \cite{nh4}. Similar results on the contribution of the gravitational waves signal within the PPN formalism for binary black holes in the EMd gravity were reported in  Refs. \cite{nh5}, \cite{nh6}.

Another appealing reason for exploring the  EMd gravity is that the theory does not seem to disobey the weak cosmic censorship conjecture.  In other words,   the electrically charged black hole with a non-zero dilatonic charge can not be destroyed (overcharged or overspun) due to the second-order perturbations introducing a backreaction effect, preventing this way the exposure of a naked singularity \cite{wc1}, \cite{wc2}.  The viability of the EMd theory at the Solar system level and its contrast with the four classical experiments of GR were reported in Refs.\cite{ss1}, \cite{ogun}. As an extension of the dilatonic black hole solutions,  we can consider the case of rotating black holes with an axion field. For instance, the emission of X-rays coming from the accreting material around black holes can be used as a complementary tool to constrain the spacetime around the X-ray binary system with low mass \cite{sxr}. \footnote{Additionally, the shadow of the supermassive black hole located  Messier 87*  can be used to constrain fundamental physics, for instance, a breakdown of the no-hair theorem due to the existence of hairy rotating black holes. \cite{sxr2}}.

 It is worth stressing that there are alternative ways to understand the properties of the spacetimes associated with black hole solutions. For example, Matzner explored how a massless scalar field hit a black hole;  a fraction of the initial wave is absorbed, and the other fraction is scattered off \cite{matzner1968scattering}. In his seminal work,  Matzner obtained the scattering cross-section in the zero-frequency  limit  showing that the total cross-section for the absorption process vanishes \cite{matzner1968scattering}. A central concept in deriving the result described earlier is the determination of those null geodesics near the black hole which are absorbed or scattered off; those possibilities depend on the value given to the initial impact parameter. The deflection angle by which the null geodesics are deflected or not around the black hole does not have a closed-form, and for that reason,  this quantity is calculated in the weak-field limit. As a consequence of this approach, an approximated expression for the differential cross-section is obtained  \cite{matzner1968scattering}. The main contribution for small-angle depends quadratically on the black hole's mass \cite{matzner1968scattering}, \cite{futte}.  For the  Schwarzschild black hole, a detailed analysis of the absorption cross-section associated with different energy and angular momentum values was carried out by Sanchez several years later 
 \cite{sanchez1978absorption}. The scattering of gravitational radiation by the Schwarzschild black hole was also explored in Ref.\cite{vishveshwara1970scattering}. In 2004, Jun and Park performed a complete analysis of the massive scalar field's role in the absorption and emission spectra of the Schwarzschild black hole \cite{jung2004effect}.

 An interesting phenomenon that appears as a massless scalar wave hits a black hole and then scatters off is the glory effect, leading to the formation of a bright halo around this compact object. In 1985, Matzner, Dewitt-Morette, Nelson, and Zhang explored the glory effect of a black hole \cite{matzner1985glory}. If the outgoing scattered wave is close to the direction of the initial incoming wave, an interference process occurs, creating a pattern in which a bright halo surrounds the black center\cite{matzner1985glory} \footnote{ In Nature, glory is an optical phenomenon that arises due to wave interference of light internally refracted within tiny droplets}. Further studies showed that the forward glory is a considerably small effect in black holes \cite{ande95}. Interestingly, the backward glory effect analysis was extended to other black hole solutions. In 2009,  Crispino, Dolan, and Oliveira explored the scattering and absorption of the massless scalar field by a Reissner-Nordstr\"om black hole \cite{crispino2009scattering}.  They applied the partial wave method to calculate cross-sections and showed good agreement with the numerically-calculated total cross-sections. One interesting point is that the effects coming from the charge are not always subdominant with respect to the black hole's mass. For example, the cross-section at a large angle and the width of the glory peak are both quite sensible with the charge of the black hole \cite{crispino2009scattering}.The same authors explored the scattering of electromagnetic waves by the Schwarzschild black hole \cite{crispino2009electromagnetic}. Since these pioneering works, the typical pattern produced by the scattering of a wave with a black hole has been used as a powerful tool to extract the properties of a black hole. The  scattering/absorption process of wave by a black hole  was examined in the case  a canonical acoustic black hole \cite{dolan2009scattering},  a charged black hole \cite{benone2014absorption}, \cite{benone2017absorption}, dirty black holes \cite{macedo2016absorption}, Kerr black holes \cite{leite2019scattering}, charged dilatonic black hole \cite{huang2020scattering}, and   for black hole in  Hovara-Lifshitz gravity 	\cite{liao2012scattering}. The absorption of $s$-waves by a remanent black hole in a metric-affine theory was explored in Ref.\cite{del19}.

These facts motivate the exploration of both the process of scattering and absorption of a massless scalar wave by a charged dilaton black hole, extending previous analysis  \cite{huang2020scattering}. In Sec. II, we introduce the EMd gravity and the charged dilatonic black hole, including the Penrose diagram for the charged black hole,  which later is fully explained in Appendix A. One of the goals of this study is to determine the classical differential cross-section for massless scalar field and to explore the backward glory effect numerically for different values of the dimensionless charge, $q$. The latter part is covered in Sec. III and in Appendix B, a semi-analytic approach to examine the glory effect contrasts with the complete numerical differential cross-section and the differential cross-section of the backward glory effect. Sec. IVA is devoted to obtaining the absorption cross-section through the partial wave method.  In Sec. IVB, we numerically compare the pattern associated with the total absorption cross-section in terms of the dimensionless coupling $M\omega$  \cite{crispino2009scattering}  for different values of  $q$. In addition to that, we employ a numerical integration method that allows us to reconstruct the differential scattering cross-section for different values of $M\omega$ and $q$, as is shown in Sec. IVC. One way to improve the previous analysis is by looking at the behavior of the absorption cross-section of a massless scalar field in the limit of high energy \cite{decanini2003complex}, \cite{decanini2010regge}, \cite{decanini2011universality}, \cite{decanini2011fine}.   For that reason, in Sec. IVD, we aim to obtain the absorption cross-section in a high-frequency limit.  The numerical analysis reveals the existence of different complex structures in the total absorption cross-section, also known as the fine and hyperfine structures  \cite{decanini2011fine}. In Sec. V, we extend the studies on the absorption of a charged massive scalar field by a charged dilatonic black hole, including a complete analysis of the behavior of the cross-section in the limit of low and high-frequency. In Sec. VA, the findings in the low-frequency regime are confirmed by numerical simulations, which point out the existence of two different phases and a critical velocity parameter, $v_c$, indicating the transition between both stages. To measure the relevance of these numerical analysis, we inspect several models of dark matter--with their typical average velocity $v_{X}$ and masses $m_{X}$-- and determine for different kinds of black holes (stellar, intermediate, supermassive,  and primordial) the critical velocity parameter, and by doing so, we conclude whether these models are compatible with a subcritical phase $v_{X}\ll v_{c}$  or a supercritical phase $v_{X}\gg v_{c}$.   In Sec. VB,  we employ numerical simulations to obtain the cross-section in the high energy limit, but only for the massive scalar field case, showing how some oscillatory behavior essentially corrects the geometric contribution. In Sec. VC, the numerical simulations for the differential scattering cross-section are displayed. In Sec. VIA, the  reflection coefficient is obtained, finding that there is a superradiant phenomenon and extending  previous results on  the literature   \cite{huang2020scattering},\cite{superr}, \cite{benone2016superradiance}.  In Sec. VIB, we explore the superradiant instability, showing that the unstable modes lead to instability if there is a potential well outside the horizon where these modes remain enclosed. Moreover, we obtain a lower bound on the $eq$ quantity when instability happens, where $e$ is the scalar charge, and $q$ denotes the charge-to-mass ratio. The reflecting-mirror boundary condition allows the black hole plus the scalar field to become a charged black hole bomb. In Sec. VII, another face of the superradiant instability is examined by taking into account the dilaton perturbations of the original model while the other fundamental fields (metric and abelian gauge field) remain frozen. Finally, the conclusions are stated in Sec. VIII. Throughout this paper we use geometric units so that $c=1$, $\hbar=1$, $G=1$. The metric signature is mostly positive, $(-,+,+,+)$. 

\section{The EMd theory}
 As we stated before, the EMd gravity is an effective theory appearing in the low-energy limit of the heterotic superstring model \cite{di1}, \cite{di2}.  Besides the graviton as part of the particle content of the theory, there  is a gauge field and a scalar field, whereas the abelian gauge field and dilaton are coupled minimally to gravity. They both coupled to each other in a non-trivial way controlled by an exponential term,  as is shown in the following action,
\begin{equation}\label{Acao}
	S=\frac{1}{16\pi}\int d^4 x\sqrt{-g}\Big(R-2\partial_\mu \phi \partial^\mu \phi- e^{-2\phi}F_{\mu\nu}F^{\mu\nu}\Big),
\end{equation}
where $F_{\mu\nu}=\partial_\mu A_\nu -\partial_\nu A_\mu$ is the electromagnetic tensor. The variation of the action (\ref{Acao}) with respect to $ g ^ {\mu \nu} $, $ A_\mu $ and $ \phi $ results in the following equation of motion,
\begin{eqnarray}\label{R}
	&& R_{\mu\nu}=2\partial_\mu \phi \partial_\nu \phi+2e^{-2\phi}\left(F_{\mu\beta}F_\nu^\beta-\frac{1}{4}g_{\mu\nu}F^2\right),\\
	&&\nabla_\nu(e^{-2\phi} F^{\mu\nu})=0,\\
	&&\nabla_\mu \nabla^\mu \phi=-\frac{1}{2}e^{-2\phi} F^2\label{box}.
\end{eqnarray}
This theory allows black hole solutions, which are asymptotically flat. These solutions described regular black holes with a non-degenerate event horizon and were obtained by Gibbons and Maeda  about 30 years ago\cite{gibbons1988black}. In the case of an electrostatic solution, the metric and matter fields are given by 
\begin{eqnarray}\label{EMD}
	ds^2&=&-f(r)dt^2+\frac{1}{f(r)}dr^2+r^2g(r)d\Omega^2,\\
	e^{2\phi}&=&g(r),\\
	F&=&\frac{Q}{r^2}dr\wedge dt,\\
	g(r)&=&\left(1-\frac{2D}{r}\right),\\
	f(r)&=&\frac{(r-2M)}{r}.
\end{eqnarray}
The ADM mass is $M$ and the ADM electric charge is $Q$.  The dilaton field introduces a new global quantity called dilatonic charge \cite{garfinkle1991charged},
\begin{eqnarray}
	D&=&\frac{1}{4\pi}\lim_{r\to\infty}\oint d\Sigma^\mu\partial_\mu\phi\approx\frac{Q^2}{2M},
\end{eqnarray}
where the 2-volume form is $d\Sigma_{\alpha} = \eta_{\alpha} \sqrt{-h} d ^ 3x $ and $ \eta_{\alpha} = f ^ {1/2} (0,1,0,0) $ is the normal vector to the surface $ \Sigma$ ( $||\eta ^{\mu}||^2 = -1 $). 

The case  with $M> 0$ represents a black hole with a spacelike singularity hidden by the event horizon which is located at $ r_{\rm{h}} = 2M $. Notice that this charged black hole solution does not present two horizons, as with Reissner-Nordstr\"om black hole. Additionally, the area of the sphere goes to zero for $ r =2D $, so this surface becomes singular. In $D\simeq M$ limit, a relationship between the electric charge and mass parameter is obtained, namely, $Q^2 = 2M^2$. In this way, it is admissible to re-write the equations only in terms of the parameters $Q$ and $M$. The singular surface area is reached if the condition $r = Q^2/M$ is met.  In the  $Q<\sqrt{2}M$ case, the event horizon hides the singularity; see the Penrose diagram depicted in Fig.(\ref{fig:pen0}). It is explained in more detail in  Appendix A. We will consider the black hole with the parameters satisfying the condition $Q^2\leq 2M^2$. In the extreme case, $ Q = \sqrt{2}M $, the singularity coincides with the horizon \cite{li2013stability}, \cite{garfinkle1991charged}. As is expected, the zero electric charged condition ($Q=0$) leads to  the Schwarzschild solution. However, the comparison  with Reissner-Nordstr\"om  is  elusive provided there is no values  of $Q$  which gives back the  RN metric \cite{garfinkle1991charged}.  Regarding this last point, we must emphasize that the comparison with the charged black hole in GR must be done with some care otherwise we will end up  misinterpreting  our findings. As is expected, the zero electric charged condition ($Q=0$) leads to the Schwarzschild solution. However, the comparison with Reissner-Nordstr\"om  is elusive provided there are no values of $Q$  which gives back the  RN metric \cite{garfinkle1991charged}. Regarding this last point, we must emphasize that the comparison with the charged black hole in GR must carry out with care; otherwise, we will end up misinterpreting our findings.
\begin{figure}
	\includegraphics[width=3.2in]{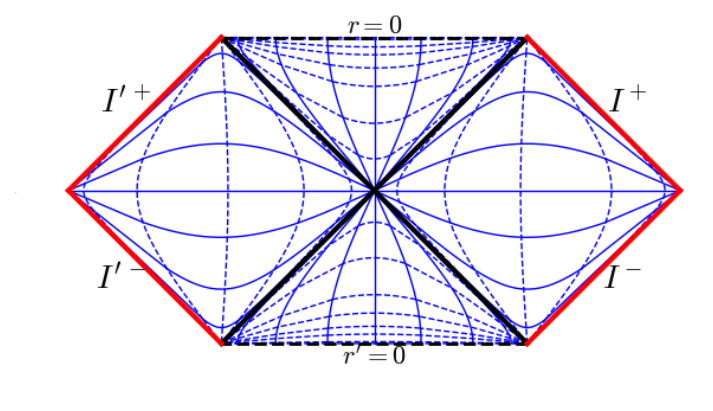}
	\caption{The Penrose diagram for the charged black hole in the EMd theory is shown. The bold black line corresponds to the event horizon, and the dashed black line corresponds to the spacelike singularity. Dashed lines correspond to time-like geodesics, whereas continuous lines denote spacelike geodesic.}
	\label{fig:pen0}
\end{figure}
\section{Classical analysis}
\begin{figure}
\includegraphics[width=2.8in]{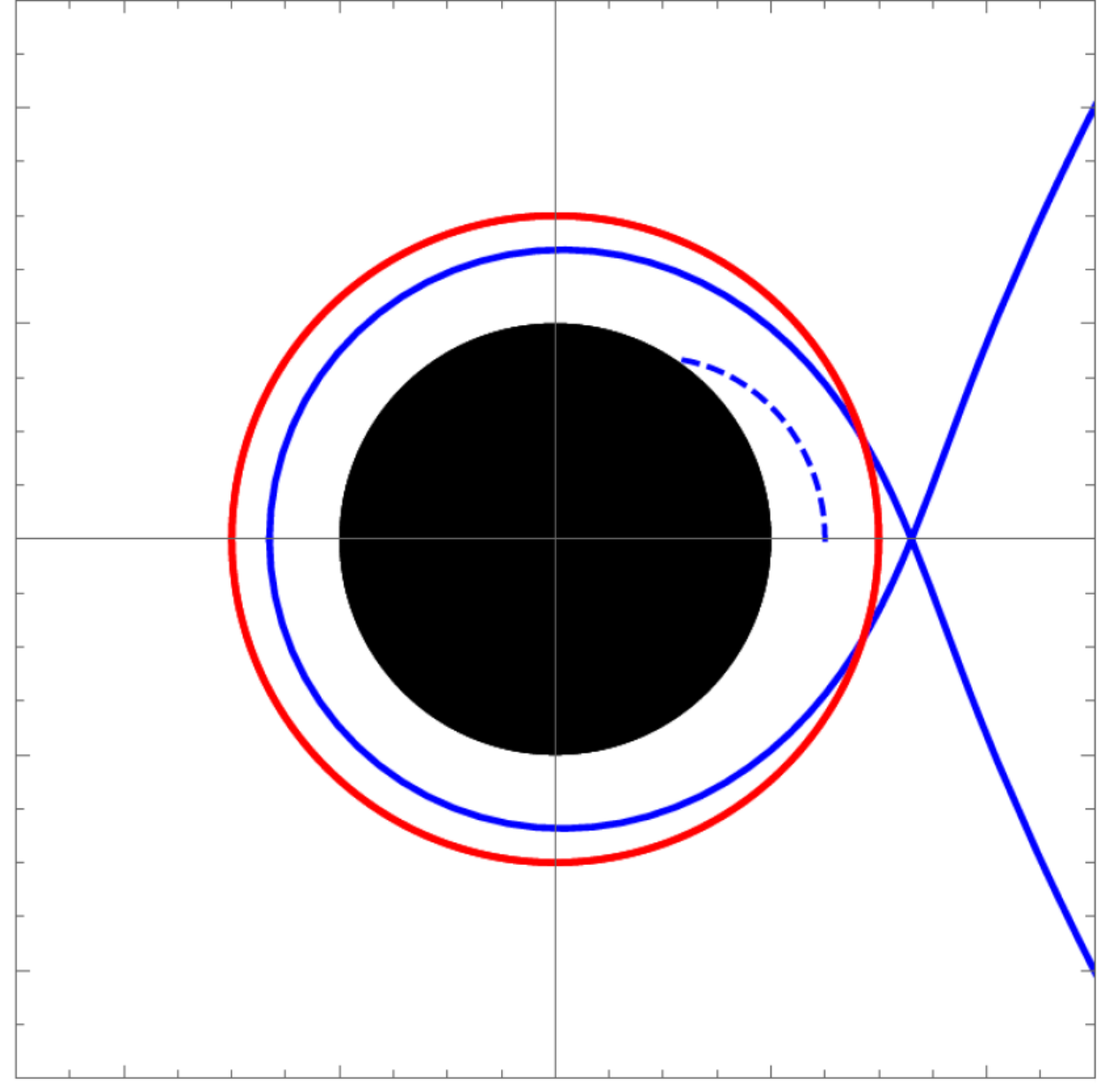}
\caption{Null geodesics around a dilatonic black hole with $ M = 1 $ and  $q=0.3$ are depicted. The black region represents the black hole horizon.}
  \label{fig:36}
\end{figure}

In this section, we will determine the absorption differential cross-section working within the short-wavelength approximation $(\lambda\ll r_{h})$, so it is sufficient to visualize plane waves as rays associated with the null geodesics \cite{huang2020scattering}. To do so, we begin by exploring the behavior of null geodesic in terms of the impact parameter. Our first task is to obtain the critical impact parameter as the distance within photons is captured by the compact object \cite{book0}. As usual,  the null geodesics for the metric (\ref{EMD}) are described by 
\begin{equation}\label{Eq:Lagrangiangeodesic}
g_{\mu\nu}\dot{x}^\mu\dot{x}^\nu=0,
\end{equation}
where $ \dot {x} ^ \mu = dx ^ \mu / d \lambda $ is the four-velocity and $ \lambda $ stands for the affine parameter.  Since the static spacetime has two Killing vectors, $\partial_t$ and $\partial_\phi$, this implies the existence of two conserved quantities, namely the energy ($E=f(r)dt/d\lambda$) and angular momentum ($L=r^2\sin^2 \theta d\varphi/d\lambda$).  A full analysis of the geodesics for dilatonic black holes was reported in \cite{gonzalez2018motion}.   Replacing the metric (\ref{EMD}) in (\ref{Eq:Lagrangiangeodesic}) for null geodesics in the equatorial plane,
we arrive at an effective problem with a radial equation and  angular velocity given by
\begin{eqnarray}\label{gr}
	\dot r^2 &=& E^2-\frac{L^2}{r^2}f,\nonumber\\
	\Omega&=&\frac{\dot{\varphi}}{\dot{ t}}= \frac{L}{E}\frac{f}{r^2},
\end{eqnarray}
so that the orbital equation can be recast as
\begin{eqnarray}\label{Eq:orbiteq}
	\left(\frac{du}{d\varphi}\right)^2&=&h(u),\label{Eq:orbiteq1}\\
	h(u)&=&\left(1-2q^2u\right)^2\left[\frac{1}{\bar b^2}-\left(\frac{1-2u}{1-2q^2u}\right)u^2\right],\nonumber\label{Eq:orbiteq2}
\end{eqnarray}
where $ u =M /r $,  $b=L/E $,  and $\bar b = M/b $ is the dimensionless inverse impact parameter. The critical radius, $ u_c =M/r_c $, can be found by solving $ h '(u) = 0 $. We then substitute the value of $u_c$ into Eq.(\ref{Eq:orbiteq}) and solve $h(u_c)=0$ to find the possible values of $b_c$  corresponding to a non-deflected geodesic. Applying this procedure we end up  with the following critical value:
\begin{equation}\label{Eq:criticalimpactparam}
	\frac{b_c}{M}=\sqrt{\frac{27-q^4-18 q^2+\left(9-q^2\right)\sqrt{\left(1-q^2\right)\left(9-q^2\right)}}{2}},
\end{equation}
where $q^2=Q^2/(2M^2)$ represents the charge-to-mass ratio.A photon will be scattered off for $b> b_c$ otherwise, it will be absorbed by the black hole [see Fig. (\ref{fig:36})].  
In fact,  the critical impact parameter can be obtained by taking into account a physical  situation where the  photon coming from the infinity is  trapped in an unstable circular orbit with radius $r_c$ around the black hole;  this case is achieved if the following conditions are met (\ref{gr}):1-$E^2/L^2=f_c/(g_{c}r^2_c)$ , 2-$2f_cg'_{c}=r_c g_{c} f'_{c}$,  and 3-$2f_c<r^2_c (f''_c/g''_c)$. Consequently,  the photon, which is scattered by the compact object and goes back to infinity, has an angle of deflection\cite{huang2020scattering} given by
\begin{equation}\label{nfs}
	\theta(b)=2\int_{0}^{u_0}\frac{du}{\sqrt{h(u)}}-\pi.
\end{equation}
It is possible to solve the above expression analytically for $ \theta(b) $ in the weak-field limit ($ b \gg M $),
\begin{equation}\label{Eq:wdef}
	\theta\approx\frac{4M}{b}+ \frac{\wp{(q)}M^2}{b^2},
\end{equation}
where $\wp{(q)}=(3\pi/4)(5-4q^2-\frac{4}{3}q^4)$. For $q=0$, we recover the well-known Einstein formula for the deflection of light, $\theta\approx 4M/b$ \cite{book0}, at the leading order within the weak-field approximation. However, this expression is different from  the one obtained for the Reissner--Nordstr\"om black hole \cite{crispino2009scattering}; where   $\wp{(q)}$ becomes  a quadratic polynomial at order $(M/b)^2$.

Having obtained the critical impact parameter for the deflected null geodesics, we are in a position to calculate  the differential cross-section \cite{futte}, \cite{crispino2009scattering}, which reads   
\begin{equation}\label{Eq:csrc}
	\frac{d\sigma}{d\Omega}=\frac{b}{\sin\theta}\bigg|\frac{db}{d\theta}\bigg|,
\end{equation}
where $\theta$ stands for the scattering angle \cite{crispino2009scattering}. To obtain (\ref{Eq:csrc})  we make use of the approximated deflection angle (\ref{Eq:wdef}). The classical differential cross-section for small angles becomes
\begin{eqnarray}
	\frac{d\sigma}{d\Omega}\simeq \frac{16M^2}{\theta^4}+\frac{M^2}{\theta^3}\wp{(q)}.
\end{eqnarray}
The dominant term in the classical scattering formula neither depends on $q$ nor the deflection angle. The situation also happened with the Reissner--Nordstr\"om  metric. Once again, the comparison with the Reissner--Nordstr\"om black hole leads to a mismatch even if the quartic term is neglected in the above formula. We conclude that the null geodesics are less sensible to $q$  for small $\theta$   provided they only see a black hole with mass $M$ \cite{crispino2009scattering}. Instead, the null geodesics notice a charged black hole with mass $M$ and charge $Q$ at intermediate deflection angles.

\begin{figure}
		\includegraphics[width=1.5in]{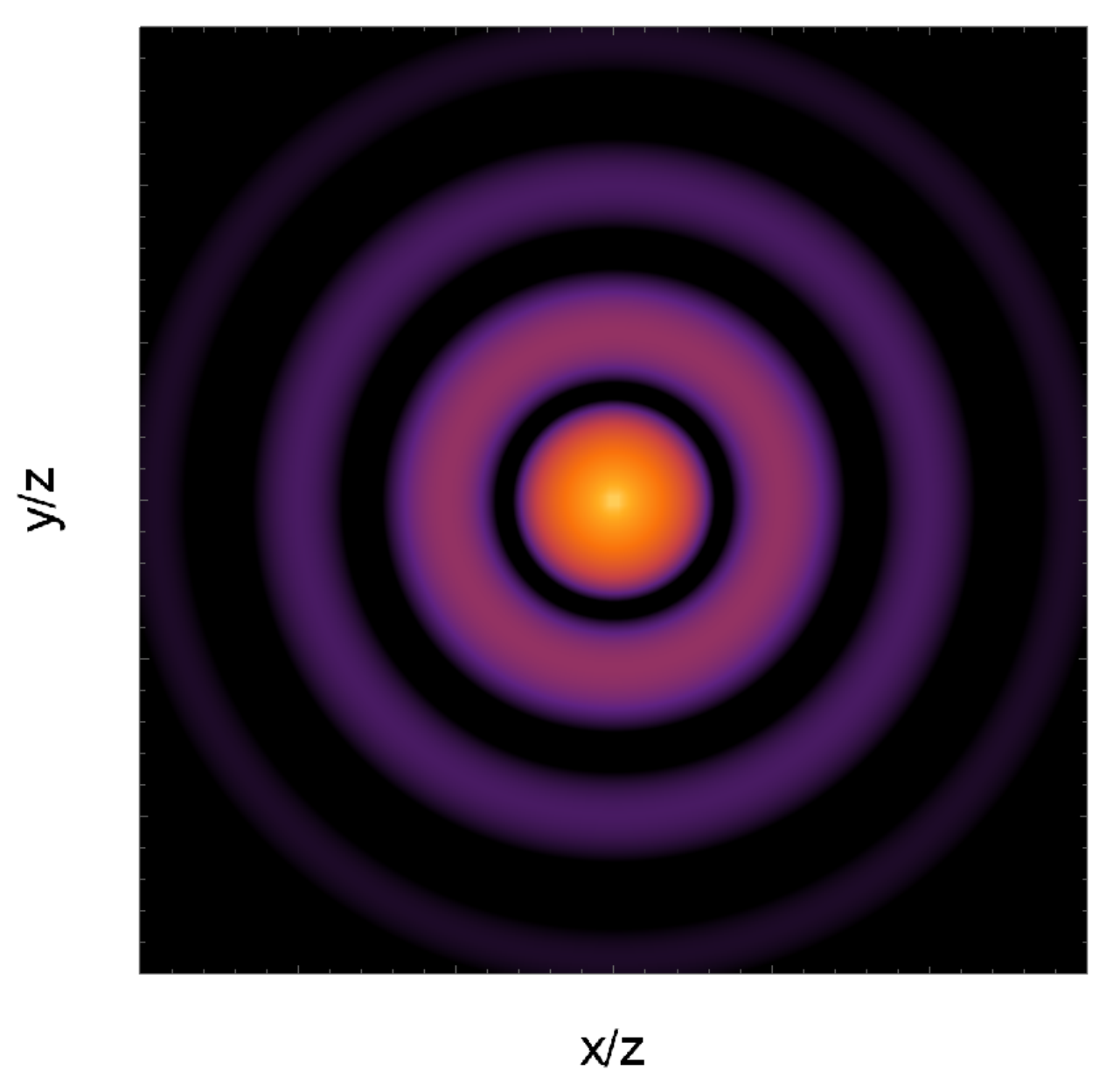}
		\includegraphics[width=1.5in]{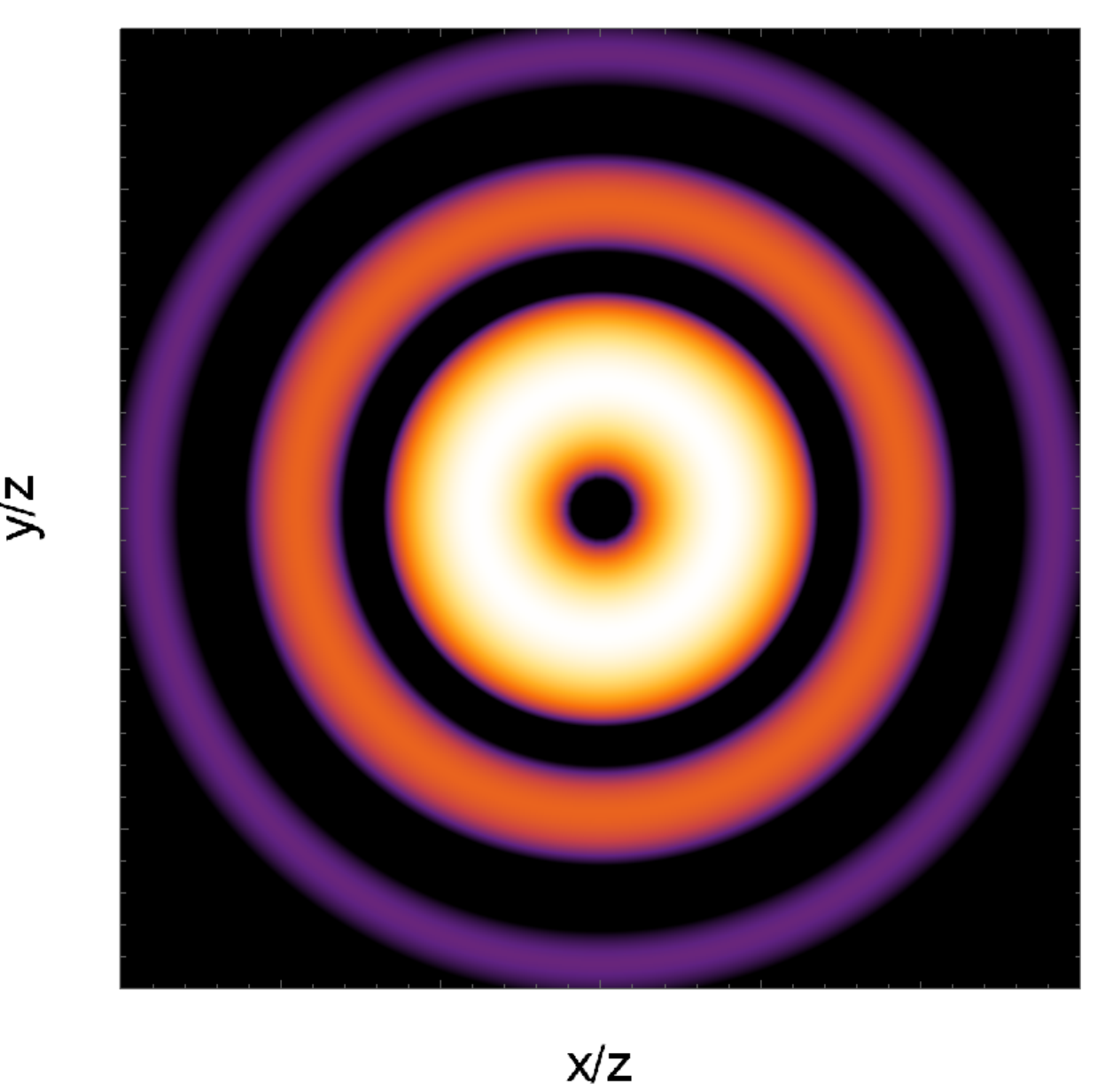}
		\includegraphics[width=1.5in]{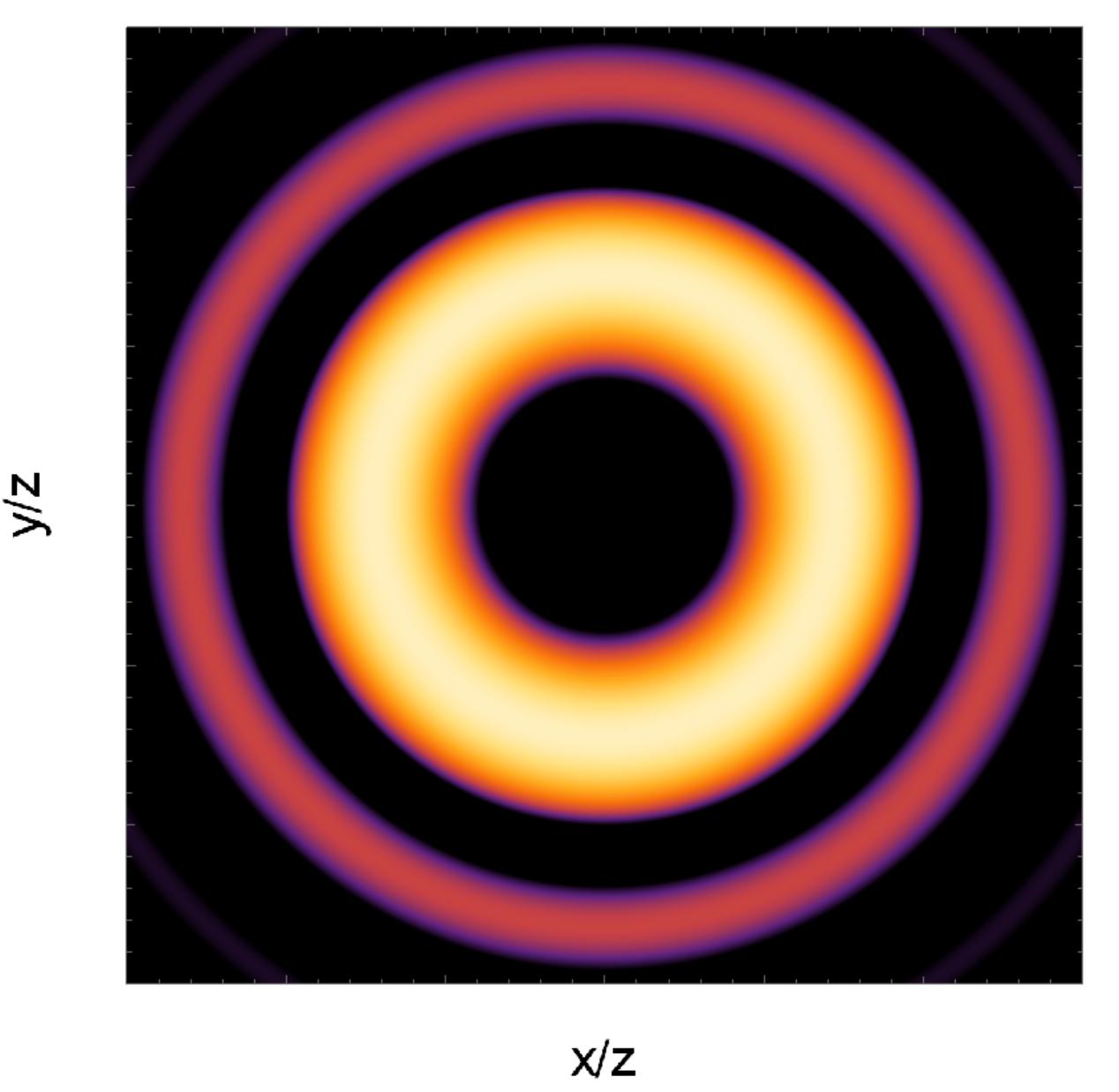}
	\caption{The interference pattern produced by the superposition of waves associated with the glory effect is illustrated--see Eq.  (\ref{eq:gloria})--. The  spin parameters correspond to   $s=\{0, 1, 2\}$ and  the decoupling parameter is fixed at $M\omega=15$.}\label{bweg}
\end{figure}

\begin{figure}
	\includegraphics[width=3.2in]{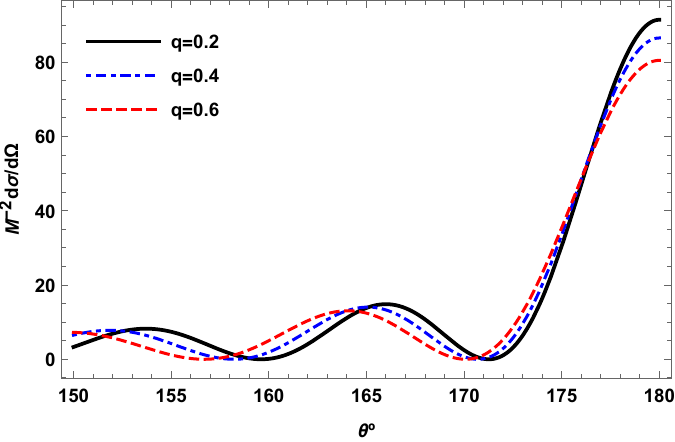}
	\caption{Differential cross-section of the dilatonic black hole with $ M \omega = 3 $ taking into account the  glory effect is depicted}
	\label{fig:gloria}
\end{figure}

\begin{figure}
	\includegraphics[width=3.2in]{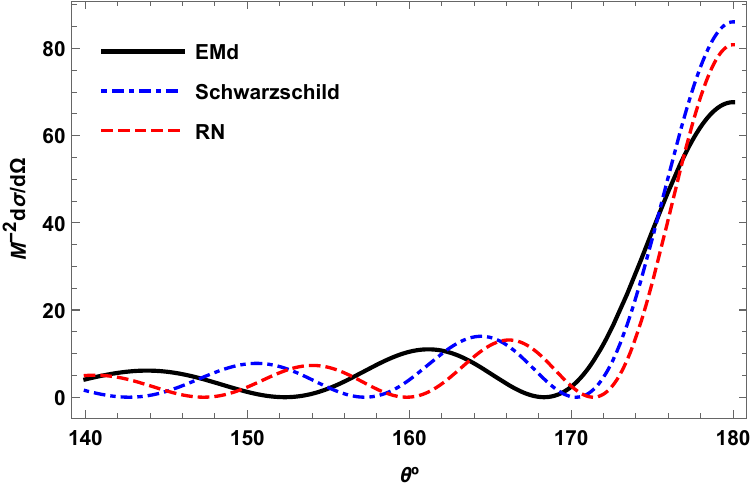}
	\caption{Differential scattering cross-sections for the  Schwarzschild,  the Reissner-Nordstr\"om, and the dilatonic charged black holes with $q=0.3$ and $M\omega=3$ are displayed. The glory effect is more relevant for the chargeless black hole.}\label{fig:a_c3}
\end{figure}

Since part of the incoming wave is scattered off by the black hole,  at some point along the line-of-sight,  it is expected that the outgoing wave interferes with the incoming wave,  generating a pattern of bright fringes around a black center \cite{matzner1985glory}. This mechanism is known as the glory effect and can be used to extract physical properties of the black hole near the horizon \cite{matzner1985glory}, \cite{crispino2009scattering}.  A complete derivation of the differential cross-section can be obtained by using a path integral approach \cite{matzner1985glory}, and it reads: 
\begin{eqnarray}\label{eq:gloria}
	\frac{d\sigma}{d\Omega}\simeq 2\pi\omega b^2_g\Big|\frac{db}{d\theta}\Big|_{\theta\simeq \pi}[J_{2s}(\omega b_g\sin\theta)]^2,
\end{eqnarray}
where $ b_g $ is the impact parameter for the geodesics that undergo a $\pi$ deflection and $ J_{2s} (x) $ is the Bessel function of first type whereas its order $ 2s $ indicates the particle's spin, $s$. There are several values of $ b_g $ corresponding to the number of times the photon has completed a loop around the black hole, i.e., for each  angle $ \theta = \pi + 2 \pi n $, where $n$ is an integer. This approximation remains valid as long as the high-frequency regime ($M \omega \gg 1 $) holds. Fig.(\ref{bweg}) illustrates a typical profile describing the backward glory effect as detected by a distant observer from the charged black hole. For zero spin, constructive interference happens for  $ \theta = \pi$, which a white center represents. This feature is also noted in the Schwarzschild black hole \cite{matzner1985glory}. The case of an electromagnetic wave impinging upon a black hole was explored in detail by  Crispino, Dolan, and Oliveira \cite{crispino2009electromagnetic}. By doing so, they obtained the complete numerical differential cross-section for electromagnetic waves, compared to the approximated version calculated for the glory effect  \cite{crispino2009electromagnetic}.
In Fig. (\ref{fig:gloria}), we show the differential cross-section associated with the glory effect of a charged dilatonic black hole which is asymptotically flat within the EMd model. For different values of the dimensionless charge, $q$, the pattern corresponds to an oscillatory function which increases its amplitudes as the angle reaches the critical value $\theta = \pi$.  The exhibited cases reflect a series of oscillations produced by the interference of waves in opposite directions very close to each other. Depending on the value of $n$, the pattern exhibits a bright or black fringe—smaller values of $q$ lead to larger amplitudes.   The comparison of the glory effect among the Schwarzschild black hole, Reissner--Nordstr\"om metric and dilatonic black holes are depicted in Fig.(\ref{fig:a_c3}).   The Reissner--Nordstr\"om  and dilatonic black holes have very near curves, and they both have a smaller differential cross-section concerning the Schwarzschild black hole, especially the amplitude peak around $\theta = \pi$ \footnote{To be able to compare our findings with the literature, we plot differential cross-section in terms of the deflection angle $\theta[\rm{deg}]$ whereas in most of Sec. III with work with radians.}.
\begin{table}
	\centering
	\begin{tabular}{c|c|c}
		$q$ & $b_g/M$ & $\frac{b_g^2}{M^2}\Big|\frac{db}{d\theta}\Big|_{\theta\simeq\pi}$\\
		\hline  \hline  
		0&  5.36 & 4.89 \\
		\hline  
		0.2 & 5.28 & 4.85 \\
		\hline  
		0.4 & 4.96 & 4.59\\
		\hline  
		0.6 & 4.65 & 4.27\\
		\hline  
		0.8 & 3.96 & 3.59\\
		\hline  
		1 & 2.06 & 2.21\\
	\end{tabular}
	\caption{The impact parameter associated with the glory effect is listed for different values of $q$.}
	\label{tab:tab2}
\end{table}
In order to arrive at an order of magnitude for the differential cross-section in the case of  glory effect (\ref{eq:gloria}), we proceed to determine first  the  impact parameter  for the glory effect called $ b_g $ and the rate $ \frac{db}{d \theta} $. We start by  fixing $M$ and $Q$, or equivalently $q$, and then we obtain $b_g$  by solving   numerically the condition $ \theta (b_{g}) = \pi $ with the help of   (\ref{nfs}). 
In table (\ref{tab:tab2}), we show different values of  $ b_g $ and $ \frac {db} {d \theta} $ for several values of $q$. Larger values of $q$ yield to smaller value of $b_g/M$, and consequently, smaller value of dimensionless cross-section.  Further, it is possible to obtain an approximated expression for the deflection angle which leads to a semi-analytic formula for  estimating the magnitude  and width of the glory peak for the dilatonic charged black hole (see Appendix B).

\section{Massless Scalar Field}
\begin{figure}
	\includegraphics[width=3.1in]{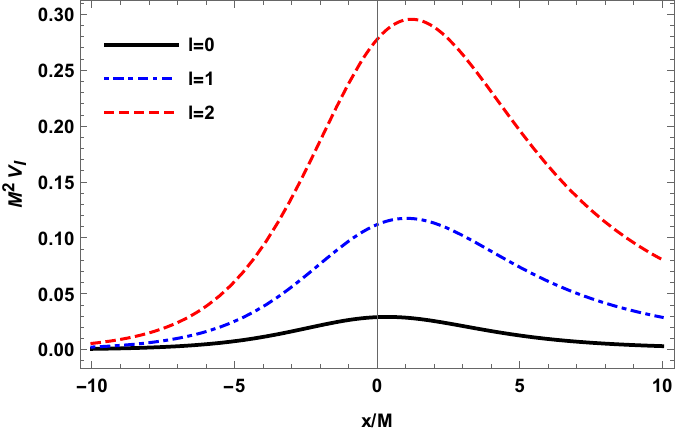}
	\caption{The effective potential for the massless scalar field in the dilatonic charged black hole background as a function of the tortoise coordinate. The charge-to-mass ratio is $q=0.5$  and  the angular momentum , $\ell $, varies. }\label{fig:11}
\end{figure}

\begin{figure}
	\includegraphics[width=3.1in]{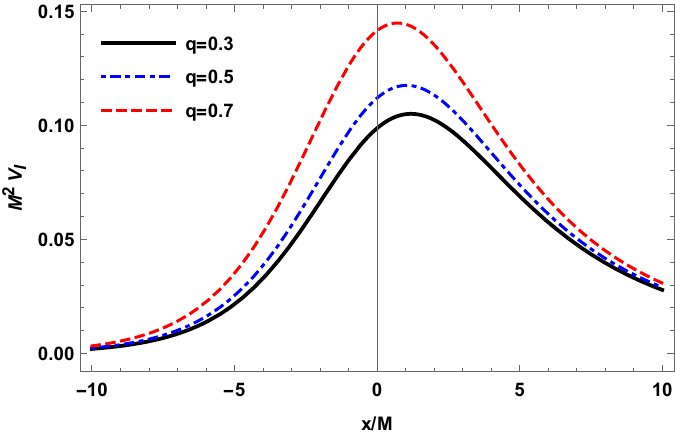}
	\caption{The effective potential associated with a massless
scalar field living  in the dilatonic charged black hole  background is displayed case as a function of the tortoise coordinate. The angular momentum is  fixed at $\ell=1$ whereas the charge-to-mass ratio varies from $ q=0.3$ to $q=0.7$.}\label{fig:12} 
\end{figure}

So far, we have been working within the short-wavelength approximation $(\lambda\ll r_{h})$ because it was useful for visualizing the behavior of waves as rays (geometric optic), represented by null geodesics. This section will apply a different approach to calculate the differential cross-section for the scattering/absorption process. This requires to employ the so-called partial wave method for intermediate wavelengths $(\lambda \simeq  r_{h})$.  Hence,  we consider an incoming wave from infinity, which is partially transmitted through the event horizon and partially scattered to infinity  \cite{crispino2009electromagnetic}, \cite{leite2019scattering}. To study that, we write down  the Klein-Gordon equation of a massless  scalar field propagating on the dilatonic black hole background,
\begin{equation}\label{CE}
	\Box \Phi=\frac{1}{\sqrt{|g|}}\partial_{\mu}\Big(\sqrt{|g|}g^{\mu\nu}\partial_{\nu} \Phi \Big)=0,
\end{equation}
where $\Box$ is the Laplace-Beltrami operator on curved spacetime $\mathcal{M}=\mathbb{R}\times \Sigma_{3}$ and $g$ is the metric determinant. It is convenient to mention that the solutions of (\ref{CE})  have associated an inner Klein-Gordon product defined by \cite{book1},\cite{book3}
\begin{equation}\label{pkg}
	\Omega(\Phi, \Psi)=\int_{\Sigma_{3}}d\Sigma_{\mu}J^{\mu},
	\end{equation}
where the  invariant volume form is defined as  $d\Sigma_{\mu}=\sqrt{|h|} n_{\mu}\Sigma_{3}dx^3$, $h$ is the determinant of the induced metric on the Cauchy surface $\Sigma_{3}$, and $n_{\mu}$ a unit vector time-like. There is a  conserved current \cite{book3} 
\begin{equation}\label{current}
	J_{\mu}=\frac{i}{2}[\Phi^{*}\nabla_{\mu}\Psi-\Psi^{*}\nabla_{\mu}\Phi],
	\end{equation}
its components include the number density of the field, $J^{t}$,  along with the current densities, $J^{i}$. The general solution to equation (\ref{CE}) for the spherically symmetric dilatonic charged black hole background can be found using the separation of variables method; that is, we propose the following ansatz 
\begin{equation}\label{separation0}
	\Phi=\sum^{\infty}_{\ell=0}\sum_{|m|\leq \ell}\int_{\mathbb{R}}{\frac{\psi_{\ell m \omega}  (r)}{r\sqrt{g(r)}}Y_{lm}(\theta,\varphi)e^{-i\omega t}d\omega},
\end{equation}
where $ Y ^ m_l (\theta, \phi): \mathbb{S}^2\rightarrow\mathbb{C} $ are the spherical harmonics and $ \omega $ is the frequency. The spacetime variables cover the  exterior of black hole, namely, $ 2M <r <\infty $, $ 0 <t <\infty $, $ 0 \leq \theta \leq 2 \pi $ and $ 0 \leq \phi \leq \pi$. Substituting (\ref{separation0}) in (\ref{CE}) we obtain the master equation for the radial solution $\psi_{\ell m \omega}$ 
\begin{eqnarray}\label{eq:EMD}
	&&f(r)\Big(f(r)\psi'_{\ell m \omega}\Big)'+(\omega^2-V(r))\psi_{\ell m \omega}=0,
\end{eqnarray}
where the prime stands for derivatived w.r.t. the radial coordinate.  We introduce a tortoise coordinate defined by the relation $ dx =f (r) ^ {- 1} dr $, so that the map between these radial coordinates reads
\begin{eqnarray}\label{tot}
	x = \int \frac{dr}{f(r)} =r+2M\ln\Big|\frac{r}{2M}-1\Big|.
\end{eqnarray}
The tortoise coordinate covers the real line interval; that is, the boundaries of the radial interval correspond to the following asymptotics:  $x(2M)\rightarrow -\infty$ and  $x(\infty)\rightarrow +\infty$. 
Replacing (\ref{tot}) in (\ref{eq:EMD}), we arrive at a more appealing master equation which resembles the Schr\"odinger one for stationary states  with energy $E=\omega^2$:
\begin{equation}\label{Eq:theradialeqpsi}
	-\frac{d^2}{dx^2}\psi_{\ell m \omega}(x)+V(r)\psi_{\ell m \omega}(x)=\omega^2\psi_{\ell m \omega}(x).
\end{equation}
The effective potential  is given by
\begin{eqnarray}\label{eq:potencial}
	M^2 V(\hat{r})&=&\Big(1-\frac{2}{\hat{r}}\Big) \Big[\frac{\ell (\ell+1)}{\hat{r}^2-2q^2 \hat{r}}-\frac{q^4 (1-\frac{2}{\hat{r}})}{ \hat{r}^4 \left(1 -\frac{2q^2}{\hat{r}}\right)^2}\nonumber\\&+&\frac{2q^2}{\hat{r}^4 \left(1-\frac{2q^2}{\hat{r}}\right)}+\frac{2}{\hat{r}^3}\Big],
\end{eqnarray}
where the radial variable with a hat denotes $\hat{r}=r/M$. As the radial coordinate approaches at the horizon, the effective potential (\ref{eq:potencial}) vanishes, and the same happens at infinity. In Fig.(\ref{fig:11}),  we show that the effective potential exhibits a bell-shaped curve in the tortoise coordinate for different values of the angular momentum and a fixed charge-to-mass ratio $q=0.5$.  It is clear from Fig.(\ref{fig:12})  that for a given $\ell$, the height of the peak increases as $q$ takes larger values.

\subsection{Absorption cross-section}
Since we are dealing with a boundary value problem for the Schr\"odinger equation (\ref{Eq:theradialeqpsi}) with an effective potential (\ref{eq:potencial}), a key point is the selection of the boundary conditions for the radial solution $\psi_{\ell m \omega}(x)$ which are compatible with the problem at hand. Close to the horizon, $r = 2M$, the effective potential  vanishes, and the boundary condition is  an ingoing wave,  
\begin{eqnarray}\label{eq:2.11b}
	\psi_{\omega \ell}\simeq T_{\omega \ell}e^{-i\omega x},
\end{eqnarray}
where $ |T_{\omega \ell}| $ is called the transmission coefficient.The index $m$ is omitted for simplicity.  As we consider the opposite limit, the dominant contribution of the effective potential is the angular barrier, $\ell(\ell+1)/r^2$. Therefore, (\ref{eq:EMD}) can be written as
\begin{eqnarray}
	\Big[r^2\frac{d^2}{dr^2}+r\frac{d}{dr}+\omega^2r^2-\Big(\ell+\frac{1}{2}\Big)^2\Big]\Big(\frac{\psi_{\omega \ell}}{\sqrt{\omega r}}\Big)=0,
\end{eqnarray}
which coincides with the  Bessel equation of order $(\ell+1/2)$, whose solution is given  in terms of  third-order spherical Bessel functions $h_l(\omega r)$,
\begin{eqnarray}\label{eq:35}
	\psi_{\omega \ell}\simeq (\omega r)\Big[(-i)^{i+1}h_l^*(\omega r)+i^{\ell+1}R_{\omega \ell}h_l(\omega r)\Big].~~
\end{eqnarray}
The factor $|R _{\omega \ell}|$ represents  the reflection coefficient. The relationship between the transmission and reflection coefficients is the usual one: $|R_{\omega \ell}|^2+|T_{\omega \ell}|^2=1$.  Using that, the leading order of the spherical Bessel function is \cite{olver2010nist}-\cite{abramowitz1988handbook},
\begin{eqnarray}
	h_l(\omega r)\simeq (-i)^{\ell+1}\frac{e^{i\omega r}}{\omega r}, 
\end{eqnarray}
then the asymptotic boundary condition becomes
\begin{eqnarray}\label{eq:t2}
	\psi_{\omega \ell}(x) = \Big(R_{\omega \ell}e^{i\omega x}+ e^{-i\omega x}\Big)
\end{eqnarray}
as long as we are in the region $\omega x \geq \ell(\ell+1)/2$. Notice that the above approach for imposing the proper boundary conditions is consistent with scattering a massless scalar field with a reflective boundary condition at infinity and the one-way membrane at the horizon. In other words, our approach is consistent with previous  works on this topic \cite{crispino2009scattering},  \cite{macedo2016absorption},  \cite{del19}.

It is instructive, at this point, to determine the behavior of the radial modes in the limit of low-frequency, $ \omega = 0$, also known as zero modes. Replacing the latter condition in Eq. (\ref{eq:EMD}), the general solution can be recast as  \cite{olver2010nist}-\cite{abramowitz1988handbook}
\begin{eqnarray}\label{szm}
	\psi_{\omega \ell}(z) = D^I_{\omega \ell} P_l(z)+D^{II}_{\omega \ell} Q_l(z),
\end{eqnarray}
where the new variable is 
\begin{eqnarray}
	z = \frac{ 1+ q^2-\frac{r}{M}}{q^2-1}
\end{eqnarray}
In Eq. (\ref{szm}), $ P_\ell (z) $ and $ Q_\ell (z) $ are Legendre functions of first and second type \cite{olver2010nist}-\cite{abramowitz1988handbook}, respectively. At the  event horizon ($ z = 1 $) the asymptotic behavior of the Legendre functions are  characterized by $Q_\ell (1)\to \infty $ and $ P_\ell (1) = 1 $ \cite{olver2010nist}. The radial modes at low-frequency that come from infinity must represent the main contribution in the region and a subdominant fraction at the horizon; this means that the modes coming from the infinity are given by the $ P_\ell (z)$  function,  
 \begin{equation}\label{eq:96}
	\psi_{\omega \ell}(z) = D^I_{\omega \ell}r\sqrt{2g(r)}P_\ell(z).
\end{equation}
To fully determine this solution, we must fix the constant $ D^I_{\omega \ell} $. We compare  the asymptotic solution obtained before (\ref{eq:35}) with the above solution at low-frequencies (\ref{eq:96}).   We are gluing the solution in the intermediate region with the boundary condition at infinity. To this end, we use the asymptotic form of the Legendre polynomial for a large argument ($|z|\gg 1$),
\begin{eqnarray}
	P_\ell(z)\simeq \frac{(2\ell)!}{2^\ell(\ell!)^2}z^\ell,
\end{eqnarray}
in order to obtain the radial modes at this level of approximation,
\begin{eqnarray}\label{eq:96b}
\psi_{\omega \ell}(r) \simeq D^I_{\omega \ell} \frac{(2\ell)!}{2^{(\ell+1)}(\ell!)^2}\Big(\frac{r/M}{1-q^2}\Big)^{\ell+1}.
\end{eqnarray}
Expanding  (\ref{eq:35}) around $\omega=0$--or $\omega r\ll 1$--,  we arrive at the following asymptotic form  
\begin{eqnarray}\nonumber
	\psi_{\omega \ell}&\simeq& [(-i)^{1+\ell}+i^{1+\ell}R_{\omega \ell}]\frac{2^{\ell}\ell!}{(2\ell+1)!} (\omega r)^{(\ell+1)}\\,
	&&+ A_{\omega \ell}i[(-i)^{1+\ell}-i^{1+\ell}R_{\omega \ell}]\frac{(2\ell)!}{2^{\ell}\ell!}(\omega r)^{-\ell}.~~\label{eq:35b}
\end{eqnarray}
Putting together (\ref{eq:96b})  with (\ref{eq:35b}), we arrive at the following coeffficient  
\begin{eqnarray}\nonumber
	 D^I_{\omega \ell}&=&(-i)^{(1+\ell)} [1+(-1)^{(1+\ell)}R_{\omega \ell}]\times\\
	  &&\frac{\ell!^3 2^{2\ell}} {(2\ell)!(2\ell+1)!}(\omega M)^{(\ell+1)}(1-q^2)^{(\ell+1)}\label{fc1}
  \end{eqnarray}
which leads to a reflection coefficient $R_{\omega \ell}=(-1)^{(1+\ell)} +\mathcal{O}(\omega)$ at low-frequency. As a result, the radial modes are almost fully reflected by the effective potential (\ref{eq:potencial}). For $q=0$, we return to the same result for the Schwarzschild metric  \cite{crispino2009scattering}. Inserting (\ref{fc1}) in (\ref{eq:96}) leads to the radial modes at low-frequency. With the latter solution at hand, we can calculate the absorption cross-section within this limit; in fact, the cross-section will mostly depend on the frequency of the incoming wave and the black hole parameters.   Furthermore, close the black hole ($r \simeq 2M$) in the limit of low-frequency, the scalar field can be written as 
\begin{eqnarray}\label{eq:96c}
\psi_{\omega \ell}\simeq (-i)^{(1+\ell)} \frac{\ell! 2^{(\ell+2)}} {(2\ell+1)!}{[\omega M\sqrt{1-q^2}}]^{(\ell+1)}.~~
\end{eqnarray}
 In order to compute the cross-section with the above solution  (\ref{eq:96})-(\ref{fc1}) we must provide a  general definition. As an incoming  wave approaches a target, the probability of an absorption process happening is indicated by the absorption cross-section \cite{futte}. The absorption cross-section is defined as
\begin{eqnarray}\label{eq:5.1}
	\sigma = -\frac{\mathcal{F}}{J_i}
\end{eqnarray}
where $\mathcal{F}$ stands for  the total absorbed flow.  The ingoing flux is calculated by integrating the radial current (\ref{current}) through a closed sphere $S_{r}$,
\begin{eqnarray}\label{eq:5.10}
	\mathcal{F} =\int_{S_{r}} J^r r^2d\Omega.
\end{eqnarray}
The incident current associated to  a plane-wave is given by $J^{z}=\omega$. Combining the incoming current with (\ref{eq:5.10}) along  with $|R_{\omega \ell}|^2+|T_{\omega \ell}|^2=1$ \cite{futte},  we obtain that the partial absorption 
cross-section reads
\begin{eqnarray}\label{eq:numerical}
	\sigma_{\ell} = \frac{\pi}{\omega^2}(2\ell+1)|T_{\omega \ell}|^2.
\end{eqnarray}
Eq. (\ref{eq:numerical}) tells us how to compute the partial cross-section whether we are interested in the low-frequency limit or for a different value of $\omega$. The above expression can also be used to compute $\sigma_{\ell}$ numerically and get the total  cross-section by summing over the different $\ell$-terms,
\begin{eqnarray}\label{acs}
	\sigma_{abs} = \sum_{\ell=0}^\infty 	\sigma_{\ell}.
\end{eqnarray}

Eq. (\ref{eq:2.11b})  can be recast as $\psi_{\omega \ell} = T_{\omega \ell} (1+ \mathcal{O}(\omega))$. Comparing the latter expression with (\ref{eq:96c}), we arrive at the transmission coefficient at low-frequency,
\begin{eqnarray}\label{tcbn}
|T_{\omega \ell}|=\frac{\ell! 2^{(\ell+2)}} {(2\ell+1)!}{[\omega M\sqrt{1-q^2}]}^{(\ell+1)}.	
\end{eqnarray}
Replacing (\ref{tcbn}) in (\ref{eq:numerical}) and keeping the lowest order in $\omega$, which corresponds to  $\ell=0$, the absorption cross-section reduces to
\begin{eqnarray}\label{tcbn2}
 \sigma= 16\pi M^2 (1-q^2),	
\end{eqnarray}
for which it follows that is proportional to the area of the spherical horizon, $\sigma= 4\pi r^2_{\rm{h}}g(r_{\rm{h}})$, as we could expect.  This result is consistent with the absorption cross-section for a massless scalar field in the Schwarzschild geometry with $q=0$

\subsection{Numerical results for absorption}
\begin{figure}
	\includegraphics[width=3.2in]{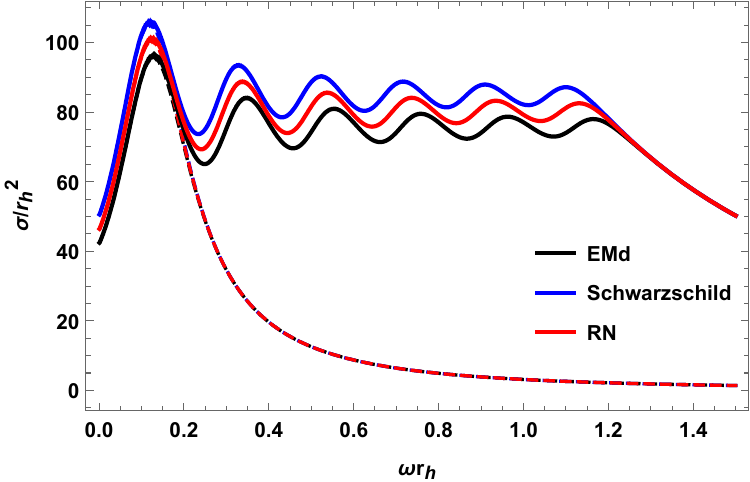}
	\caption{Total absorption cross-sections for EMd, Schwarzschild, and RN black holes are shown. The blue line is  Schwarzschild black hole, and the red line is the  RN case, whereas the black line denotes the dilatonic charged black hole. The angular momentum varies from $\ell=0$ to $\ell=20$. The dashed line denotes the lowest multipole,  $\ell=0$.}
	\label{fig:abs_comp}
\end{figure}

In this section, we are going to present our main findings of the absorption cross-section obtained with the numerical implementation of the partial wave method described above for \textit{arbitrary frequency}. To numerically recover the partial and total absorption cross-section  over the interval $x \in (-\infty, \infty)$ we must  impose at the horizon and  infinity the following boundary conditions 
\begin{equation}\label{bcn}
	\psi_{\omega \ell} = \left\{
        \begin{array}{ll}
             T_{\omega \ell} e^{-i\omega x}, & \quad  x \rightarrow -\infty  \\
            e^{-i\omega x} +  R_{\omega \ell} e^{+i\omega x},  & \quad x \rightarrow +\infty
        \end{array}
    \right.
\end{equation}
and simultaneously solve the master equation (\ref{Eq:theradialeqpsi}) subject to the latter boundary condition (\ref{bcn}). By taking the modulus of (\ref{bcn})and its derivative, we can show that the transmission coefficient becomes
\begin{eqnarray}\label{eq:numerica}
	|T_{\omega \ell}|^2 = \Big\{\frac{1}{4}\Big[|\eta_{\omega \ell}|^2+\frac{1}{\omega^2}\Big|\frac{d\eta_{\omega \ell}}{dx}\frac{dx}{dr}\Big|^2_{x=x_\infty}\Big]+\frac{1}{2}\Big\}^{-1},
\end{eqnarray}
where $\eta_{\omega \ell} = e^{-i\omega x}+R_{\omega \ell}e^{i\omega x}$. The transmission coefficient, and hence the absorption cross-section, is entirely determined by (\ref{eq:numerica}), which is solved as we stated before. 
\begin{figure}
	\includegraphics[width=3.1in]{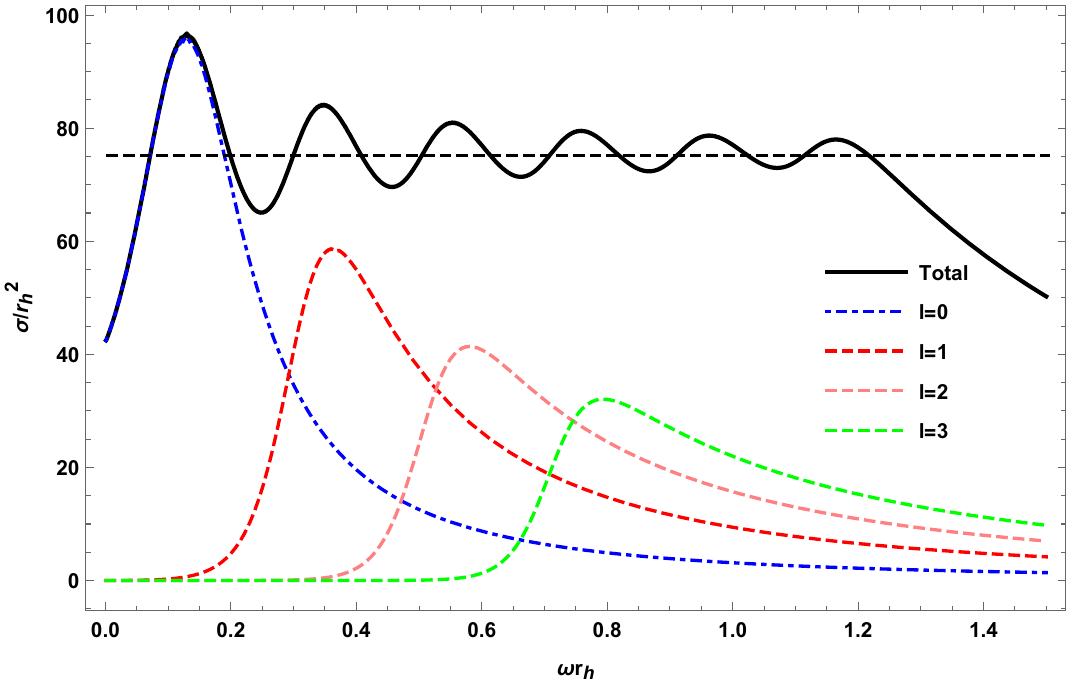}
	\caption{Absorption cross-section for a massless scalar on a dilatonic charged black  hole with mass $M$ and charge $Q$ in terms of $M\omega$.  The dashed line is the geometric cross-section, $\sigma \propto M^2$. The dashed lines show the different partial  contributions  at a given $\ell$ whereas the total cross-section from $\ell=0$ to $\ell=5$ corresponds to the continuous black line.}
	\label{fig:abs}
\end{figure}
Fig.(\ref{fig:abs}) shows the absorption cross-section in terms of the  decoupling parameter $M\omega$. For each value of $\ell$, the cross-section pattern has a main peak and then decays to zero. As a result, the total cross-section ends with a series of peaks and then vanishes for large $M\omega$.The  $\ell=0$ term is the main contribution for small values of $M\omega$ provided the geometric-optic limit suppresses the other partial terms. Notice that each width of the peak is of order $ \mathcal{O}(M\omega)^{-1}$ \cite{sanchez1978absorption}. Having a vanishing cross-section for large values of $\ell$ is easy to understand; the greater the value of $\ell$, the stronger the potential barrier becomes. Besides, Fig.(\ref{fig:abs_comp}) illustrates the comparison among total absorption cross-sections for dilatonic, Schwarzschild, and RN black holes. It turns out that the amplitudes of the first peak differ a little bit,  in the sense that RN and Schwarzschild have almost the same order of magnitudes, but the dilatonic black hole exhibits considerably lower amplitude. Concerning the Schwarzschild case, the peaks associated with the  RN and EMd black holes are shifted to the right. However, all the profiles tend asymptotically to the same value as we stated before. 
\subsection{Scattering cross-section}
In this section, we will compute the scattering cross-section numerically for dilatonic black holes. To do so, let us begin by recalling how the differential cross-section is computed. The total wave function for the Schr\"odinger equation in a scattering problem with a central potential can be written as $\Phi_{t}=\Phi_{0}+\Phi_{s}$, indicating the contribution of a free incoming wave $\Phi_{0}$ along with the scattered wave, $\Phi_{s}$ \cite{futte}. For the sake of simplicity, the ingoing wave has a wavenumber vector along the $z$-axis, $\Phi_{0} \propto e^{i\omega z}$. The free wave term can be written in terms of the spherical Bessel function and the Legendre polynomial as follows \cite{futte}:     
\begin{eqnarray}\label{pwave}
	e^{i\omega z}=\sum^{\infty}_{\ell=0}(2\ell+1)i^{\ell}j_{\ell}(\omega r)P_{\ell}(\cos \theta).
\end{eqnarray}
The spherical Bessel function in (\ref{pwave}) includes an ingoing term and an outgoing spherical wave, so we must isolate the contribution which respects the boundary condition at infinity. Using the asymptotic expansion of $j_{\ell}(\omega r)$ at infinity  \cite{olver2010nist}-\cite{abramowitz1988handbook}, we are able to obtain the asymptotic form for the scattered wave as $\Phi_{s}= f_{\omega}(\theta)e^{i\omega r}/r$, where the scattering amplitude is given by 
\begin{eqnarray}\label{eq:campo_forte}
	f_{\omega}(\theta) = \frac{1}{2i\omega}\sum_{\ell=0}^\infty(2\ell+1)[e^{2i\delta_\ell(\omega)}-1]P_\ell(\cos\theta),
\end{eqnarray}
and the  complex-valued phase shift of the scattered wave is expressed in terms of reflection coefficient, 
\begin{eqnarray}
	e^{2i\delta_\ell(\omega)} =e^{i(\ell+1)\pi}R_{\omega \ell}.
\end{eqnarray}
Using (\ref{eq:campo_forte}) the differential scattering cross-section reads $d\sigma_{s}/d\Omega=|f_{\omega}(\theta)|^2$, so the total scattering section is obtained by integrating over the solid angle \cite{futte}, \cite{crispino2009scattering}, \cite {dolan2009scattering}, namely, the scattering cross-section is the resummation over the partial cross-section, 
\begin{eqnarray} \label{sfs}
	\sigma^{s}= \sum_{\ell \geq 0} \sigma^{s}_{\ell}=\frac{\pi}{\omega^2}\sum^{\infty}_{\ell=0} (2\ell+1)|1-e^{2i\delta_\ell(\omega)}|^2.
\end{eqnarray}
In other words, the main physical traits of the scattering cross-section  is contained in  the scattering amplitude (\ref{eq:campo_forte}) or equivalently in the  phase shift (\ref{sfs}).   Putting together (\ref{acs}) and (\ref{sfs}), we arrive at the total cross-section  $\sigma_{t}=\frac{2\pi}{\omega^2}\sum_{\ell \geq 0}[1-\Re(e^{2i\delta_\ell(\omega)})]$, which yields to the optical theorem,  $\sigma_{t}=\frac{4\pi}{\omega}\Im{f_{\omega}(0)}$ \cite{futte}. In fact, we can determine the  phase shift  within the wave partial method (valid at intermediate scales, $\lambda \simeq r_{h}$) under certain conditions. Within the approximation of large angular momentum, we define the deflection angle as a function of a continuous $\ell$ variable \cite{dolan2009scattering}, \cite{whee}, and its relation with the phase shift becomes
\begin{eqnarray} \label{defa}
	\theta(\ell)=2\frac{d}{d\ell}[\Re{\big(\delta_{\ell}\big)}].
\end{eqnarray}
Considering that the impact parameter can be written as  $b=(\ell + \frac{1}{2})/\omega$, and comparing (\ref{Eq:wdef}) with (\ref{defa}), we pick up an expression for the derivative of the shift phase \footnote{Starting from the condition $b=\frac{\ell + 1/2}{\omega}$ and the weak-field limit condition $b\gg M$, this expression remains valid as long as $\ell\gg M\omega$.}
\begin{eqnarray} \label{defa2}
\frac{d}{d\ell}[\Re{\big(\delta_{\ell}\big)}]\simeq \frac{2M\omega}{(\ell + \frac{1}{2})}+ \wp(q) \frac{(M\omega)^2}{2(\ell + \frac{1}{2})^2},
\end{eqnarray}
where $\wp(q)$ is just a form factor which depends on  the charge-to-mass ratio $q$. Integrating (\ref{defa2}), it yields an estimation of the phase shift within the geometric limit, where the first two  orders are
\begin{eqnarray} \label{defa3}
\delta_{\ell}\simeq 2M\omega\ln (\ell + \frac{1}{2})- \frac{\wp (q)(M\omega)^2}{2 (\ell + \frac{1}{2})}+...  
\end{eqnarray}
The second possibility corresponds to the Born approximation \cite{futte}; however, this approach does not apply to the lowest order in (\ref{defa3}) as was mentioned in \cite{whee2}. Any other contribution in $\delta_{\ell}=\mathcal{O}[(M\omega)^{n}]$ with $n\geq 2$ can be obtained from the phase shift definition, which involves the  Born potential.

\begin{figure}
	\includegraphics[width=3.2in]{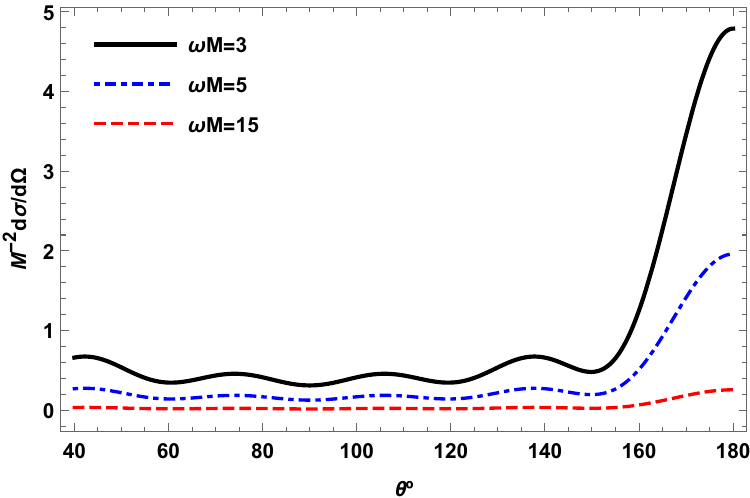}
	\caption{Scattering cross-sections for a massless scalar plane wave with
angular frequency  incident $\omega$ on a dilatonic black hole. Fixing $q=0.5$, the typical profile is shown  for  different values of $M\omega$. On the horizon axis, a degree $\theta$ variable is used. }\label{fig:a_c2}
\end{figure}

\begin{figure}
	\includegraphics[width=3.2in]{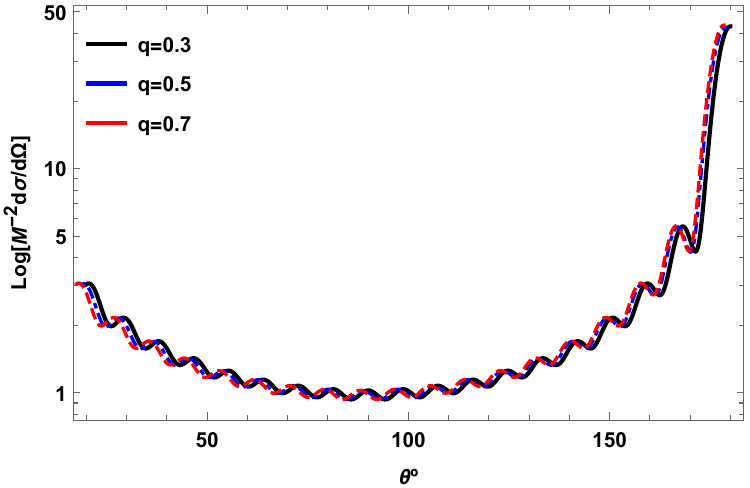}
	\caption{ The logarithmic scattering cross-section as a function of the dispersion angle is depicted for different $q$  with a fixed decoupling parameter, $M\omega=3$.}\label{fig:a_c5}
\end{figure}

As we did previously for the absorption case,  the scattering amplitude can be computed numerically by solving the equation (\ref{Eq:theradialeqpsi}) and extracting the reflection coefficient. We  determined the differential scattering cross-section for different values of the decoupling parameter $M\omega=3$, $M\omega=5$ and $M\omega=15$  at a  given value of $q$ [see Fig.(\ref{fig:a_c2})]. The profile exhibits a series of oscillations whose amplitudes increase as the angle approaches $\pi$, signaling the relevance of the glory effect \cite{crispino2009scattering}. The typical pattern of the differential scattering cross-section, in logarithmic scale, is shown in Fig. (\ref{fig:a_c5}) for different values of $q$ at a given $M\omega$ fixed; increasing the value of $q$  does not change the pattern significantly. For smaller angles, we obtain that the higher the $q$ value, the lower the amplitude, whereas the situation is reversed for larger angles. 

\begin{figure}[h!]
	\centering
		\includegraphics[width=3.0in]{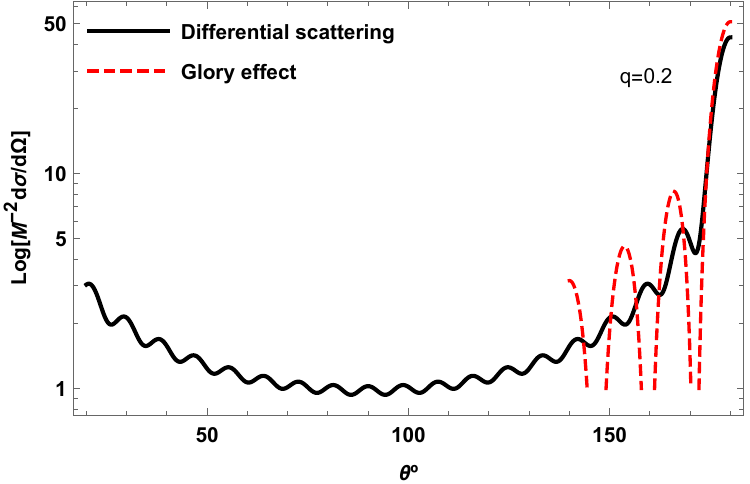}
	\caption{Differential  scattering cross-sections as a function of the dispersion angle  for  $M\omega=15$ in a logarithmic scale. The charge-to-mass ratio is $q=0.2$. Similar patterns are obtained for  $q=0.4$ and $q=0.6$.} \label{fig:a_c6}
\end{figure}
Having performed all the above numerical analysis, we can go even further and compare the scattering cross-section with the corresponding curve obtained under the backward glory effect approximation for several values of $q$ but with fixing decoupling parameter, $M\omega=15$.  Fig. (\ref{fig:a_c6}) illustrates this point; for angle  smaller than  $\theta=\pi$, both curves do not agree. However,  both findings seem to converge around  $\theta=\pi$. The latter fact does not seem to be affected by the different values of $q$ considered.sidered.

\subsection{Absorption cross-section for high energies}
\begin{figure}
	\includegraphics[width=3.2in]{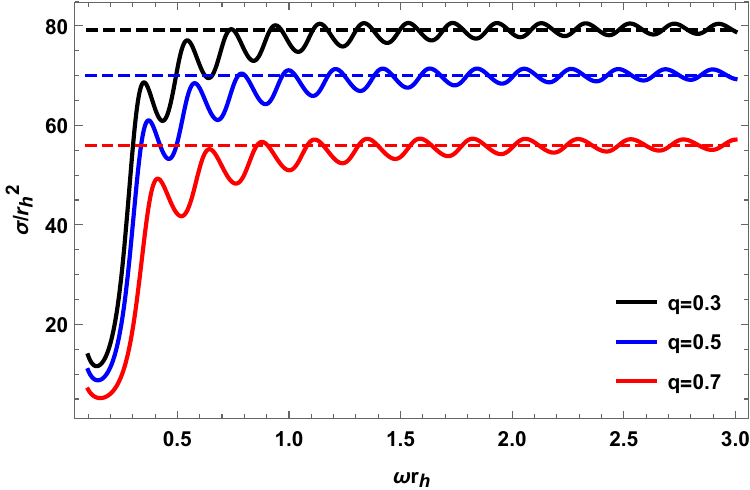}
	\includegraphics[width=3.2in]{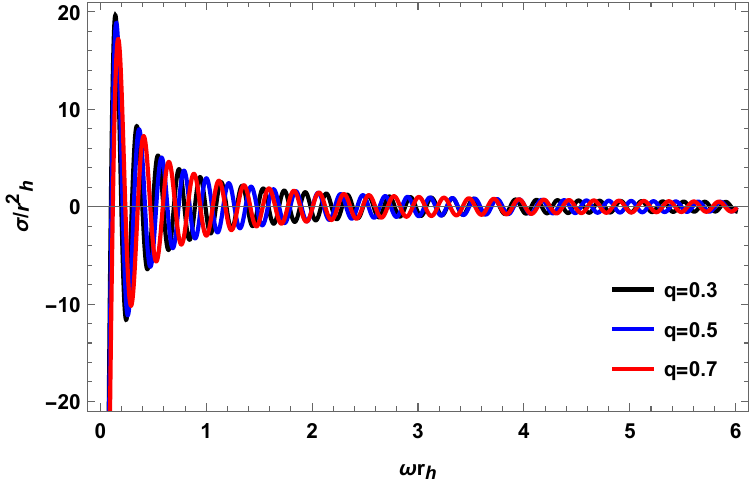}
	\caption{Top panel: Total absorption cross-sections for a dilatonic black hole with different $q$ are displayed; the horizontal dashed line indicates the corresponding geometric contributions to the cross-section. Low panel: The absorption cross-sections for dilatonic black holes in the sinc approximation for different $q$ are shown, representing the oscillatory contribution to the absorption cross-section.}\label{fig:abs_np2}
\end{figure}

\begin{figure}
\centering
	\includegraphics[width=3.2in]{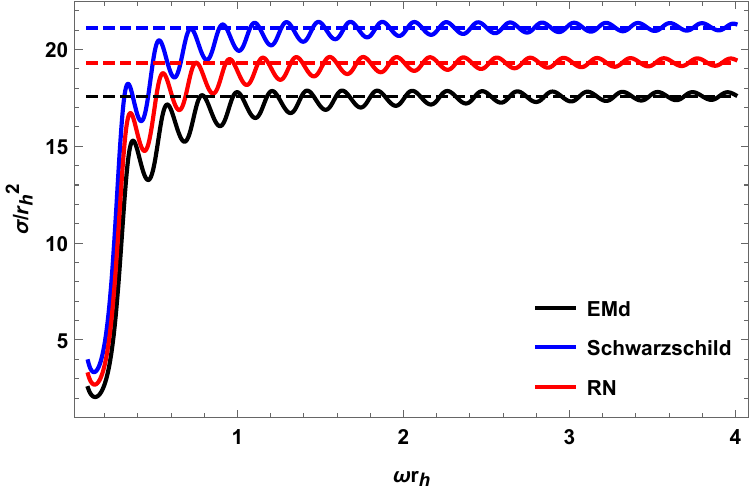}
	\caption{The absorption cross-sections of three types of black holes (Schwarzschild, RN, and dilatonic) are displayed. The dashed line represents the geometric contribution to the absorption cross-section.}\label{fig:abs_np3}
\end{figure}

\begin{figure}
\centering
	\includegraphics[width=3.2in]{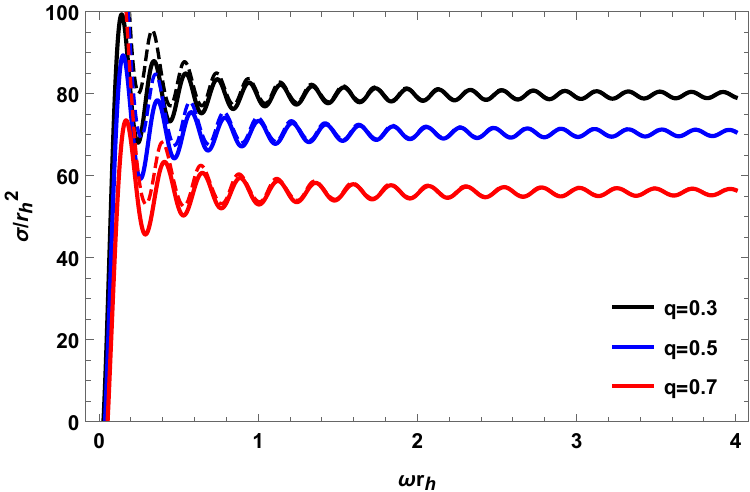}
	\caption{It shows the total absorption cross-section versus  $\omega M$ for a massless scalar field. The charge-to-mass ratio varies from $0.3$ to $0.7$.  The continuous line represents the numerical simulation, whereas the dashed line denotes the sinc approximation (oscillatory part of the cross-section). The latter contribution differs from the numerical simulation for a small $\omega M\ll 1$.}\label{fig:sinc4}
\end{figure}

\begin{figure}
\centering
	\includegraphics[width=3.2in]{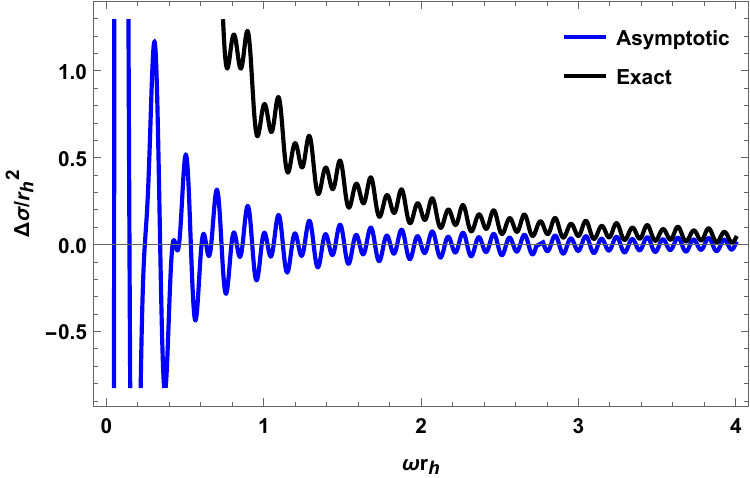}
	\caption{Difference between the numerical absorption cross-section and the high-frequency absorption cross-section is shown. The black line represents $\Delta \sigma_{abs}=\sigma^{num}_{abs}-\sigma_{hf}$ (fine structure) and the blue line indicates the fluctuation $\Delta \sigma^{hfine}_{abs}=\sigma^{h_{o}}_{abs}-\sigma_{hf}$, where $\sigma^{ho}_{abs}$ has higher order contributions and for that reason the supra-index $h_{o}$ is added. Here $q=0.2$.}\label{fig:sinc2}
\end{figure}

Until now  we have been dealing with the  analytic expression of the absorption cross-section in the limit of low-frequency, where the $\ell=0$ mode represents the main contribution \cite{futte}, \cite{crispino2009scattering}, \cite {dolan2009scattering}.  Of course, the general case is included in  (\ref{eq:numerical}) and is achieved numerically by solving (\ref{Eq:theradialeqpsi}) with the proper boundary conditions.  However, this procedure is time-consuming as one approaches the regime of high frequencies. Nevertheless, it is still possible to obtain an analytic expression that captures the main traits of the absorption cross-section in the opposite part of the spectrum (high-energy). To do so,  we are going to follow the method designed by Decanini and collaborators \cite{decanini2003complex}, \cite{decanini2010regge}, \cite{decanini2011universality}, \cite{decanini2011fine} several years ago based on the complex angular momentum method. These authors showed that the main contributions for $\sigma^{a}$  are two terms of different nature. The optical contribution comes from the area associated with the photon sphere, whereas the oscillatory part is modeled by the sinc function, which involves the orbital frequency and the Lyapunov exponent of the null unstable geodesics  \cite{decanini2011universality}. Interestingly, the studies of determining an analytic formula in the high-frequencies limit go back to the seminal paper of Sanchez, more than forty years ago \cite{sanchez1978absorption}. However, the recent method developed by Decanini \emph{et al.} is relatively straightforward  \cite{decanini2011universality}. The starting point is to recall that the Greybody factor is related with the transmission coefficient, $\Gamma_{\ell}(\omega)=|T_{\ell\omega}|^2$, and then to tackle the issue of computing the oscillatory part of absorption cross-section by using a complex angular method. This procedure requires isolating the Regge poles in the transmission coefficient with the help of the residues theorem, see \cite{decanini2011universality} for further details. They  proposed that the transmission coefficient  which accounts for the decaying waves near the photon sphere
\footnote{The existence of a photon surface plays a  key role in observations of astrophysical objects, as pointed out by K.S.Virbhadra and his collaborators  \cite{ks1}, \cite{ks2}, \cite{ks3}.} (or the critical  unstable null geodesic) is easily modeled by the following formula,
\begin{equation}
	|T_{\omega \ell}|^2 = \frac{1}{1+\text{exp}\Big[\frac{-2\pi(\omega^2-V_0(\ell))}{\sqrt{-2V_0''(\ell)}}\Big]},\label{tp}
\end{equation}
where the potential (\ref{eq:potencial}) and  its second derivative in (\ref{tp})  near the maximum can be expanded as 
\begin{eqnarray}\nonumber
	&&V_0(\ell)=\frac{f_c}{r_c^2g_c}(\ell+1/2)^2
	+ \underset{\ell \to +\infty}{\cal O}(1),\\
	&&V^{''}_0(\ell)=-2\left(\frac{\eta_c
		f_c}{r_c^2g_c}\right)^2(\ell+1/2)^2+ \underset{\ell \to +\infty}{\cal O}(1)\label{ersec},
\end{eqnarray}
and the critical radius $r_c$ is obtained by imposing $ V'(r_c) = 0$, 
\begin{eqnarray}
	r_c=\frac{3+q^2}{2}+\frac{\sqrt{9-10q^2+q^4}}{2}.
\end{eqnarray}
In (\ref{ersec}) the subindex $c$ indicates quantities evaluated at the  radius of the circular unstable geodesic. The maximum absorption is achieved for $\omega^2 \simeq V_0(\ell)$ \cite{decanini2011universality},\cite{benone2014absorption}. In the limit of high-frequency,  the oscillatory part of the absorption cross-section can be recast as
\begin{eqnarray} \label{siq}
	\sigma_\mathrm{abs}^\mathrm{osc}(\omega)&=&-8\pi\sigma_\mathrm{geo}\,
	\,\eta_c\, e^{-\pi \eta_c}\,
	\text{sinc}\left[2\pi
	b_c\omega\right],
\end{eqnarray}
where the eikonal cross-section reads $\sigma_ \mathrm{geo} = \pi b_c ^ {2}$.  The critical impact parameter is related  to the critical radius through  the expression $\omega^2=f_c/r_c^2g_c=b^{-2}_{c}$, and the parameter that accounts for the instabilities of  the null circular orbits is
\begin{eqnarray}\label{Eta}
	\eta_c=\Big[g_c(f_c-\frac{r_{c}^{2}}{2}f''_c)+f_c(2r_cg'_c+\frac{r_{c}^{2}}{2}g''_c)\Big]^{1/2}.
\end{eqnarray}
The total high-frequency formula for the absorption cross-section is decomposed as $\sigma_{hf}=\sigma_\mathrm{abs}^\mathrm{osc}+ \sigma_\mathrm{geo}$ \cite{decanini2010regge},  \cite{decanini2011universality}.  In Fig.(\ref{fig:abs_np2}), we first show the absorption cross-section for the dilatonic black hole for several values of $q$, which was obtained numerically. Besides the oscillatory pattern, the amplitudes of the cross-section decrease for small $\omega M$ as the parameter $q$ takes larger values. A similar effect is observed in the Reissner-Nordstr\"om case \cite{crispino2009electromagnetic}.  For each value of $q$  the curves are shifted to the right.
Fig.(\ref{fig:abs_np3}) shows the comparison among the absorption cross-section  for  Schwarzschild, RN, and dilatonic black holes. The absorption cross-section for  Schwarzschild black hole exhibits the largest amplitude, whereas the dilatonic black hole has the smallest ones, even smaller than the Reissner-Nordstr\"om black hole. This difference becomes relevant for large $M\omega$.
At this point, we must evaluate how good the sinc approximation is regarding the full numerical absorption cross-section. Fig. (\ref{fig:sinc4}) displays both curves in terms of the decoupling parameter $\omega M$ for different values of $q$.  The sinc approximation is in excellent agreement with the full numerical curve within the range $2M\omega\geq 1 $; however,  these curves can be distinguished for a minimum value of $\omega M$, showing some overestimation. The larger the value of $q$ is, the bigger the difference between the full numerical curve and the sinc approximation. 
We continue our analysis by computing  the difference between the numerical absorption cross-section and the high-frequency contribution within the sinc approximation.  The idea is to examine the absolute estimator, $\Delta \sigma_{abs}=\sigma^{num}_{abs}-\sigma_{hf}$ \cite{decanini2011fine}. It was found that in the Schwarzchild black hole there is a considerably small difference, typically below $5\%$ as pointed out by   Decanini \emph{et al.} \cite{decanini2011fine},  \cite{decanini2011universality}. In our case,  the numerical simulation reveals that the fluctuaction given by  $\Delta \sigma_{abs}$ is  really low,  $0.3$ order of magnitude for $M\omega \gtrsim 2$.  But this difference increases for $M\omega < 2$, reaching one order of magnitude [see Fig. (\ref{fig:sinc2})]. This fluctuation is known in the literature as the \textit{fine structure} of the absorption cross-section \cite{decanini2011fine},  \cite{decanini2011universality}.
\begin{figure}
\centering
	\includegraphics[width=3.2in]{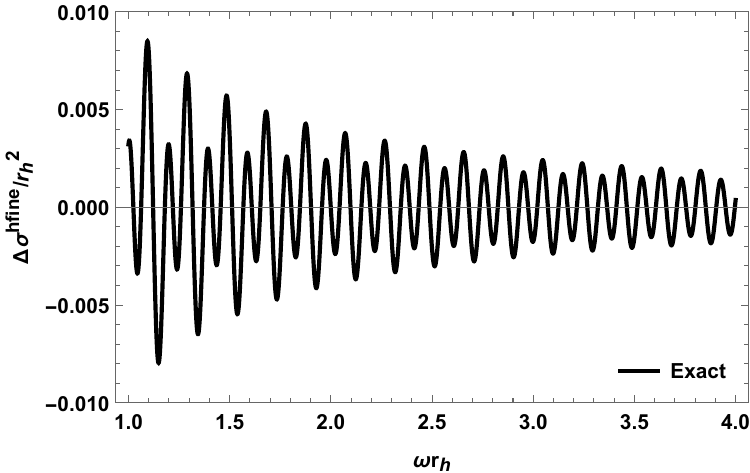}\
	\caption{Hyperfine structure of the total absorption cross-section in terms of $\omega r_{h}$  for the dilatonic charged black hole. Here $q=0.2$.  }\label{fig:sinc3}
\end{figure}
The analysis can be extended by considering higher terms in the oscillatory part of the high-frequency approximation of the absorption cross-section\cite{decanini2011fine}. The new corrections in the absorption cross-section read
\begin{eqnarray}\label{hyps}
	\sigma^{ho}_{abs}&\simeq&\sigma_{geo}\Big(1-\,
	8\pi\,\eta_c\, e^{-\pi \eta_c}\,
	\text{sinc}\left[2\pi
	b_c\omega\right]\nonumber\\&+&16\pi\,\eta_c\, e^{-2\pi \eta_c}\,
	\text{sinc}\left[4\pi
	b_c\omega\right]\Big).
\end{eqnarray}
Using (\ref{hyps}),  we can estimate the absolute difference between $\sigma^{ho}_{abs}$ and $\sigma^{hf}$. Fig. (\ref{fig:sinc2}) indicates that these fluctuations are small for large $M\omega$. In a sense, this is indicating a numerical convergence provided we are comparing two consecutive terms. Additionally, we obtain the difference $\Delta \sigma^{hfine}_{abs}=\sigma^{ho}_{abs}-\sigma^{hf} -\sigma_{0}$, being $\sigma_{0} \propto \omega^{-2}$ a soft term derived in \cite{decanini2011fine},   \cite{decanini2011universality}. This fluctuation is known as the hyperfine structure of the absorption cross-section because of the order of magnitude in the amplitude is around $0.001$ whereas in the fine structure is around $0.01$ \cite{decanini2011fine}. The latter numerical analysis reveals a complex  pattern in the high-energy absorption cross, different from the one related to the fine structure [see Fig.(\ref{fig:sinc3})].

\section{Charged  massive scalar field}\label{sec:charged}
In this section, we will explore the case of a charged massive scalar field propagating on the dilatonic black hole background. In doing so, we will extend the previous analysis based on a neutral massless scalar field by focusing on the new features introduced by the mass of the field and its charge. The absorption cross-section due to the interaction of massive but neutral scalar field in Reissner-Nordstr\"om black hole was explored in 
\cite{benone2014absorption}, where the cross-section was obtained at different frequency ranges (low, high, and intermediate). Some years later, the same authors revised their analysis, correcting some conclusions about the low-frequency limit \cite {benone2017absorption}. Benone \emph{et al.} used as a starting point a dimensionless parameter introduced by Jun and Park, which helps to identify the role played by a massive scalar field in the absorption, and emission spectra of Schwarzschild black hole  \cite{jung2004effect}. The process of absorption of a massive charged scalar field by a  Kerr-Newman black hole was reported in  \cite{bc2019}. The absorption of a massive wave of spin $s=1/2$ around a small Schwarzschild black hole was explored by Doran \emph{et al}. \cite{doran2005}. For the EMd gravity, the absorption cross-section associated with a massless scalar field was partially addressed  \cite{huang2020scattering}; however, the analysis of a charged massive scalar field impinging upon a dilatonic charged black hole was not contemplated. Our main goal is to determine the absorption cross-section and scattering cross-section for a charged massive scalar field for a dilatonic black hole.

\begin{figure}
	\centering
	\includegraphics[width=3.2in]{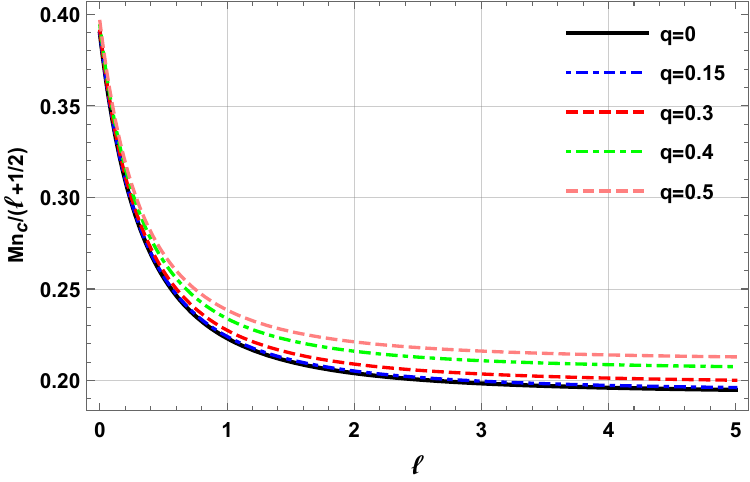}
	\caption{Critical mass coupling $Mn_c$ as a function of multipole $\ell$.}\label{fig:52}
\end{figure}

It is helpful to introduce a dimensionless parameter, say $v$, which includes the field's mass and the incoming frequency/energy. Initially introduced by Jun and Park \cite{jung2004effect}, this parameter reads
\begin{eqnarray}
	v =\sqrt{1-\frac{m^2}{\omega^2}},
\end{eqnarray}
 and  is  well-defined ($0<v\leq 1$) as long as  $\omega>m$. To carry on, we consider the dynamics of a charged massive scalar field near a dilatonic black hole (\ref{EMD}), hence the master equation is given by a generalized Klein-Gordon equation
\begin{eqnarray}
D^{\mu}D_{\mu}\Psi&\equiv& g^{\mu\nu}\left(\nabla_{\mu}-ieA_{\mu}\right)\left(\nabla_{\nu}-ieA_{\nu}\right)\Psi=m^2\Psi. \nonumber\\
	\label{eq:perturEq}
\end{eqnarray}
Here  $e$ is the charge of the scalar field, $m$ is the mass of the scalar field, and $ \nabla_ {\mu} $  stands for the covariant derivative in curved spacetime. The background electric potential is  $ A_\mu$. As usual,  we propose that the charged massive scalar field $\Psi$ admits the usual decomposition (\ref{separation0}). Replacing the latter ansatz  in the generalized KG equation (\ref{eq:perturEq}), we arrive at the Regge-Wheeler equation  for the radial modes in the tortoise coordinate,
\begin{eqnarray}\label{EMD_2}
	-\partial_{xx}\psi_{\omega\ell m}+\left(V-\Big[\omega-eV_{q}\Big]^{2}\right)\psi_{\omega\ell m}\label{eq:radEq}=0~~~~~
\end{eqnarray}
where the electric potential is  $V_{q}\equiv A_t=\sqrt{2}Mq/r+ k$  whereas the  potential $V$ reads
\begin{eqnarray}\label{eq:potencial_carregado}
	V(r)&=&f(r)\Big[\frac{r f'(r) g'(r)+f(r) \left(2 g'(r)+r g''(r)\right)}{2 r g(r)}\nonumber\\&+&\frac{f'(r)}{r}-\frac{f(r) g'(r)^2}{4 g(r)^2}+\frac{\ell (\ell+1) }{r^2 g(r)}+m^2\Big].
\end{eqnarray}
An issue regarding (\ref{EMD_2}) is the selection of the effective potential because we have $V$, $V_q$ and a mixed term between $V_{q}$ and $\omega$. Therefore, it is not clear which effective potential must be chosen. However,  we are going to name as  the effective potential all the terms which do not contain  the frequency,
\begin{eqnarray}
	V_{\rm{eff}}=V-e^2\Big(\frac{\sqrt{2}qM}{r}+k\Big)^2.
\end{eqnarray}
\begin{figure}
	\centering
	\includegraphics[width=3.2in]{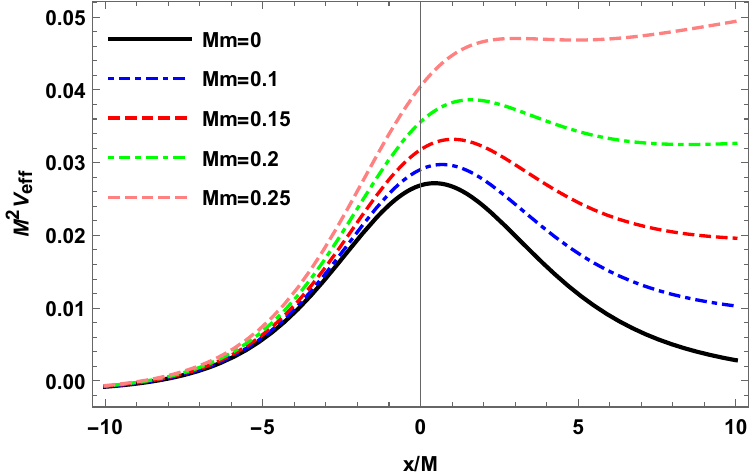}
	\caption{Effective potential  in terms of tortoise coordinate for $ \ell = 0$,  $mk=0.1$, $ q = 0.4 $, and different values of  $Mm$.}\label{fig:50} 
\end{figure}
We must summarize the main parameters of our geometry and the intrinsic parameters that characterize the scalar field. The black hole presents initially three global charges $M$, $Q$, and $D=Q^2/2M$, so it is possible to define a dimensionless parameter $q^2=D/M=Q^2/2M^2$ which assembles two of them, ending up with two parameters $M$ and $q$. On the other hand, the scalar field has mass $m$, frequency $\omega$, and charge $e$, so several combinations can be arranged to obtain dimensionless coupling, for instance,  $M\omega$,  $Mm$, and $Mn$ with $n=\sqrt{m^2-e^2 k^2}$. The latter ones allow us to study different regimes that characterize the inner or far-field approximation and how big/small the field mass is concerning the black hole's mass. In addition to that,  we introduce the concept of critical mass encoded in the $Mn_c$ parameter following the Jung, and Park prescription \cite{jung2004effect}. Indeed, the critical mass is defined as the local maximum of the effective potential: $V(r\rightarrow \infty)=n^2$. In the far-field region, we can expand the effective potential and consider only the corrections up to order $1/r^2$,
\begin{eqnarray}\label{expa3}
	&&V_{\rm{eff}}=\left(m^2-e^2 k^2\right)-\frac{2 m^2 M+2M\sqrt{2}q e^2 k}{r}\nonumber\\&&+\frac{\ell(\ell+1)-2M^2e^2 q^2}{r^2},
\end{eqnarray}
where the field's charge shifts the field mass, another effect introduced by the field charge is that standard $M$-term includes a correction, and the same happens for the standard angular momentum barrier (\ref{expa3}). Fig (\ref{fig:50}) shows that the usual bell-shape of the effective potential is drastically modified as $Mm$ increases; that is, for higher values of $Mm$ appears a local minimum besides the already existing local maximum. For values higher than the critical one, $Mm>Mn_c$,  the unbound modes are absorbed for any value of $\omega$. For instance, we obtain that the\textit{ effective }critical mass for $\ell=0$ is $Mn_c=0.195$, for $\ell=1$ is $Mn_c=0.338$, and for $\ell=2$ reads  $Mn_c=0.517$. The relation between $Mn_c$ and the angular momentum is displayed in Fig. (\ref{fig:52}); for the large value of $\ell$, the relation between both quantities becomes linear.   
We also check that the general shape of the potential does not change as $Me$ increases, at least for large $x/M$, but $V_{\rm{eff}}$ becomes negative near the horizon because the electric contribution seems to dominate over the gravitational terms.  
In order to solve the Regge-Wheeler equation (\ref{EMD_2}), we must add some boundary conditions for the radial modes at the end of the interval. To do so, we notice that the effective potential vanishes at the horizon and goes to a constant $m^2$ at infinity. The boundary condition at the horizon must be an ingoing plane wave, but at infinity, there is a superposition of the outgoing wave plus an ingoing wave,
 \begin{equation}
	\psi_{\omega\ell m}=\begin{cases}
		T_{\omega \ell}e^{-i\xi x}, & x\to -\infty\\
		e^{-i\rho x}+ R_{\omega \ell}e^{i\rho x}, & x\to \infty,
	\end{cases}\label{eq:boundary}
\end{equation}
where $\xi = \omega-eV_{q}(r_{\rm{h}})$ and  $\rho = \sqrt{\bar{\omega}^2-m^2}$. Using the definition of four-current vector (\ref{current}),  we infer the  currents for the incoming, reflected, and transmitted waves,
\begin{eqnarray}
	J^T_{x} = \xi|T_{\omega \ell}|^2,\quad J^R_{x} = -\rho|R_{\omega \ell}|^2,\quad J^I_{x} = \rho.
\end{eqnarray}
From the conservation of four-current, it follows the conservation of the flux associated with the spatial 3-current, $J^T_{x}=J^I_{x}+J^R_{x}$,  which in terms of the amplitudes reads
\begin{eqnarray}\label{cflux}
	|R_{\omega l}|^2+ \frac{\xi}{\rho}|T_{\omega l}|^2 = 1.
	\end{eqnarray}
The extra factor in front of the transmission coefficient (\ref{cflux}) is signaling an interesting effect. We consider two different situations regarding the incoming wave near the black hole horizon to understand this aspect fully. For $eV_{q}<0$ the exponent of the incoming wave in (\ref{eq:boundary}) is positive. However,   the exponent $\xi$ could take a negative value for $eV_{q}>0$, and this wave turns into an outgoing wave which amplifies the already existing outgoing waves at infinity, showing up a superradiance phenomenon \cite{superr}. We will come back later to this topic in the final section and explain in more detail, using a different point of view, the physics behind it. 

Using the same arguments mentioned in Sect. IVA for determining the partial absorption cross-section along with the change of $\omega\leftrightarrow \rho$, we obtain that the partial absorption cross-section reduces to $ \sigma_{\ell}=(\pi/\rho^2)(2l+1)(1-|R_{\omega l}|^2)$. Replacing (\ref{cflux}) in the latter expression, we arrive at the new formula  for  the partial absorption cross-section in terms of the transmission  amplitude,
\begin{eqnarray}\label{eq:RR}
	\sigma_l
	&=& \frac{\pi\xi}{\rho^3}(2l+1)|T_{\omega l}|^2,
\end{eqnarray}
where  $\xi/\rho^3$ factor accounts for the superradiance effect.  For a massless and chargeless scalar field $\rho^3=\omega^3$ and $\xi=\omega$, and consequently, the extra-factor  leads to the usual $\xi/\rho^3=\omega^{-2}$ contribution. The total absorption cross-section is derived by summing over the partial contribution of each angular momentum, $\sigma_{abs}=\sum_{\ell}\sigma_{\ell}$. 
\subsection{Low-frequency regime}
In this subsection, we tackle the absorption cross-section's computation in the low-frequency limit,  $M \omega\ll 1$. In addition to that, we are supposing  $mM\ll 1$ and $eq\ll 1$ to obtain analytical expressions \cite {benone2017absorption}. This level of approximation is based on reasonable physical grounds; namely, the low-frequency condition implies that  $mM\ll 1$  provided $m<\omega$. The second condition indicates that the electric field generated by the $e$-charge is tiny concerning the background electric field associated with the charged black hole. We will split the spatial region into three different zones, depending on whether the wave is near the black hole (region I), in an intermediate region (region II), or far away from the compact object (region III). We also will  glue each local solution in order to assemble a global solution that satisfies the boundary condition stated before (\ref{eq:boundary}). 

Let us begin by mentioning that the simplest case corresponds to the solution in the region I, where the effective potential vanishes. For simplicity, we define $\phi_{\omega\ell m}=\psi_{\omega\ell m}/r\sqrt{g(r)}$ and  write down the boundary condition listed in (\ref{eq:boundary}) as follows
\begin{eqnarray}\label{s1}
	\phi^{I}_{\omega\ell m} \approx  T_{\omega\ell}(r-2M)^{-2Mi\xi}.
\end{eqnarray}
In region II, we explicitly take the limit of low-frequency in the effective potential $M \omega\ll 1$ so we can neglect the  $\omega$ term and the whole effecive potential but keep the terms rising from the second derivative of $\phi_{\omega\ell m}$.  In doing so,  we employ the approximation $eq\ll 1$ to neglect the background electric potential.  The master equation  reduces to
\begin{eqnarray}
	\frac{d^2\phi^{II}_{\omega\ell m}}{dr^2} + \left(\frac{f'(r)}{f(r)}+\frac{g'(r)}{g(r)}+\frac{2}{r}\right)\frac{d\phi^{II}_{\omega\ell m}}{dr}= 0, 
\end{eqnarray}
and its solution is
\begin{eqnarray}\label{eq:r2}
	&&\phi^{II}_{\omega\ell m} \approx\frac{c_1}{\beta}\ln\Big(\frac{r-2M}{Mr-2M^2q^2}\Big)+c_2,
\end{eqnarray}
where $c_1$ and $c_2$ are  integration constants whereas $\beta=2M^2(1-q^2)$. To match the solution of region I (\ref{s1}) with the solution belonging to region II (\ref{eq:r2}), we need to recast  (\ref{s1}) in a different way,
\begin{eqnarray}\label{eq:r11}
	\phi^{I}_{\omega\ell m} \approx T_{\omega\ell}[1-2Mi\xi\ln(r-2M)].
\end{eqnarray}
Near the horizon the second solution (\ref{eq:r2}) reduces to 
\begin{eqnarray}\label{eq:r22}
	\phi^{II}_{\omega\ell m}\approx \frac{c_1}{\beta}\ln\Big(\frac{r-2M}{\beta}\Big)+c_2.
\end{eqnarray}
It is straighfoward  to determine the integration constants by simply equating (\ref{eq:r2}) and (\ref{eq:r22}) in the superposition region,
\begin{eqnarray}\label{r2}
	c_1 &=&-2Mi\xi\beta T_{\omega\ell},\\ c_2 &=&1-2Mi\xi\ln\beta T_{\omega\ell}.
\end{eqnarray}
We carry on by looking at the shape of the master equation in the far-field limit (region III). In this region, it is not possible to neglect the mass term neither the $\omega^2$ term; however, the $2Me^2q^2/r^2$ factor is neglected provided the condition $eq\ll 1$ holds, but the $ek$ term cannot be discarded because it will modify the effective frequency by generating a frequency shift. Proceeding in the manner as we did before, (\ref{EMD_2}) can be simplified as 
\begin{eqnarray}\label{ccu}
\frac{d^2 r\phi^{III}_{\omega\ell m}}{dr^2}+\Big(\bar{\omega}^2-m^2+\frac{\varpi}{r}-\frac{\ell(\ell+1)}{r^2}\Big)r\phi^{III}_{\omega\ell m}=0,
\end{eqnarray}
where $\varpi=[2Mm^2-4M\bar\omega^2-2M\sqrt{2}qe\bar\omega]$ and $\bar{\omega}=\omega-ek$. Eq. (\ref{ccu}) resembles the well-known  Coulomb wave equation in spherical coordinates \cite{abramowitz1988handbook}. Its solution can be recast in terms of the regular  Coulomb function $F_l(\eta,\bar{\rho} r)$  and the irregular Coulomb function  $G_l(\eta,\bar{\rho} r)$, respectively 
\begin{eqnarray}\label{exp4}
\phi^{III}_{\omega\ell m}= a\frac{F_l(\eta,\bar{\rho} r)}{r}+ b\frac{G_l(\eta,\bar{\rho} r)}{r},
\end{eqnarray}
where the  two  parameters $\eta$ and $\bar\rho$ are related in the following manner,
\begin{eqnarray}
	\eta &=& -\frac{M(2\bar\omega^2-m^2+\sqrt{2}qe\bar\omega) }{\bar\rho},\quad \bar\rho=\sqrt{\bar\omega^2-m^2}.~~
\end{eqnarray}
Using the asymptotic expansion of the Coulomb functions \cite{abramowitz1988handbook} for large argument, the solution  reduces to  
\begin{eqnarray}\label{exp5}
\phi^{III}_{\omega\ell m}= A e^{i\vartheta} +  B e^{-i\vartheta},
\end{eqnarray}
such that the phase is written in terms of the radial coordinate, angular momentum $\ell$, and the two parameters that we introduced before plus an additional parameter $\delta_{\ell}$; namely,   $\vartheta = \bar\rho r - \ell\pi/2 - \eta \ln (2 \bar\rho r)+ \delta_{\ell} $. Here,  the new parameter  $\delta_{\ell}=\text{arg}\Gamma( \ell + 1 + i \eta )$ is the so-called Coulomb phase shift \cite{abramowitz1988handbook}. Taking into account (\ref{exp4}) and  (\ref{exp5}), we can show  how the integration constants are related to each other: $B = (-a+ib)/2i$ and  $A =(a+ib)/2i$.

\begin{figure}
\centering
	\includegraphics[width=3.2in]{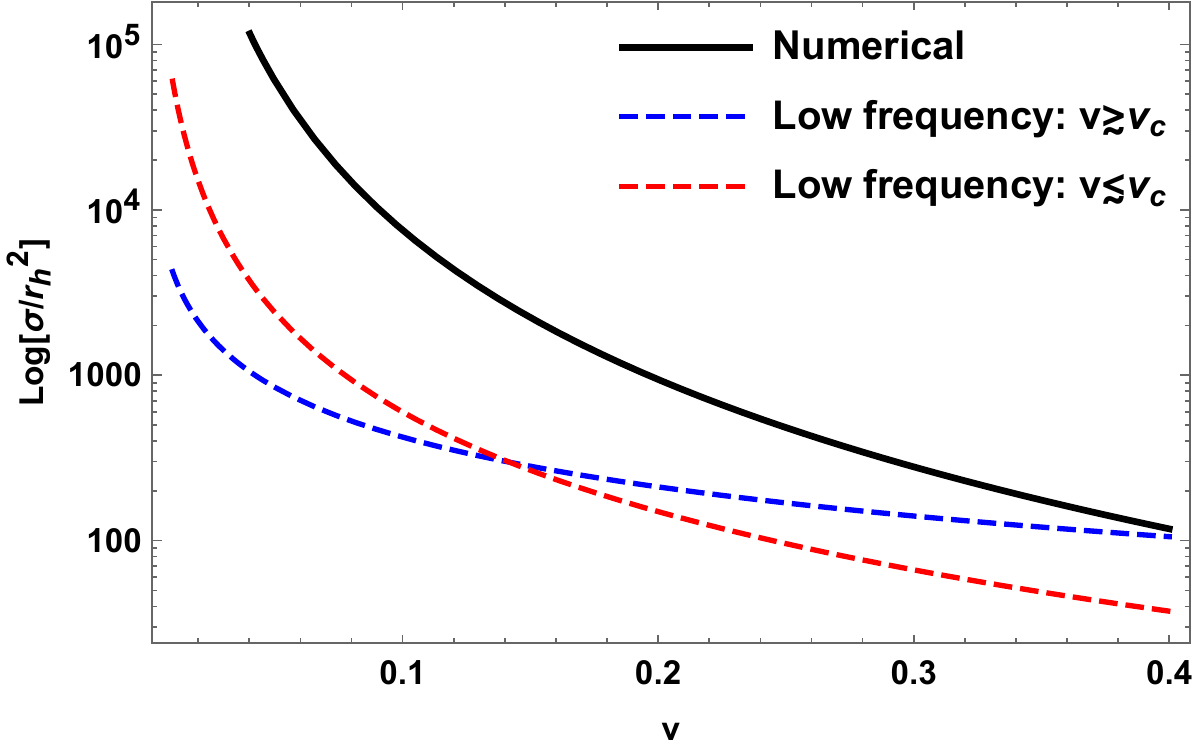}
	\caption{Comparison among the approximated absorption cross-section for low-frequency  with $v<v_{c}=0.138$,   low-frequency  with $v>v_{c}=0.138$,  and the full numerical  absorption cross-section in logarithmic scale. The fixed parameters are $Mm=0.022$, $Me=0.001$, $k=0$, $q=0.4$ and $\ell=0$. The black line corresponds to the numerical cross-section.}\label{fig:en6}
\end{figure}

In order to match continuously the solution $\phi^{II}_{\omega\ell m}$ with the far-field solution $\phi^{III}_{\omega\ell m}$, we first have to determine  a simple expression for the asymptotic expansion of (\ref{exp4}). To do so, we work in the 
low-frequency ($M\bar\omega\ll 1$, $Mm\ll 1$, $Me\ll 1$) regime  with zero angular momentum $\ell=0$. The coulomb functions  for $\ell=0$ satisfy the next two identities: i-$F_0(\eta,x) = \lambda x$ and ii-$G_0(\eta,x) = \frac{1}{\lambda}$. The $\lambda$ parameter is written in terms of $\eta$, that is, $\lambda^2 = 2\pi\eta/[e^{2\pi\eta} - 1]$. With the help of the former identities, the general solution (\ref{exp4}) admits a simple form,
\begin{eqnarray}
	\phi^{III}_{\omega\ell m}=a\lambda \bar\rho + \frac{d}{\lambda r},
	\label{s32}
\end{eqnarray}
which can be glued smoothly in the far-field region  with (\ref{eq:r22}) after having expanded in series at infinity as
\begin{eqnarray}\label{coen1}
\phi^{II}_{\omega\ell m}= -\frac{c_1}{Mr}+c_2.
	\label{rn23}
\end{eqnarray}
Combining (\ref{s32}) and (\ref{rn23}),  the integration constants in (\ref{s32}) are written in terms of the transmission amplitude,
\begin{eqnarray}
	\begin{array}{ll}
		a=\frac{1-2Mi\xi\ln\beta T_{\omega\ell}}{\lambda \bar\rho},\\
		d= 2i\xi\beta\lambda T_{\omega\ell}.
	\end{array}
	\label{crn2}
\end{eqnarray}
Replacing (\ref{crn2}) in the definition of the coefficients $A$ and $B$, we arrive a formal expression for the reflection  coefficient,  
\begin{eqnarray}\label{eq:R}
	|R_{\omega\ell}|^2=\frac{|A|^2}{|B|^2} = \frac{|1-2Mi\xi\ln(\beta)-2\xi\beta\lambda^2\bar\rho|^2}{|1-2Mi\xi\ln(\beta)+2\xi\beta\lambda^2\bar\rho|^2}.
\end{eqnarray}
The connection with the absorption cross-section (\ref{eq:RR}) is obtained by putting together (\ref{cflux}) and  (\ref{eq:R}). The absorption cross-section  with $\ell=0$ in the low-frequency regime reads
\begin{eqnarray}\label{eq:sec}
	\sigma_{abs} = \frac{\pi}{\bar{\rho}^2}\Big[\frac{8\xi\beta\lambda^2\bar{\rho}}{(1+2\xi\beta\lambda^2\bar{\rho})^2+4M^{2}\xi^2\ln^2 \beta}\Big]
\end{eqnarray}
In general, (\ref{eq:sec}) has a non-trivial behavior with the frequency. Therefore, we must look for an asymptotic expression that can easily compare with the numerical simulations. Keeping in mind that we are considering an approximated low-frequency formula, it is valid to employ that $\lambda\approx 1$. Replacing the latter approximation in (\ref{eq:sec}) and using that $v\gtrsim v_c=2\pi Mm$ we obtain  an analytic expression, 
\begin{eqnarray}\label{absf}
	\sigma_{abs} \approx\frac{ 8\pi\beta\xi}{\bar\rho},
\end{eqnarray}
which  reduces to the  Schwarzschild case, $\sigma_{abs} = 16\pi M^2/v$, for $q=0$. In addition to the previous limit, we must explore another scenario where the $v$ parameter is small, but it could be near the critical velocity value, $v\lesssim v_{c}=2\pi Mm$ \cite {benone2017absorption}. In that case,  the $\lambda$ parameter  can be calculated in the limit $\bar\rho\to 0$ again, but now, there are some corrections due to the fact the field has mass and charge, 
\begin{eqnarray}\label{approxb}
	\lambda^2\approx \frac{2\pi M(m^2+\sqrt{2}q\,em)}{\bar\rho}.
\end{eqnarray}
Replacing  (\ref{approxb}) in (\ref{absf}), the new absorption cross-section reads 
\begin{eqnarray}\label{pl}
	\sigma_{abs} \approx 32\pi^2M^{3}(1-q^2)(m^2+\sqrt{2}q\,em)\frac{\xi}{\bar\rho^2}.
\end{eqnarray}
Eq. (\ref{pl}) agrees with the Schwarzschild case for $e=0$;that is, $\sigma_{abs} \approx A^{2}(r_{h})Mm/(2v^{2})$ for $v<v_c$ \cite {benone2017absorption}.  In general, we will to use $v=\bar{\rho}/\xi$ for ploting the absorption cross-section in terms of $v$ provided this definition agrees with the case $e=0$ employed before.
\begin{figure}
\centering
	\includegraphics[width=3.2in]{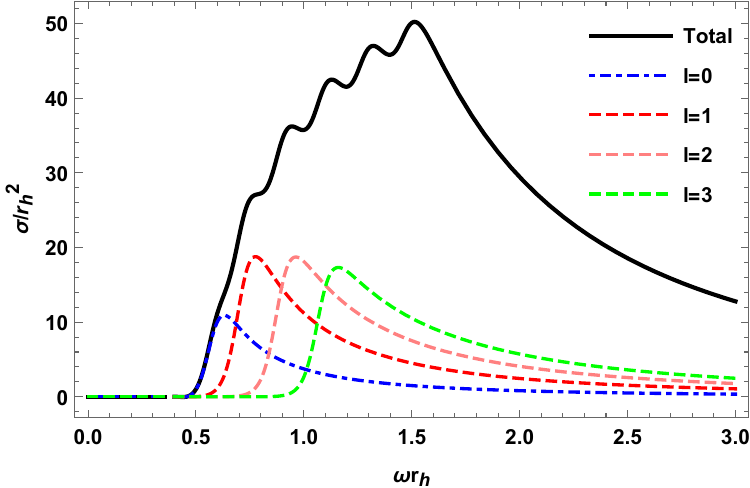}
	\caption{Partial (dashed lines) and total absorption cross-sections (black line) in terms of $\omega r_{h}$  for a charged massive scalar field and a charged dilatonic black hole.  Some parameters are fixed  as $Me=1.6$, $Mm=0.4$ and $q=0.4$ whereas  the angular momentum $ \ell$ varies. Due to the time-consuming issues in computing $\sigma^{num}_{abs}$ the resummation was performed using a low value of $\ell_{max}$.}\label{fig:p_c}
\end{figure}
\begin{figure}
\centering
	\includegraphics[width=4.3in]{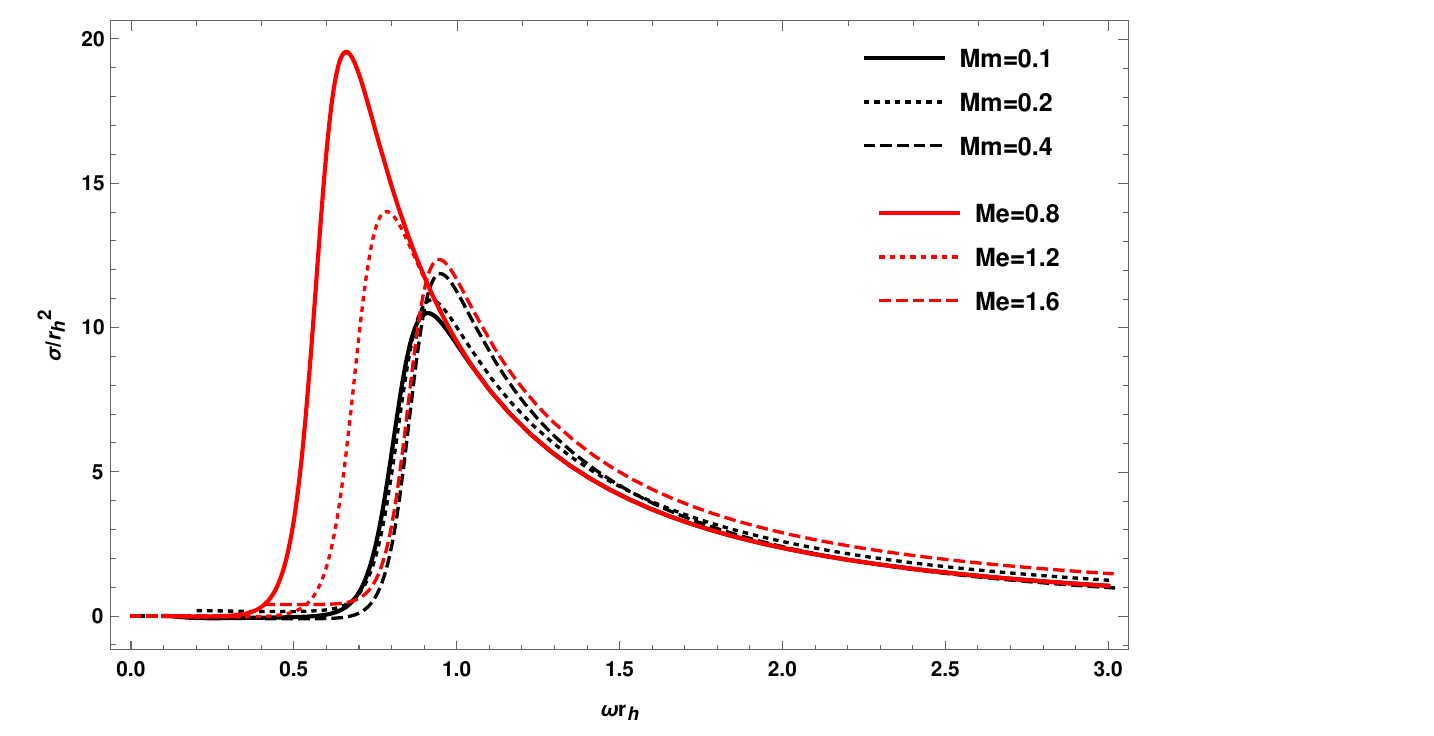}
	\caption{Partial absorption cross-section  in terms of $\omega r_h$ for a charged dilatonic  black holes and a massive charged scalar field. The red  lines correspond to different values of $Mm=\{0.1, 0.2, 0.4\}$  while  the other paramater are fixed ($e=1.6$, $\ell=1$ and $q=0.5$), whereas the black line is associated with  the partial absorption cross-section  for different values of $Me=\{0.8,1.2,1.6\}$  but fixing other parameters as follows, $Mm=0.1$, $\ell=1$ and $q=0.5$.}\label{fig:p_c2} 
\end{figure}
\begin{figure}
\centering
	\includegraphics[width=3.2in]{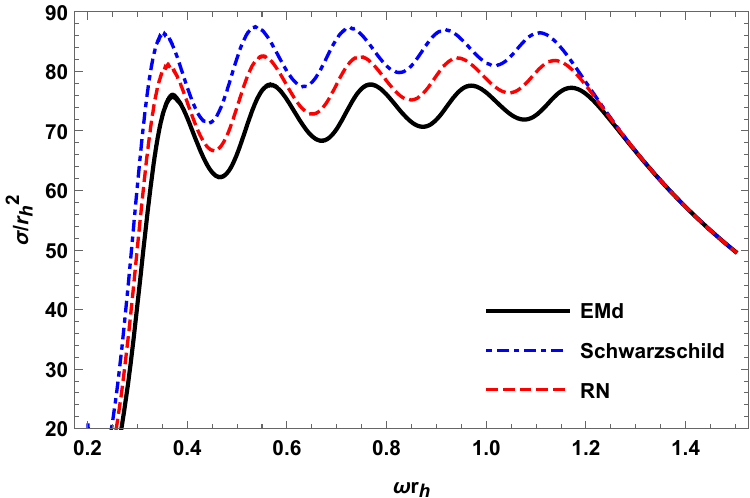}
	\caption{The total absorption cross-section for EMd,  RN, and Schwarzschhild black holes. The fixed parameters are $Me=0.2$, $Mm=0.2$, $q=0.4$, $ek=0.1$, and $\ell=1$.} \label{fig:abs_esp_comp_2}
\end{figure}
Now, we are in a position to present the exact/numerical results regarding the absorption cross-section. We begin by showing the numerical absorption cross-section for different values of angular momentum and its total resummation. Fig.(\ref{fig:p_c})  shows the partial absorption cross-section  for $ \ell = \{1,2,3\}$ and the total absorption cross-section for a dilatonic charged black hole in the case of a charged massive scalar field. The general pattern is the main peak for mildly frequency, $\omega r_{h}\simeq 1.2$,  plus a series of oscillations of lower amplitudes for small frequency and then decays to zero for large $\omega r_{h}$.  For a given  $\ell$, the partial cross-section is much more sensitive to the variation of $Me$ than the variation of $Mm$ as is shown in Fig. (\ref{fig:p_c2}). Comparing the numerically-obtained total cross-section for the dilatonic black hole with the RN black hole, we obtain that the latter one has a much larger amplitude (similar to the Schwarzschild case) for $0.3\leq \omega r_{h}\leq 1.3$. In contrast, the three numerical cross-sections converge to the same value for a larger value of $\omega r_{h}$ [cf. Fig. (\ref{fig:abs_esp_comp_2}).
Another interesting point is to compare how good is the approximated absorption cross-section for low-frequency in relation with the full numerical solution [see Fig. (\ref{fig:en6})].   The relative error $\log\Delta \sigma_{abs}/\sigma^{num}=\log |\sigma^{num}-\sigma^{low}|/|\sigma^{num}|$  is really large at $v\simeq 0$ due to some numerical instabilities around that point; however,   $\log\Delta \sigma_{abs}/\sigma^{num } \leq 1$ for $v \in (0, 0.4)$. It reaches its minimum value of $0.0042$ at $v \simeq 0.42$ and then continuous to grow  until $v=0.8$  where its amplitude remains below the unity; therefore, this is a good approximation only on the interval $v \in (0.4, 0.8)$. In the opposite limit ($v<v_{c}$), we obtain that the approximated cross-section only described the problem qualitatively because $\log [\Delta \sigma_{abs}/\sigma^{num}]$ is less than the unity on the interval $v \in (0.1, 0.8)$.  We then conclude that there is a transition in the cross-section at low-frequency around $v=0.138$,  as is seen in Fig. (\ref{fig:en6}), and that both approximated cross-sections are good enough to describe the main behavior of the full numerical cross-section.

Before closing our analysis on the absorption cross-section, we would like to explore the physical consequences on the critical values obtained for $v_{c}$. We can proceed in three different manners. First, we consider that dark matter is accommodated as a chargeless scalar field with an ultralight mass $m_{X} \approx 10^{-22} \, \text{eV} / c^2$ and with really low speed, $v_{X} \approx 4 \times 10^{-4}$ \cite{Hui:2016ltb} \footnote{Here we are considering that dark matter can be described by a  massive scalar field  \cite{Hui:2016ltb};  but also, it can be represented by a massive millicharged scalar field \cite{mc1}}. For a black hole  such as Cygnus X-1,  recent measurements indicate a stellar mass of $M_{BH}=(21.2\pm 2.2)M_{\odot}$ \cite{cy1}. Then, the critical velocity becomes $v_{c}\simeq 10^{-11}$, a much smaller value than $v_{X}$, and for that reason we can use the distribution in the region $v\gg v_{c}$.  In the case of black hole with intermediate masses $10<M_{BH}/M_{\odot}<10^{5}$ \cite{ibm}, the critical velocity  varies over the interval $10^{-12}<v_{c}<10^{-8}$ , so it is clear once again that $v_{X}\gg v_{c}$.  The recent observation of  a supermassive black hole called  $IZw1$  using the X-ray telescopes NuSTAR13 and XMM-Newton14 in the hard band of X-ray  made possible the determination of its mass, $M_{BH}=(3.1\pm0.5) 10^{7}M_{\odot}$ \cite{smbh}. Using the latter value, the critical velocity reads $v_{c}\simeq 2.3 \times 10^{-5}$,  hence we must continue to use the region where $v\gg v_{c}$, or equivalently, the ultralight dark matter model is properly described by the $v_{X}\gg v_{c}$ phase. In order to break this tendency we must consider compact object supra-massive, for instance,  the supermassive black hole located in the supergiant elliptical galaxy called Virgo A--Messier 87-- has a mass $M_{BH}=(6.4\pm 0.5) 10^{9}M_{\odot}$ \cite{let2}. Using the latter value, we obtain that the critical velocity $v_{c}=5.3 \times 10^{-3}$, and therefore, it is appropiate to use the cross-section in the interval $v_{X}\ll v_{c}$. In the E4 supergiant elliptical galaxy called NGC 4889, a supermassive black hole was detected with a bigger mass,  $M_{BH}=(2.1\pm 1.6) 10^{10}M_{\odot}$ \cite{suprabh}, which leads to a critical velocity two order of magnitudes below  $v_{X}$ \footnote{Regarding the value of $v_{X}$,  we may say that represents the average velocity (rms) for a given dark matter particle distribution in different physical scales.}. Second, we could consider a different scenario where the dark matter particle is accommodated as a heavier particle with a millicharge.  This corresponds to a range of masses and charges given by $m_{X} \simeq 1 \rm{Gev}$ and $e\epsilon \leq 10^{-14}e(m_{X}/\rm{Gev})$ \cite{mc1}.  A more recent  rough analysis revealed that $q_{X}/e< 10^{-13\pm 1} (m_{X}c^2/\rm{Gev})$ \cite{mc2}. Other estimations  based on the physics of the early universe place different constraints on the mass belonging to the $\rm{Mev}$ scale, $m_{X} \in [1,80]\rm{Mev}$, and a charge $q_{X}/e<10^{-6}$, respectively \cite{mc3}. Either way, we are going to start  by considering a conservative estimation on  the millicharged dark matter mass, say $m_{X}=\rm{Gev}$, and a low velocity $v_{X}=10^{-3}/3$ \cite{mc1}. For stellar black hole \cite{cy1}, the critical velocity is $v_{c}\simeq 1.6 \times 10^{20}$, for intermediate black hole with $M_{BH}=10 M_{\odot}$ \cite{ibm} it leads to   $v_{c}\simeq 7.6 \times 10^{19}$, and for a supermassive black hole  \cite{smbh} we arrive at  $v_{c}\simeq 7.6 \times 10^{25}$; namely, in all the case considered, the millicharged dark matter is well-described by the supercritical phase with $v_{c}\gg v_{X}$. Notice that the bounds obtained in \cite{mc3} lead to mildly lower values of $v_{c}$. The range of critical values of $v_c$ is not considerably improved in more sofisticated frameworks. For instance,  taking into account primordial black holes with masses $0.2<(M_{PBH}/M_{\odot})<5$ \cite{pbh}, the critical velocity continues to lie on the interval   $1.5\times 10^{15}<v_{X}<3.8 \times 10^{16}$. A third option is to employ our previous results on the critical velocity  ($v_{c}=0.138$) to determine what range of black hole masses is favored by the constraints on each dark matter model. As an example, we examine the  ultralight chargeless dark matter scenario \cite{Hui:2016ltb} and use the relation  $m_{X}M_{BH}=2.1 \times  10^{-2}$ along  with $m_{X}\lesssim 10^{-22}/c^2$, and $M_{BH}=f_{BH}M_{\odot}$.  It turns out that the favored range for black hole mass reads $f_{BH}\gtrsim 1.73 \times 10^{11}$. The supermassive black hole located in the elliptical galaxy NGC 4889  is the closest compact object to satisfy this relation provided   $M_{BH}=(2.1\pm 1.6) 10^{10}M_{\odot}$ \cite{suprabh}. We will come back to these constraints by considering the perturbation  in the dilaton channel along with the latest LHC constraints on the dilaton mass in order to evaluate what range of black hole masses is more relevant (cf. Sec. VII)
\subsection{High-frequency}
\begin{figure}
	\includegraphics[width=3.2in]{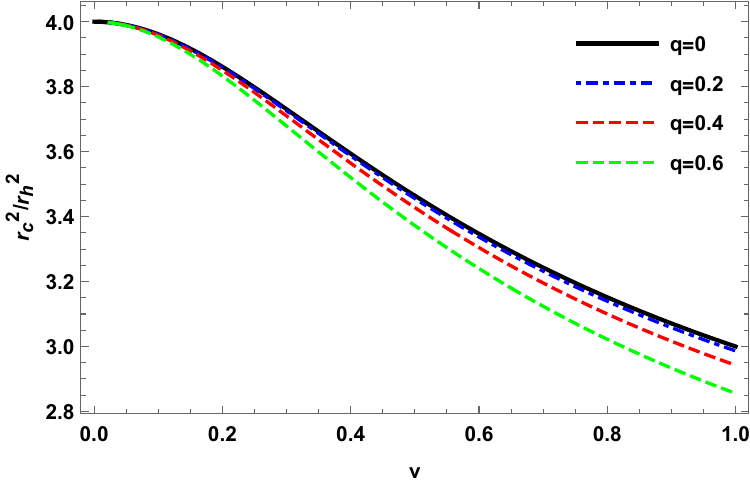}
	\caption{Critical radius of the unstable orbit in terms of the incident speed $v$ for different values of charge-to-mass ratios q. The scalar field charge is $e=0$. }\label{fig:p_F}
\end{figure}

\begin{figure}
	\includegraphics[width=3.2in]{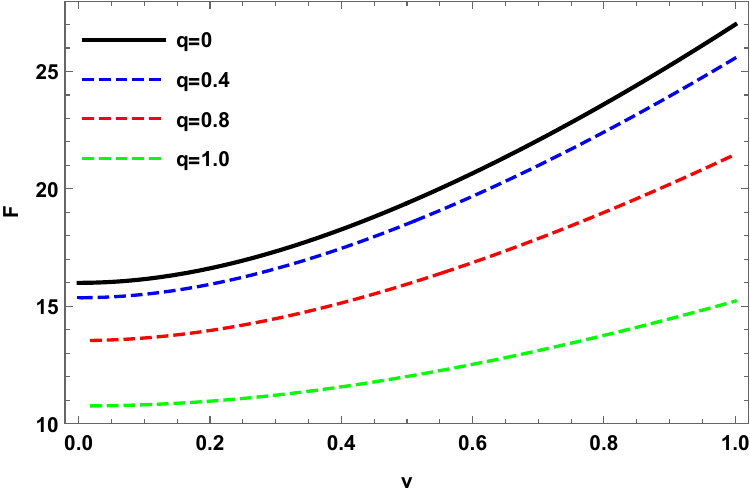}
	\caption{The form absorption function is depicted in terms of the incident speed $v$ for
		various charge-to-mass ratios q. The scalar field charge  is $e=0$.}\label{fig:p_F2}
\end{figure}

In this subsection,  the absorption cross-section in the high-frequency limit is contemplated in the opposite part of the spectrum. We begin by analytically obtaining an estimation of the cross-section. We restrict ourselves for the sake of brevity to a massive but chargeless scalar field ($e=0$). Our first task is to determine the critical impact parameter of the unstable geodesic within the eikonal approximation since the absorption cross-section reduces to $\sigma_{hf} = \pi b_c^2$. The geodesic equation for massive particles at the equatorial plane  ($\theta = \pi/2$)  with initial energy that correspond to an unbound orbit is
\begin{eqnarray}
	\dot{r}^2 = E^2 - f(r) \Big(m^2 +\frac{L^2}{r^2g(r)}\Big).\label{eq:energy}
\end{eqnarray}
Here,  $\dot{r} = dr/d\tau$ denotes the radial velocity in terms of the proper time, $\tau$  stands for the proper time,  $E =  f \dot{t}$ refers to the conserved energy,  and the conserved angular momentum reads $L = r^2g(r) \dot{\phi}$. It is convenient to introduce the  impact parameter  in terms of $L$, $E$, and the dimensionless velocity $v=1-m^2/E^2$, namely,  $b \equiv L / (E v)$ \cite{benone2014absorption}.  Replacing the definition of $b$ and $v$ in (\ref{eq:energy}), the radial velocity becomes,
\begin{eqnarray}
	\frac{\dot{r}^2 }{L^2} = \frac{1}{b^2 v^2} - f(r) \left( \frac{1-v^2}{b^2 v^2} + \frac{1}{r^2g(r)} \right).
	\label{phf}
\end{eqnarray}
The dilatonic charged black hole will absorb geodesic curves with an impact parameter  $b <b_c$ coming from the infinity, but those geodesics with a bigger impact parameter, $b> b_c$, will continue to go back to infinity after having passed near the black hole. To proceed further, we find the critical impact parameter of the unstable orbits by imposing the condition $\dot{r}(r_{c})=0$, yielding
\begin{eqnarray}
	b_c= \frac{r_c \,g_c^{1/2}}{v f_c^{1/2}}\left[1 - (1-v^2)f_c  \right]^{1/2},  \label{bc-def}
\end{eqnarray}
where the sub-index $c$ denotes evaluation at the critical radius of the unstable orbits, which by the way, can be found by demanding that $[d\dot{r}/dr](r_c)=0$.  It implicit expression reads, 
\begin{eqnarray}
	r^2_c= \frac{g_{c}}{f_{c}}\frac{1}{b^2_{c}v^{2}\big(\frac{f_c}{g_c}\big)'} \Big((1-v^2)f'_{c}-\frac{2f_c}{r_c} [1- (1-v^2)f_c]  \Big).\label{rc-def}~~~~~~
\end{eqnarray}
In order to isolate the divergency of impact parameter as $v\to 0$, it is useful to define the form absorption function as $F(v,q)=v^2b_c^2/M^2$ \cite{benone2014absorption} since it remains finite in the former limit. The absorption cross-section then becomes
\begin{eqnarray}
	\sigma_{hf} =\frac{\pi F(v,q)M^2}{v^2}.
\end{eqnarray}
Once the critical radius is obtained numerically by solving (\ref{rc-def}) and the critical impact parameter is extracted from (\ref{bc-def}),  we can determine the form absorption function. Fig.  (\ref{fig:p_F}) illustrates the behavior of $r_c$ in terms of $v$ for different values of $q$, indicating that the lower the values of $v$ correspond to higher critical radii. By increasing $q$, we end up with a lower critical radius. A similar situation is obtained by exploring the typical profile of the form absorption function in terms of  the incident velocity $v$ for several values of the charge-to-mass ratio[cf. \ref{fig:p_F2}].

\begin{figure}
	\includegraphics[width=3.2in]{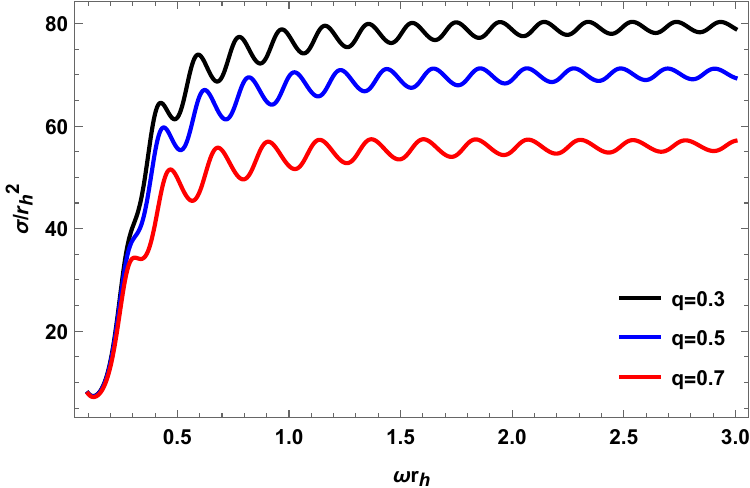}
	\caption{High-energy absorption cross-section in terms of $\omega r_{h}$ for different values of $q$   for a dilatonic black hole. The other parameters are fixed at $Mm=0.4$ and $e=0$.}\label{fig:abs_mass2}
\end{figure}

\begin{figure}
	\includegraphics[width=4.4in]{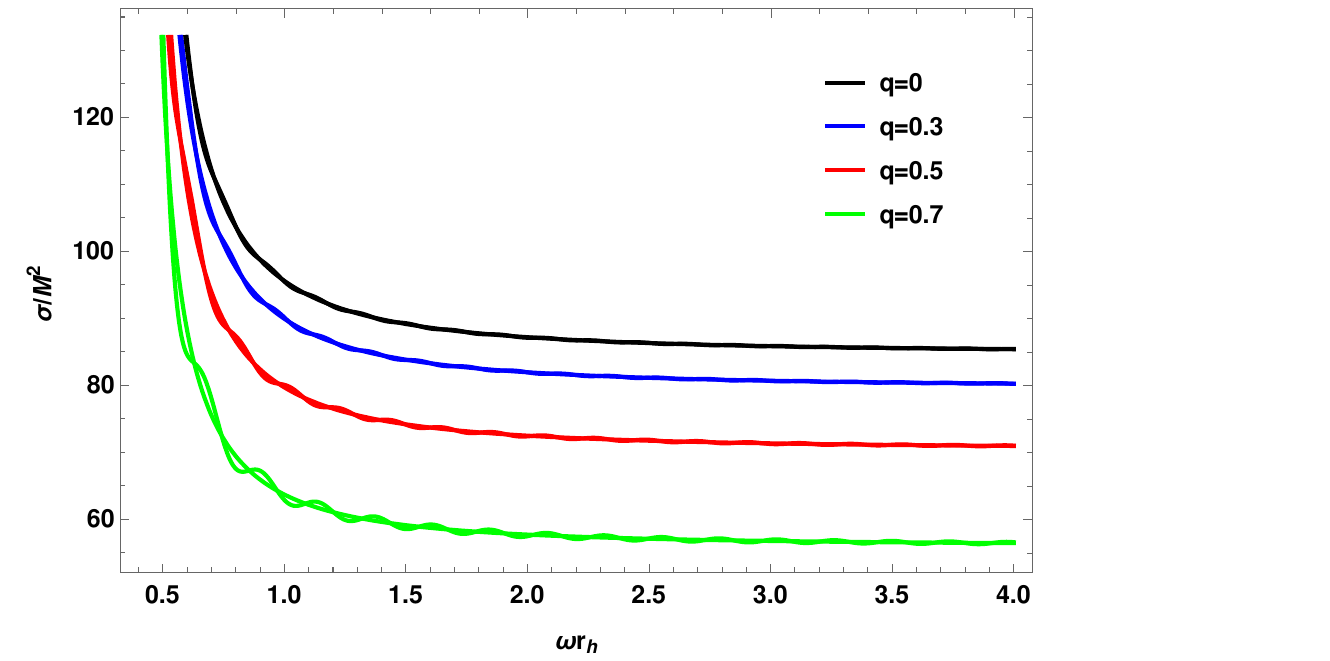}
	\caption{For a charged dilatonic black hole, it is shown the oscillatory absorption cross-section (wiggly curve)  along with the high-frequency cross-section in terms of $\omega r_{h}$ for different values of $q$. The other parameters are fixed at $Mm=0.4$ and $e=0$.}\label{fig:abs_mass3}
\end{figure}

To account for the entire absorption cross-section, we need to add the high-frequency contribution to the oscillatory part as we did before \cite{decanini2011universality}, \cite{benone2014absorption}.  Indeed, the oscillatory part is obtained by isolating the Regge poles\cite{decanini2011universality}, and it reads
\begin{eqnarray} \label{xz1}
	\sigma_\mathrm{abs}^\mathrm{osc}(\omega)&=&-\,
	8\pi\,vb_c\Lambda\, e^{-\pi vb_c\Lambda}\,\sigma_\mathrm{geo}
	\text{sinc}\left[2\pi v
	b_c\omega\right].
\end{eqnarray}
Eq. (\ref{xz1}) tells us that the oscillatory part is linked with the Lyapunov exponent, $\Lambda$,   as it is related to the unstable circular orbit.  The Lyapunov exponent is a measure of the average rate at
which nearby trajectories converge or diverge in the phase space \cite{cardoso2009geodesic}; it simply reads $\Lambda=[V''(r_c)/2 \dot{t}^2(r_c)]^{1/2}=\eta_{c}/vb_{c}$. This exponent is evaluated at the potential barrier peak. The numerical analysis shows that the greater $q$, the lower the values of $\Lambda$ are, or equivalently, the unstable circular orbits become more and more unstable as $q$  continues to take lower values. The total absorption cross-section $\sigma_{abs}=\sigma_{hf}+\sigma_\mathrm{abs}^\mathrm{osc}$ is depicted in terms of $\omega r_{h}$ for some values of $q$ [see Fig. (\ref{fig:abs_mass2})]. It is possible to show how the oscillatory part is mounted on the high-frequency absorption cross-section, as shown in Fig.(\ref{fig:abs_mass3}).
\subsection{ Numerical results for scattering}
\begin{figure}
	\centering
	\includegraphics[width=3.2in]{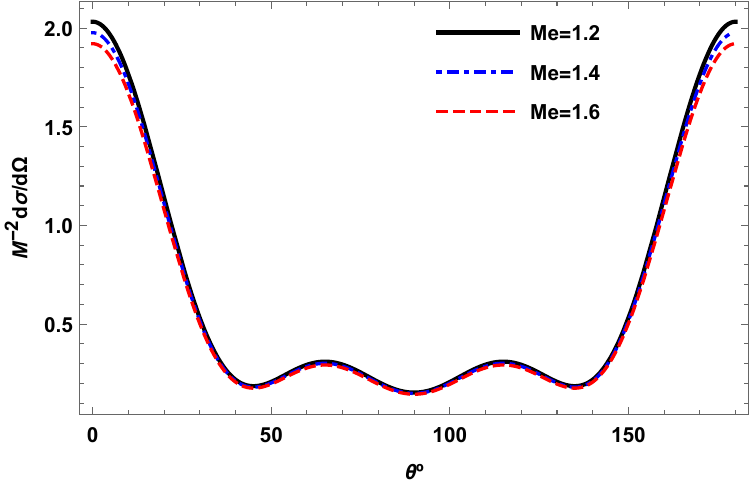}
	\caption{Differential scattering cross-section in terms of the angle for a dilatonic black hole  with $Mm=0.4$, $q=0.5$, $M\omega=3$ while varying $Me$.}\label{fig:f_carr}
\end{figure}

The main results for the scattering cross-section in the case of a charged massive scalar field are mentioned in the following. The new expressions can be derived by applying the same approach of Sec. IVD. By doing so, the frequency $\omega$ must be changed accordingly in order to account for the presence of the  charged field, in a similar fashion as it was done in Sec.V. The boundary conditions used for running the numerical simulation are listed in (\ref{eq:boundary}) and the transmission coefficient which later on is replaced in (\ref{eq:RR}) to get the partial scattering cross-section can be recast as 
\begin{eqnarray}\label{eq:numerica2}
	|T_{\omega \ell}|^2 = \Big\{\frac{1}{4}\Big[|\psi_{\omega \ell}|^2+\frac{1}{\bar{\rho}^2}\Big|\frac{d\psi_{\omega \ell}}{dx}\frac{dx}{dr}\Big|^2_{x=x_\infty}\Big]+\frac{\xi}{2\bar{\rho}}\Big\}^{-1}.~~~~~~
\end{eqnarray}
For  $M \omega = 3$, we numerically obtained the differential scattering cross-section in terms of the angle for some values of $Me$ [cf. Fig. (\ref{fig:f_carr})]. The general feature is a series of peaks with two maximums at the end of the interval. The variation of each height for different selections of $Me$ is tiny; in fact, the higher $Me$  is, then the lower values of $d\sigma_{s}/d\Omega$ are. The major distinction among these curves is close the critical angle $\pi$ due to the backward glory effect, as is expected for massive s-waves. A similar situation is achieved by taking diverse values of $Mm$. However,  higher amplitudes correspond to bigger values of the  $Mm$ parameter.
\begin{figure}
	\centering
	\includegraphics[width=3.2in]{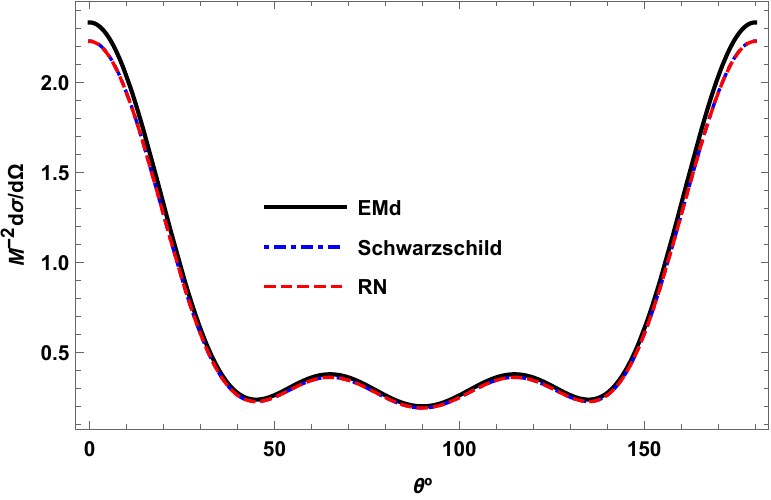}
	\caption{Differential scattering cross-section for Schwarzschild, RN, and dilatonic black holes with $Me=0.8$, $q=0.4$, $Mm=0.1$, $M\omega=3$. The selected maximum multipole was taken at $\ell_{max}=3$ due to the time-consuming effect of the numerical simulation.}\label{fig:f_c}
\end{figure}
Fig.(\ref{fig:f_c}) displays the differential scattering cross-section for Schwarzschild, RN, and dilatonic black holes.  The highest amplitude over the angle interval corresponds to the dilatonic black hole, even close to the critical angle $\pi$, whereas the  Schwarzschild and  RN black holes have almost identical amplitudes. 
\section{Superradiant scattering}
\subsection{Numerical results}
If a  charged massive scalar field is scattered off by a charged black hole, under certain conditions, the energy of the outgoing wave likely becomes greater than the energy of the incoming wave. As a result of that, the outgoing modes at spatial infinity are amplified \cite{superr}, \cite{benone2016superradiance}. The latter effect is known as the superradiant scattering \cite{superr}; a process by which the outgoing wave extracts energy from the black hole, in a similar fashion as it happens with Penrose's mechanism for extracting energy/angular momentum from a rotating black hole \cite{book3}, \cite{superr}\footnote{TThe seminal works of Penrose \cite{pe1}, and Christodoulou  \cite{pe2}  showed that energy can be extracted from a rotating black hole using orbiting particles. However, the possibility of extracting rotational energy by using a test field impinging on a rotating black hole was suggested by Misner in \cite{pe3}.} 
\footnote{Notice that the superradiance phenomenon is not restricted to a black hole with event horizon\cite{saa1}.  Other systems can exhibit this effect; see \cite{saa2} for more details.}.  A simple way to visualize the superradiant scattering effect is by considering that the event horizon acts as a membrane that loses energy. Numerical studies of the superradiance effect were carried out  for different kinds of black holes; namely,  static black holes \cite{benone2016superradiance}, stringy black holes \cite{li2013stability},  and Kerr-Newman  black hole \cite{bc2019}(see  \cite{superr}  more details).
\begin{figure}
	\includegraphics[width=3.2in]{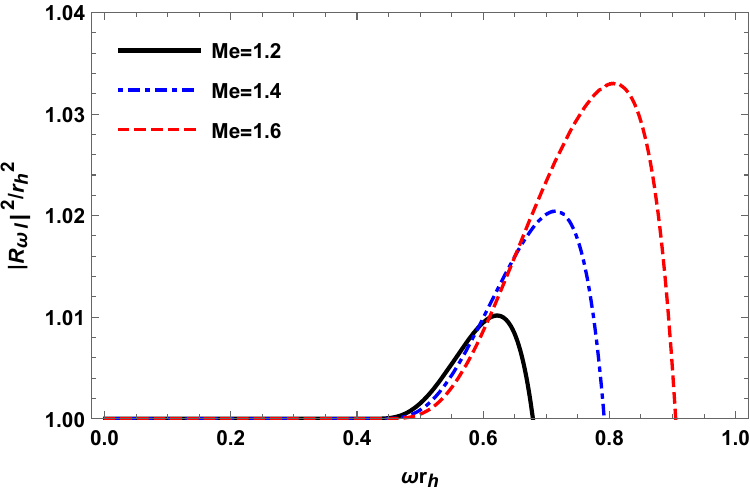}
	\caption{Reflection coefficient for a dilatonic black hole as a function of $\omega r_{h}$ while keeping fixed  $Mm=0.4$, $q=0.8$, $\ell=0$ but varying  $Me$.}\label{fig:sup_EMD}
\end{figure}
\begin{figure}
	\includegraphics[width=3.2in]{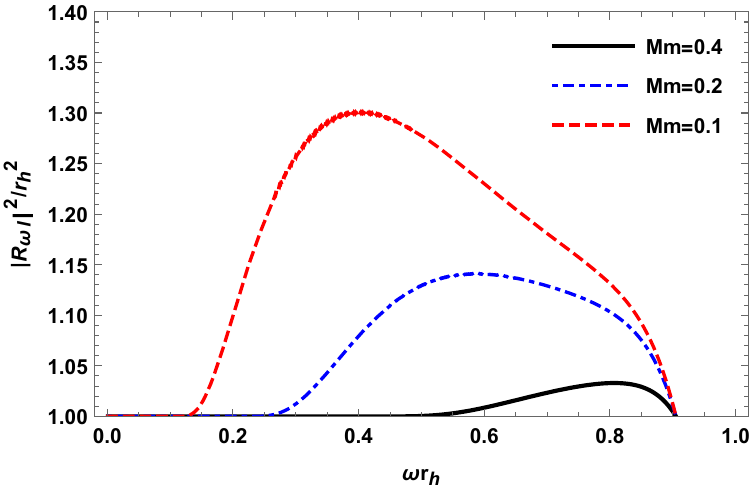}
	\caption{Reflection coefficient for a dilatonic black hole as a function of $\omega r_{h}$ while keeping fixed  $Me=1.6$, $q=0.8$, $\ell=0$ but varying  $Mm$.}	\label{fig:sup_EMD_carr}
\end{figure}
\begin{figure}
	\includegraphics[width=3.2in]{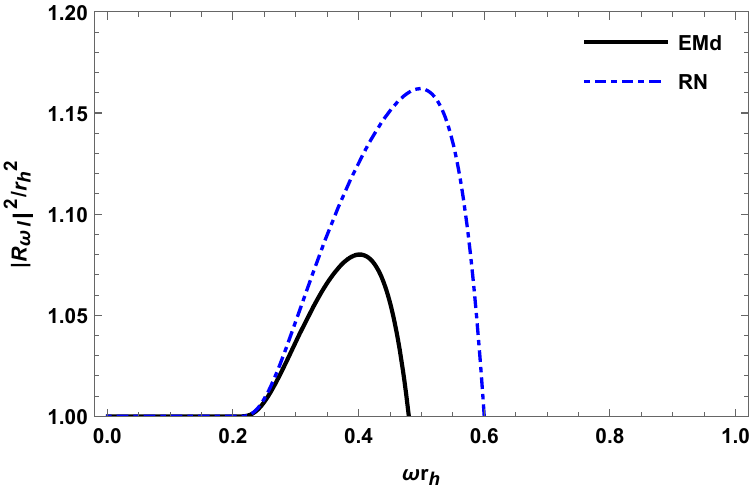}
	\caption{The reflection coefficient for RN black hole and dilatonic charged black hole are displayed for fixed $e=1.2$, $q=0.8$, and $\ell=0$.}\label{fig:sup_comp_carr} 
\end{figure}
Numerically speaking,  we can study the superradiance effect by solving (\ref{eq:numerica}) subject to the boundary conditions (\ref{eq:2.11b})-( \ref{eq:t2}). We look under which conditions the reflection coefficient  $|R_{\omega\ell}|^2>1$, signaling an amplification effect of the scattered waves.   In Fig.(\ref{fig:sup_EMD}), we illustrate the superradiant effect for a dilatonic charged black hole in terms of $\omega r_{h}$ by varying  the $Me$ parameter. These results indicate that higher values of $Me$ lead to higher amplitudes in the  $|R_{\omega\ell}|^2$ coefficient; however, we obtain the opposite effect by increasing the values of $Mm$ as is shown in Fig. (\ref{fig:sup_EMD_carr}). We end up with this section by comparing the superradiance effect between two charged black holes, the RN black hole, and the dilatonic black hole. It turns out the superradiance effect for a dilatonic black hole is less powerful than the RN case, at least, for the case of a massive charged scalar field impinging upon a dilatonic black hole [cf. Fig. (\ref{fig:sup_comp_carr})].
Different perturbation schemes could potentially enhance the superradiance effect for a dilatonic black hole. For instance,  the reflection coefficient could be more significant in the dilaton channel or the electromagnetic sector. As can been, for mild decoupling parameter $\omega r_{\rm{h}} \in (0.2, 0.6)$, the relative difference $||R|^{2}_{EMd}-|R|^2_{RN}|/|R|^{2}_{RN}$ increases up to $0.08$ near their peaks.

\subsection{Superradiant instability and the mirror mechanism}
Bardeen, Press, and Teukolsky \cite{sro1} studied the process of extracting energy from a rotating black hole along with its impact at the astrophysical level. Press and Teukolsky \cite{sro2} showed that the outgoing wave gained energy from the rotating black hole and could admit modes that change their main feature by developing a superradiant instability. By having a finite potential well outside the black hole, these amplified modes remain trapped,  generating the so-called  \textit{ black hole bomb}  mechanism \cite{sro2}. Later, Starobinsky explored the process of scattering of a scalar, electromagnetic and gravitational waves impinging on a Kerr black hole \cite{aas1}. Starobinsky and Churilov   focused on  electromagnetic and gravitational waves
 \cite{ac1}, reporting the existence of the superradiant effect with
classical waves, along with an estimation of the particle creation rate.  
For a massive scalar field,  Damour, Deruelle, and Ruffini  showed
that  a superradiant instability arises   since  the mass term acts as a reflecting mirror which  permits the formation 
of bound states (scalar clouds) around rotating compact objects
\cite{ddr}, \cite{ze}, \cite{de1}.

Nowadays, several mechanisms lead to a black hole bomb configuration. This configuration appears  in a reflecting mirror around a rotating black hole \cite{lemos1}, \cite{dego}, a massive vector field plus a rotating black hole \cite{pani}, and  with the inclusion of a magnetic field \cite{dubo}, \cite{kono}.  Regarding the case of non-rotating charged black hole impinging by a massive charged scalar field, the possibility of creating a black bomb configuration was explored by Hod in several seminal papers \cite{hod1}, \cite{hod2}, \cite{hod3}. The Reissner-Nordstr\"om black hole ( extremal or non-extremal ) do not meet the two necessary conditions for creating a charged black hole bomb; that is,  having a trapping potential outside the black hole and the existence of growing trapped modes \cite{hod1},  \cite{hod2},  \cite{hod3}. These findings would seem to support the conjecture that charged black holes remain stable against charged massive scalar perturbation. However, a year later, the same author showed that a charged black hole bomb exists as long as a lower bound on the charge-to-mass ratio is fulfilled. If this bound is held, then the superradiant instability will be triggered provided the trapping modes are confined due to a mirror-like boundary condition \cite{hod4}.

Several numerical pieces of evidence showed that a charged black hole in a cavity develops a superradiant instability for a massive charged scalar field; in fact, the final state of such configuration corresponds to a hairy black hole plus a boson condensate \cite{gual}. However,  the charged  RN  case impinged by a massive charged Dirac field does not develop a superradiance instability \cite{huang1}. For a massive charged scalar field, the superradiant instability developed around the Kerr-Newman background would seem more favorable than the one associated with a Kerr black hole \cite{huang2}.

East and Pretorius studied the development of a superradiant instability for massive vector bosons around a nearly extremal spinning black hole by including the full non-linear evolution/backreaction effects into the analysis  \cite{east1}. It was realized that instability is an efficient process to extract energy from a black hole. All the differential energy or angular momentum borrowed from the compact object help to form a massive vector cloud around it \cite{east1} for a short time. East showed the existence of four different phases:1- initial instability, 2-black bomb regime, 3-saturation regime,  and 4-final end-state \cite{east2}. Interestingly,  it was found that the production rate of gravitational waves is a subdominant process concerning the power released by the Proca field during the exponential instability  \cite{east2}.

Besides, the existence of a rotating black hole with a tiny massive scalar hair was reported in Ref. \cite{her1}; the transition from the initial superradiant instability until the final equilibrium state associated with a scalar cloud around the spinning black hole was also addressed  \cite{her1}. Finally, at the astrophysical level, the imprints left by scalar and vector fields near a rotating black hole undergoing a  superradiant instability were contemplated in Ref. \cite{wi1}. The production of gravitational waves generated by the ultralight vector field was examined in Ref. \cite{brito2}.

A charged rotating black hole with an AdS boundary and a massive scalar field also exhibit a superradiant instability\cite{reall}. Green, Holland, Ishibashi, and Wald presented a complete treatment of the superradiant instability for black holes with an asymptotically AdS boundary within the context of GR \cite{green1}. The non-linear numerical evolution of the charged black hole with an AdS boundary plus a scalar field revealed that the final system corresponds to a hairy black hole \cite{green2}. In fact, after the initial unstable state, the system exhibits a transition towards an equilibrium configuration characterized by a fundamental frequency, different from the initial one, which triggered the superradiant instability. During the dynamical evolution, the exchange of mass and charge reduces the black hole mass, but the final mass will always be greater than the initial one because the area theorem must hold. The latter statement can be confirmed through the first and second law of black hole thermodynamic\cite{beke}. In  $D$-dimensions,  Ads charged black holes can  develop superradiant instability  not only in GR  \cite{huang3},\cite{ros1},  but also in  $f(R)$-gravity \cite{ros2}. Numerical simulations showed that a rotating AdS$_{4}$ black hole can trigger superradiant instabilities in several channels \cite{ros3}.

 
In order to a massive charged scalar field exhibits a bound states $(\omega^2<m^2)$ and a superradiance effect ($\omega<eA_{t}(r_{\rm{h}})$), the frequency must lies on the following interval  \cite{hod4}, \cite{li1}, \cite{li2}:
\begin{equation}
\label{bic}
  0<\omega<{\rm{Min}}\{eA_{t}(r_{\rm{h}}),m\}.
\end{equation}
However, this result (\ref{bic}) does not prove necessarily that the charged massive scalar field is trapped outside the black hole, generating a superradiant instability. To achieve that goal a mechanism  based on the  mirror-like boundary condition must be imposed \cite{dego}, \cite{hod4}, \cite{li2}.  In other words,  we place a reflecting mirror for $r>r_{\rm{h}}$ at the location $r=r_{\rm{m}}$, demanding that none part of the scattered wave could be transmitted through it, $\psi_{\omega}(r=r_{\rm{m}})=0$. A similar argument was used by Hod \cite{hod4} and Li \cite{li2}, independently.

Our next step is to consider the Schr\"odinger equation (\ref{EMD_2}) along with the total effective potential $V_{\rm{total}}=V-[\omega-eV_{q}]^{2}$, where $V$ is given in (\ref{eq:potencial_carregado}). 
For practical reasons, it makes sense to define $V=f \mathcal{H}/r^2$ \cite{hod4}. Let us begin by studying the local behavior of $\psi_{\omega \ell m}$ near the horizon; namely, we wish to determine whether  $\psi_{\omega \ell m}$ has a maximum or minimum outside the horizon. To do so, we introduce a new variable $\mathfrak{Z}=(r/r_{\rm{h}})-1 \in (0, \infty)$. Using the superradiance condition, that is $\omega\ll [eq\sqrt{2}/2+ek]$, we  get rid of the terms which depend on the frequency in  $V_{\rm{total}}$, so we ended up with a potential proportional to the $\mathcal{H}$-term,
\begin{eqnarray}\label{pfx}
	V_{\rm{total}}=\frac{\mathcal{H}(r_{\rm{h}})\mathfrak{Z}}{r^{2}_{\rm{h}}}+\mathcal{O}(\mathfrak{Z}^2),
\end{eqnarray}
where
\begin{eqnarray}
\mathcal{H}(r_{\rm{h}})=1+  m^2 r^2_{\rm{h}} +\frac{\ell (\ell+1)}{(1-q^2)}+\frac{q^2}{2 \left(1-q^{2}\right)}.
\end{eqnarray}

First, we note that  $\mathcal{H}(r_{\rm{h}})$ is positive defined provided $0<q<1$ for a dilatonic black hole. As a consequence of this, $V_{\rm{total}}$ is also positive near the event horizon. Now, we replace (\ref{pfx}) in the Schr\"odinger equation (\ref{EMD_2}), yielding  to the next differential equation in terms of $\mathfrak{Z}$-variable
\cite{abramowitz1988handbook},
\begin{eqnarray}\label{eq:S1}
	\Big(\mathfrak{Z}^2\frac{d^2}{d\mathfrak{Z}^2}+ \mathfrak{Z}\frac{d}{d\mathfrak{Z}}- \mathcal{H}(r_{\rm{h}}) \mathfrak{Z}\Big)\psi_{\omega \ell m}=0,
\end{eqnarray}
which has as a general solution a linear combination of the modified Bessel function of order zero ($\nu=0$),
\begin{eqnarray}\label{sols}
\psi_{\omega \ell m}=c_{1}I_{0}[\sqrt{2\mathcal{H}(r_{h})\mathfrak{Z}}] + c_{2}K_0[\sqrt{2\mathcal{H}(r_{h})\mathfrak{Z}}]. 
\end{eqnarray}
In order to keep the solution regular as $\mathfrak{Z} \rightarrow 0$, the coefficient  $c_{2}$ must vanish. The solution $\psi_{\omega \ell m}$  is then a positive and convex function close to the horizon. To carry on, we need to investigate whether it is possible to impose a reflecting mirror-like boundary condition at $r_{\rm{m}}$. In other words,  we must prove that that the function   $\psi_{\omega \ell m}$ has at least a maximum located at  $r_{\rm{max}}$ such that  $r_{\rm{h}}<r_{\rm{max}}<r_{\rm{m}}$. This is equivalent to show that  $\partial_{xx}\psi_{\omega \ell m}(r_{\rm{max}})=V_{\rm{total}}(r_{\rm{max}})<0$ \cite{hod4}. Taking into account the total effective potential evaluated at $r_{\rm{max}}$,  and replacing the value of the critical frequency, $\omega_{c} = e(q\sqrt{2}/2+k)$,  it follows that the total potential becomes
\begin{eqnarray}
	&& V_{\rm{total}}=\Big(\frac{r_{\rm{max}}-r_{\rm{h}}}{r_{\rm{max}}}\Big)\frac{\mathcal{H}(r_{\rm{max}})}{r_{\rm{max}}^2}-\frac{e^2 q^2}{2}\Big(\frac{r_{\rm{max}}-r_{\rm{h}}}{r_{\rm{max}}}\Big)^2~~~~~~~
\end{eqnarray}
which can be translated in the following inequality,
\begin{eqnarray}\label{eq:eQ}
	\frac{eq}{\sqrt{2}}>\sqrt{\frac{r_{\rm{max}}}{(r_{\rm{max}}-r_{\rm{h}})}\frac{\mathcal{H}(r_{\rm{max}})}{r_{\rm{max}}^2}}.
\end{eqnarray}
However,  we can improve the latter lower bound (\ref{eq:eQ}) by using the following second condition $r_{\rm{max}}<r_{\rm{m}}$. Indeed,  we may  employ  the former  inequality to change the constraints on $r_{\rm{max}}$ by one  inequality which involves $r_{\rm{m}}$,
\begin{eqnarray}\label{sc2}
	\frac{r_{\rm{max}}}{(r_{\rm{max}}-r_{\rm{h}})}>\frac{r_{\rm{m}}}{(r_{\rm{m}}-r_{\rm{h}})},\quad \frac{\ell(\ell+1)}{r_{\rm{max}}^2}>\frac{\ell(\ell+1)}{r_{\rm{m}}^2}\nonumber\\
\end{eqnarray}
Replacing  (\ref{sc2}) in (\ref{eq:eQ}), we arrive at an inequality in terms of $eq$ and the position where the  reflecting mirror is located, $r_{\rm{m}}$,
\begin{eqnarray}\label{eq:eQ2}
	\frac{eq}{\sqrt{2}}>\sqrt{\frac{r_{\rm{m}}}{(r_{\rm{m}}-r_{\rm{h}})}\frac{\mathcal{H}(r_{\rm{m}})}{r_{\rm{m}}^2}}.
\end{eqnarray}
To this extent, we obtained a lower bound for the quantity $eq$ such that the unstable modes remain trapped in a well potential for $\mathcal{H}(r_{\rm{m}})>0$, indicating that the perfect mirror condition is a suitable mechanism for the creation of a charged black hole bomb as happens in the RN case \cite{hod4}. Nevertheless, we must verify that the $\mathcal{H}$ function is always positively defined outside the horizon. To show this numerically, it is useful to  introduce some new variables called $0<s=r_{\rm{h}}/r_{\rm{m}}<1$, $0<q<1$, and $\bar{m}=2Mm>0$. Then, Eq. (\ref{eq:eQ2}) can be recast as follows,
\begin{eqnarray}\label{eq:eQF}
	\frac{e}{m}>\sqrt{\frac{2s^2}{(1-s)}\frac{\mathcal{H}}{q^2 \bar{m}^2}},
\end{eqnarray}
where the $\mathcal{H}$ function  depends on dimensionless variables/parameters,
\begin{eqnarray}\label{eq:HH}
	\mathcal{H}=s+\frac{\ell(\ell+1)}{(1-q^2 s)}+ \frac{s^2q^2}{2(1-sq^2)}-\frac{s^2 q^4(1-s)}{4(1-sq^2)^2}+\frac{\bar{m}^2}{s^2}.~~~~
\end{eqnarray}
Eq. (\ref{eq:HH}) indicates that the $\mathcal{H}$ function remains positive on the interval of interest. For instance, we can check this point by taking  $\ell=1$ and small scalar field mass, $\bar{m}=0.6$,  while plotting this function over the interval $0<s<1$ and $0<q<1$.  We also consider the   $\ell=1$ case with large scalar field mass, $\bar{m}=1.6$. In all the cases explored, we obtain that (\ref{eq:eQF}) satisfies the well-known lower bound $e/m\gtrsim 1$; in agreement with the RN case \cite{hod4} and the numerical simulation reported in \cite{gual}. Even though these charged black holes are not found in Nature, they represent an excellent toy model for helping us understand better the theoretical side of them, enlarging our views on black holes in extended gravity theories.\footnote{An important fact concerning the superradiance/superradiant instability for charged dilatonic black holes is that the reflection amplitude is lower than the Reissner-Nordstr\"om case. It would seem that adding a new degree of freedom does not necessarily help enhance the superradiance effect, at least for perturbation with a scalar field. One way to understand this point is by noticing that both theories are not equivalent. Starting from $\phi=\phi_{0}=cte$ in the EMd theory, we obtain that $F^2=0$ which can be satisfied by    $F_{\mu\nu}=0$,  leading to $R_{\mu\nu}=0$.}. 

\section{Dilaton perturbation and superradiant instability}
Whereas in Sec.VIA, the superradiant scattering topic, was partially addressed by considering that the perturbation of the dilaton field can be accommodated as a charged massive scalar field; it is time to study the case in which the dilaton perturbations of the original Lagrangian are taken into account. As the dilaton field is part of the fundamental fields which generate the black hole is not evident if this procedure is well-defined. To the best of our knowledge, under certain conditions, this procedure was applied for studying the matter near black holes within the context of scalar-tensor theory \cite{cac} and for exploring the existence of different types of scalar field instabilities in the EMd model \cite{nh3}. 
Hence, we write the perturbed dilaton field as $\phi=\phi_{{\rm{bg}}}+ \Pi$ with $\Pi\ll \phi_{{\rm{bg}}}$ and consider a quadratic perturbed Lagrangian for the dilaton field as follows,
\begin{eqnarray}\label{pdd}
\Delta {\mathcal{L}}_{d}=-{\mathcal{D}}_{\mu}\Pi\big({\mathcal{D}}^{\mu}\Pi\big)^{\dagger} - \mu^{2}_{\Pi}\Pi \Pi^{\dagger}-\lambda_{g}(\phi_{\rm{bg}})\Pi\Pi^{\dagger} F^{2}_{\rm{bg}},~~
\end{eqnarray}
where the covariant derivative is defined as ${\mathcal{D}}_{\mu}=\nabla_{\mu}-iq_{d}A^{{\rm{bg}}}_{\mu}$ being $q_{d}$ the dilaton charge. The coupling function is $\lambda_{g}(\phi_{\rm{bg}})=\lambda_{0} e^{-2\phi_{\rm{bg}}}$ with $0<\lambda_{0}\leq 1$. The index ${\rm{bg}}$ refers to background quantities that are not perturbated. In addition to the kinetic term for the charged dilaton field, there is a canonical mass term and a quadratic coupling between the dilaton field and the background Maxwell kinetic term. However, first, let us explain the physical reason for selecting the Lagrangian (\ref{pdd}). In order to have superradiance scattering, the field must have a well-defined asymptotically mass, so this property explains the mass term. Second, perturbating the original field equations by considering only the dilaton field and keeping the metric and gauge field frozen, we obtain that the field equations for the background remain unaltered as long as the mixed terms  $\mathcal{O}(\partial\phi_{\rm{bg}}\partial \Pi)$ and $\mathcal{O}(\Pi T^{em}_{\mu\nu})$ can be neglected \cite{cac}. In fact, setting $q_{d}=0$ and $\lambda_{0}=1$, the field equation for the chargeless dilaton  is recovered; namely, $\Box \Pi=\Pi  e^{-2\phi_{\rm{bg}}}F^{2}_{\rm{bg}}$. We go beyond this particular case by including a well-defined mass term or potential for the perturbed dilaton with charge $q_d$  along with an arbitrary coupling $\lambda_{0}$. \footnote{Notice that the perturbed dilaton field can be considered complex,  yielding two equivalent field equations. Besides, the quadratic coupling $\Pi\Pi^{\dagger} F^{2}_{\rm{bg}}$ leads an effective/runaway mass, so it cannot be identified as the mass of the perturbed dilaton. Moreover, the dilaton model can generally admit radiative corrections, which also neglected. Besides, we  employed the gauge condition:
 $\nabla^{\mu}A^{{\rm{bg}}}_{\mu}=0$.}

\begin{figure}
	\includegraphics[width=3.4in]{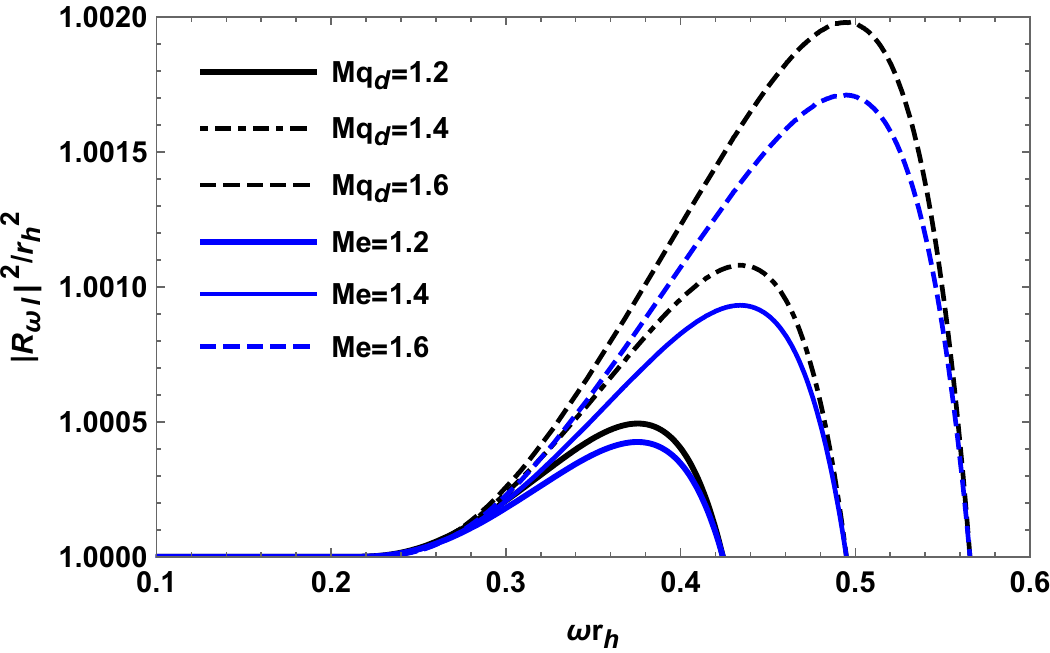}
	\caption{The reflection coefficient in terms of  $\omega r_{\rm{h}}$  for the dilatonic black hole is associated with two different perturbation methods. The black curves correspond to the perturbation method mentioned in (\ref{pdd}) whereas the blue  curves  stand for the perturbation encoded in (\ref{eq:perturEq}). The parameters are fixed at  $M\mu_{\Pi}=0.2$, $q=0.5$, and  $\ell=1$.}\label{compa1x} 
\end{figure}
\begin{figure}
	\includegraphics[width=3.4in]{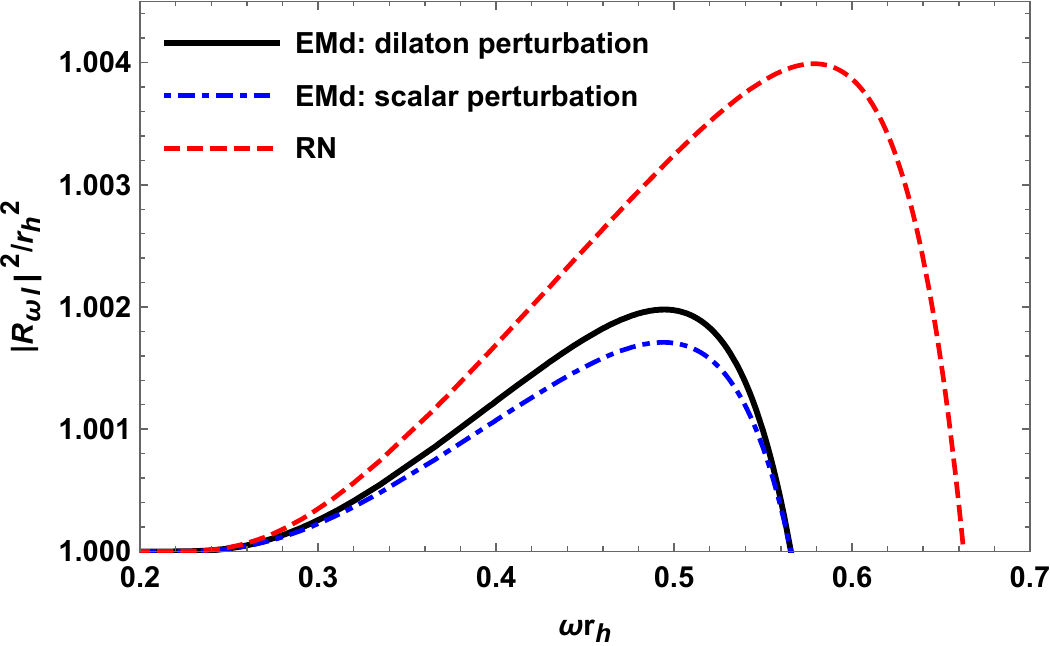}
	\caption{It is shown the reflection coefficient in terms of  $\omega r_{\rm{h}}$  for the dilatonic black hole associated with two different perturbation methods and the  Reissner-Nordstr\"om  case (red curve). The black curve corresponds to the perturbation method mentioned in (\ref{pdd}) whereas the blue  curve  stands for the perturbation encoded in (\ref{eq:perturEq}). The other parameters are fixed at  $M\mu_{\Pi}=0.2$, $q=0.5$, $Mq_{d}=1.6$, $Me=1.6$, and  $\ell=1$.}\label{compa2x} 
\end{figure}

As we mentioned above,  we need to inspect the impact of this perturbation in the superradiance scattering of the dilaton field. Therefore, we followed the same approach and method developed in Sec. VIA, computing the reflection coefficient for several physical situations. The numerical simulations indicate that the reflection coefficient increases for larger values of $Mq_{d}$ and smaller values of $M\mu$. We contrast two different perturbation methods, the one mentioned in Sec. VIA and the other related to Eq. (\ref{pdd}). To do so, we consider the same values of $Mq_{d}$ and $Me$ and plot the reflection coefficients in terms of $\omega r_{\rm{h}}$ as is depicted in Fig. (\ref{compa1x}). The second perturbation scheme seems to enhance the reflection coefficient by increasing its amplitude as $Mq_{d}$ varies from $1.2$ to $1.6$. Another point of the debate is related to the impact of the second perturbation scheme (\ref{pdd}) concerning the first perturbation method (\ref{eq:perturEq}), and also, in comparison to the  Reissner-Nordstr\"om  case. Fig. (\ref{compa2x}) displays these three possibilities. Both perturbation schemes lead to lower amplitudes in the reflection coefficient concerning the  Reissner-Nordstr\"om  case. However, the second perturbation scheme produces an enhancement in comparison to the first method (\ref{eq:perturEq}). In a way,  the latter fact tells us that the second scheme provides a better consistent treatment for dealing with the dilaton perturbations. 

We examine the impact of the second perturbation method for the existence of unstable modes which remain trapped outside the black hole once the reflecting mirror boundary condition is imposed. We will stress the main results as a similar method was employed in Sec. VIB. For instance, we arrive at a new lower bound for the charge-to-mass ratio,
\begin{eqnarray}\label{compaxy2}
	\frac{q_{d}}{\mu_{\Pi}}\gtrsim \frac{0.7}{q}\sqrt{\Big(1+\frac{s^2}{\bar{m}^2} J(s, q)\Big)\frac{1}{(1-s)}},
\end{eqnarray}
where
\begin{eqnarray}\nonumber
	J&=&s+\frac{\ell(\ell+1)}{(1-q^2 s)}+ \frac{s^2q^2}{2(1-sq^2)}-\frac{s^2 q^4(1-s)}{4(1-sq^2)^2}\\
	&-&\lambda_{0}q^2s^2(1-sq^2). \label{J1x}
\end{eqnarray}
The function (\ref{J1x}) is strictly positive on the intervals $0<s=r_{\rm{h}}/r_{\rm{m}}<1$ and $0<q<1$ regardless the values of $0<\lambda_{0}\leq 1$.  Yet another effect of incluiding the dilaton perturbations into our analysis is that the new lower bound satisfies the following relation,
\begin{eqnarray}\label{compaxz1}
	\frac{q_{d}}{\mu_{\Pi}}<\frac{e}{m},
\end{eqnarray}
for different values of $\lambda_{0}$, $\ell$, and $\bar{m}$. In the superradiance scattering analysis or superradiant instability, we considered $M\mu_{\Pi}=\mathcal{O}(0.1-1)$. Let us restore the fundamental constant to obtain the typical values of $M_{BH}$ as a proper estimation for future analysis and to contrast with the exploratory analysis presented in Sec. VA. Recently,  a light dilaton scalar field was considered as a new window that potentially could appear as part of the UV completion of the standard model, and the Higgs boson at the LHC \cite{lhc}. The latter analysis put the dilaton mass in the interval $10{\rm{Gev}}/c^2\leq \mu_{\Pi}\leq 300 {\rm{Gev}}/c^2$. Taking into account this constraints with the expression  $(GM/\hbar c)\mu_{\Pi}=\mathcal{O}(1)$, we obtain that the black hole mass associated with the second perturbation scheme for the dilaton must live in the following interval,  $ 4.45 \times 10^{-22}M_{\odot}\leq M_{BH}\leq 1.23 \times 10^{-22}M_{\odot}$. This mass range corresponds to primordial black holes; the dilaton channel leads to black hole masses with three orders of magnitude below the standard RN case \cite{hodf}. The superradiant instability in the dilaton channel will happen as long as its typical time-scale remains several order the magnitudes greater than the lifetime of the light dilaton, $\tau_{\rm{ins}}\gg \mathcal{O}(10^{-27}) s$ for $\mu_{\Pi}=300 {\rm{Gev}}/c^2$.   A further analysis based on the quasinormal modes will corroborate this estimation, as mentioned elsewhere.  

\section{Conclusions}
In this paper, we have given an overview of the process of scattering and absorption of a massless scalar field impinging on a charged dilatonic black hole within the framework of EMd gravity. While some of the results of Sec. III appears elsewhere; we motived these results by confronting the analytical formulae with several numerical simulations. We obtained the differential cross-section in three different ways. First, we compared it with the one associated with the backward glory effect and the semi-analytic estimation of the glory effect based on the logarithmic approximation, which seems to differ by $1\%$ from the standard estimation of the glory effect. Near the critical angle, $\theta=\pi$, we found that the larger the charge-to-mass ratio $q$, the smaller the amplitude of the differential cross-section becomes. The dilatonic charged black hole has smaller amplitude, near the critical angle than the Reissner-Nordstr\"om  case. Using the partial wave method for a massless scalar field, we numerically integrated the Schr\"odinger-like equation to reconstruct the partial and total absorption cross-sections in terms of the decoupling parameter $M\omega$. For instance,  we showed that the total absorption cross-section has smaller amplitudes than Reissner-Nordstr\"om  for mild frequency $M\omega \in (0.4, 1.2)$. We carried on by considering the absorption cross-section in the limit of high-frequency and discovered the existence of two different kinds of complex behaviors, also known as the fine structure and the hyperfine structure. 
We have validated these results by estimating the relative errors among the numerical solution and the approximated absorption cross-sections.  

We have also investigated a massive charged scalar field impinging on a dilatonic charged black hole. We showed that the absorption cross-section could be derived by using a matching method to construct a global solution on the interval of interest $r \in (r_{\rm{h}}, \infty)$ in the limit of low-frequency. Moreover, by doing so, we showed the existence of two phases which are separated by a critical velocity parameter, $v_{c}$; these parameter depends on the mass of the compact object and the mass of the scalar field. We have confirmed our previous analysis by performing a numerical simulation of the total absorption cross-section for any frequency. For completeness, we have provided a complete comparison of our numerical value of $v_{c}$ and the typical velocity of dark matter in different scenarios, including black holes from stellar masses to supermassive ones. In addition to that, we obtained the differential scattering cross-section for different values of field masses and charges, concluding that a light charged scalar field enhanced the differential scattering cross-section. We have determined numerically the reflection coefficient associated with the scattered waves. We found a superradiant phenomenon because the squared of the reflection coefficient takes values bigger than the unity for different values of mass' field (or the charge's field). We have verified that smaller values of scalar field mass enhance the superradiance scattering but larger values of the scalar field charge. In the case of moderate frequency,  we have found that the superradiant effect is lessened concerning the Reissner-Nordstr\"om case. To better understand the role played by these unstable modes, we have examined the mechanism that led to superradiant instability.

The superradiant instability's success lies in whether a potential well can be formed in the region outside the event horizon to enclose the growing modes. The mechanism is known as the reflecting-mirror boundary condition. Further, we derived an analytical formula (lower bound) that determines the minimal value of the charge's field $eq$, triggering the superradiant instability. We ended up in Sec. VII  with a new perturbation scheme that contemplated the dilaton perturbations but kept the other fundamental fields frozen (\ref{pdd}). The numerical simulation showed that the amplitudes of the reflection coefficient are amplified. Moreover, the dynamical superradiant instability with the mirror-reflecting mechanism introduced a new lower bound for the ratio $q_{d}/\mu_{\Pi}$ (\ref{compaxy2}).

A point that should be addressed shortly is a fully nonlinear numerical simulation of the complete field equation of motions to reply to an appealing question: What is the system's final state composed of charged massive scalar field plus a dilatonic black hole? Recent nonlinear simulations seem to indicate that the final state could be a hairy black hole along with a condensate scalar field. Furthermore, these findings would indicate that it is possible to extract only a fraction of the total charge of the black hole \cite{panif}.

\acknowledgments
We want to thank Prof. A. Starobinsky for his valuable comments and suggestions.  M.G.R. is partially supported by  Funda\c{c}\~ao de Amparo \`a Pesquisa e Inova\c{c}\~ao Esp\'irito Santo (FAPES, Brazil) grant under the PPGCosmo Fellowship Programme. E.V.L.M is supported by  Coordena\c{c}\~ao de Aperfei\c{c}oamento de Pessoal de N\'ivel Superior(CAPES, Brazil).  J.C.F. is supported by Conselho Nacional de Desenvolvimento Cient\'ifico e Tecnol\'ogico (CNPq, Brazil) and  FAPES.

\appendix
\section{Penrose diagram}\label{Diagrama}
\begin{figure}
	\includegraphics[width=3.2in]{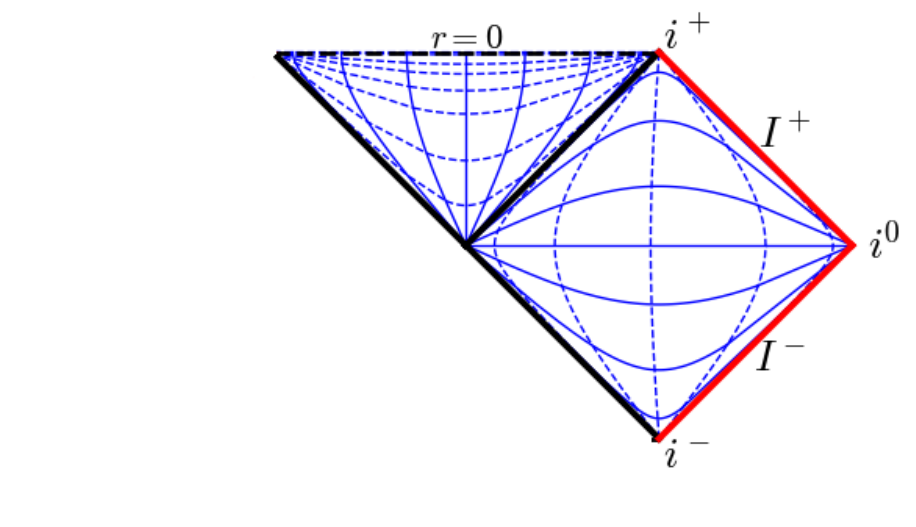}
	\caption{Penrose diagram for a dilatonic black hole, the bold red line corresponds to the event horizon whereas the dashed black line  corresponds to the spacelike singularity.}\label{fig:pen}
\end{figure}

Given the metric associated with a static dilatonic black hole,
\begin{equation}
	ds^2= -f(r)dt^2+f^{-1}(r)dr^2+r^2g(r)d\Omega^2,
\end{equation}
with $ - \infty <t <\infty $ and $2M <r <\infty $, we like to find its corresponding Penrose diagram. Let $x$ be the tortoise coordinate 
\begin{eqnarray}\label{x0}
	x=r+2M\ln\Big|\frac{r}{2M}-1\Big|,\quad\quad (0<x<\infty),
\end{eqnarray}
we can rewrite the metric as
\begin{eqnarray}
	ds^2= f(r(x))(-dt^2+dx^2)+r^2(x)g(r(x))d\Omega^2. 
\end{eqnarray}
Defining null coordinates,
\begin{eqnarray}
	\tan(V)=v=t+x,\quad\quad\tan(U)=u=t-x,
\end{eqnarray}
on the interval $ - \pi/2 <V <\pi/2 $ and $  - \pi/2 <U <\pi/2  $, we can cover the entire spacetime at once; that is, 
\begin{equation}
	ds^2=-\frac{f(r)dU dV}{\cos^2(U)\cos^2(V)}+ \left(\frac{r}{x}\right)^2 \frac{g(r)\sin^2(V-U)}{4\cos^2(U) \cos^2(V)}d\Omega^2,
\end{equation}
where $r$ is defined  in term of $u$ and $v$ as 
\begin{equation}
	\frac{1}{2}(v-u)=r+2M\ln\Big|\frac{r}{2M}-1\Big|.
\end{equation}
We now introduce a conformally related metric, $ds^2 = C_S^{-2}d\tilde s^2$, where conformal factor reads
\begin{eqnarray}
	C_S^{-2} = \frac{1}{4\cos^2(V)\cos^2(U)}.
\end{eqnarray}
The line element of conformal metric becomes
\begin{eqnarray}
	d\tilde s^2 = -4f(r)dU dV+ \left(\frac{r}{x}\right)^2 g(r)\sin^2(V-U)d\Omega^2.
\end{eqnarray}
Note that at the limit where $ r \to 2M $, the line element is positive defined. The conformal spacetime is regarded as a manifold $M$ with boundary $\partial M$: the coordinates are $(U, V, \theta,\phi)$ with
$(\theta,\phi)$ the standard angular coordinates on the unit sphere and the boundary is the union of the sets future null infinity $I^+$, past null infinity $I^-$, future timelike infinity $i^+$, past timelike infinity $i^-$ and spacelike infinity $I^0$, defined by
\begin{eqnarray}
	I^+&=&\{(U, V, \theta,\phi)|U\in(-\pi/2,\pi/2), V=\pi/2\}\nonumber\\
	I^+&=&\{(U, V, \theta,\phi)|V\in(-\pi/2,\pi/2), U=-\pi/2\}\nonumber\\
	i^\pm&=&\{(U, V, \theta,\phi)|U=\pm\pi/2, V=\pm\pi/2\}\nonumber\\
	i^0&=&\{(U, V, \theta,\phi)|U=-\pi/2, V=\pi/2\}
\end{eqnarray}
With this we can identify the different regions in the spacetime: $ i ^ 0 $, $ i ^ {\pm} $ and $ I ^ {\pm} $. $ i ^ 0 $ is the spacelike infinite that is, when $ r \to \infty $ and $ t $ finite. $ i ^ {\pm} $ are the future and past timelike infinities, that is $ t \to \pm \infty $ and $ r $ finite. Finally, $ I ^ {\pm} $ are the future null infinites ($ t \to \infty $, $ r \to \infty $ and
$ rt $ finite) and past ($ t \to- \infty $, $ r \to \infty $ and $ r + t $ finite). The Penrose diagram, with all regions, is drawn in the Fig.(\ref{fig:pen}). 

Null geodesics begin at a point of $I^-$ and end at a point of $I^+$. The timelike geodesics of the spacetime begin at $i^-$ and end at $i^+$, whereas spacelike geodesics start and end at $i^0$. Near the boundary, $r\to\infty$, then $r/x\to 1$ and consequently $f(r)\to 1$ and $g(r)\to 1$. In this way,  the metric can be approximated by
\begin{equation}
	ds^2=-4dU dV+ \sin^2(V-U)d\Omega^2.
\end{equation}
Note that the null infinity of this solution is the same as the Schwarzschild black hole \cite{book3}. A point that not always is true, specially for class of metric that are not asymptotically flat. The degenerate metric is obtained by taking $V=\pi/2$ and $dV = 0$ in above formula, 
\begin{eqnarray}
	d\tilde s^2 = \cos^2(U)d\Omega^2.
\end{eqnarray}
This shows that $I^+$ is a null surface whose cross-sections ($U$ =constant) are spheres of radii $\cos(U)$. This radius
vanishes at $i^0$ and $i^+$, showing that $i^0$ and $i^+$ (and similarly $i^-$) are points, not 2-spheres. Fig.(\ref{fig:pen}) is a standard representation (Penrose
diagram) of the conformal spacetime: the axis are
$T = (U + V )/2$ and $X = (V-U)/2$, the coordinates
$(\theta, \phi)$ are suppressed, so every interior point in the diagram
represents a sphere of radius $(\frac{r}{x}\sin(V-U))$. Points at $I^+ (I^-)$ represent
spheres of radii $\cos(U) (\cos(V ))$, whereas $i^0$ and $i^{\pm}$ represent
actual points. 

Remembering that the area of the sphere goes to zero for $r = Q^2/M$ and the surface is singular. Fig. (\ref{fig:pen}) is plotted under  the condition $Q^2<2M^2$,  ensuring that the physical singularity is hidden by the event horizon.

For a particle moving in this spacetime, the Lagrangian for free particle is written as
\begin{equation}\label{Eq:idt}
	\mathcal{L}=\frac{1}{2}g_{\mu\nu}\dot{x}^\mu\dot{x}^\nu.
\end{equation}
In this case, the motion constants in the $\theta=\pi/2$ plane are
\begin{eqnarray}
	E = f\dot t,\qquad
	L = \dot\phi r^2g(r)\label{eq:fg0}
\end{eqnarray}
Substituting (\ref{eq:fg0}) in (\ref{Eq:idt}), we arrive at one dimensional effective problem in the radial coordinate,
\begin{eqnarray}\label{eq:fg2}
	\dot r^2 &=& E^2-f(r)\Big(\frac{L^2}{r^2g(r)}+\epsilon\Big)
\end{eqnarray}
where $\epsilon = 1, 0,-1$ for timelike, null, and spacelike geodesics, respectively. Here the overdot means derivative with respect to proper time/affine parameter. The effective potential is given by 
\begin{eqnarray}
	V_{eff} = \left(\frac{L^2}{r^2 \left(1-\frac{Q^2}{Mr}\right)}+\epsilon-\frac{2ML^2}{r^3 \left(1-\frac{Q^2}{Mr}\right)}-\frac{2M\epsilon }{r} \right).\nonumber\\
\end{eqnarray}
The effective potential has a positive constant term  for $\epsilon=1$ (timelike geodesics) as $r\to \infty$. Furthermore, considering $Q^2<2M^2$, the effective potential has a similar behavior to the Schwarzschild spacetime.
	
\section{Logarithmic approximation for the backward glory effect}
\begin{figure}
	\includegraphics[width=3.2in]{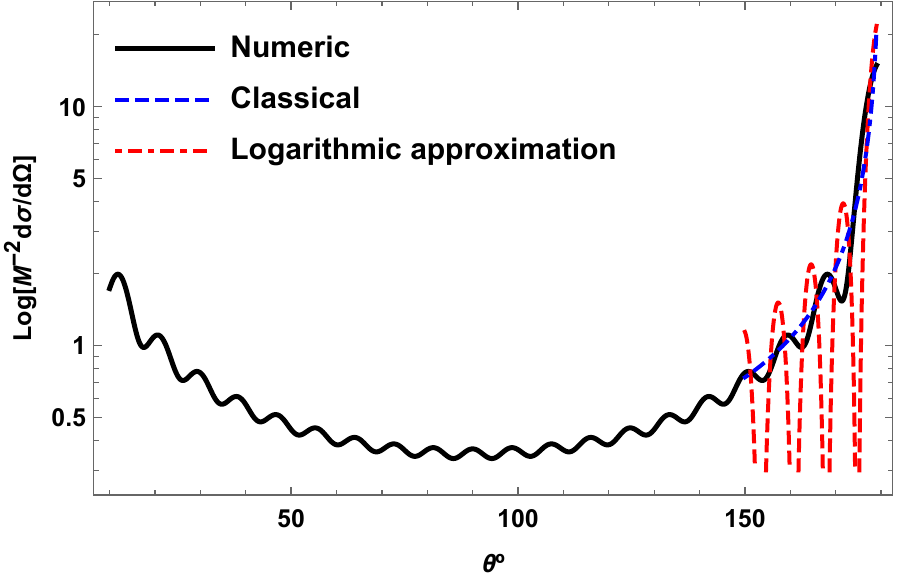}
	\caption{Numerical differential cross-section for the dilatonic black hole with $M \omega = 5$ and  $q=0.2$. The backward glory effect and the logarithmic approximation glory effect are both shown.}\label{fig:scompa1} 
\end{figure}

In this appendix, we will explore a different kind of approximation to determine the  deflection angle
(\ref{nfs}). The main idea is  to obtain a semi-analytic expression for the glory impact parameter $b_{g}$ and its derivative. To do so,  we consider the R.H.S of (\ref{Eq:orbiteq}) for  the critical impact parameter  and determine its roots, 
\begin{eqnarray}
	u_1 &=& u_c,\\ u_2 &=& \frac{4q^2u_c^2-2q^2u_c-2u_c+1+\beta}{4(1-2q^2u_c)},\\ u_3 &=& \frac{4q^2u_c^2-2q^2u_c-2u_c+1-\beta}{4(1-2q^2u_c)},
\end{eqnarray}
where the $\beta$-parameter is defined as 
\begin{eqnarray}
	\beta &=& \Big(16q^4u_c^4-16q^4u_c^3+4q^4u_c^2+16q^2u_c^3\nonumber\\&-&4q^2u_c-12u_c^2+4u_c+1\Big)^{1/2}.
\end{eqnarray}
Putting all together, the function $h(u)$ reduces to a cubic polynomial,
\begin{eqnarray}
	h(u) = (u-u_c)(u-u_2)(u-u_3)
\end{eqnarray}
Solving (\ref{nfs}), we obtain  that the deflection angle can be written in terms of elliptic integrals,
\begin{eqnarray}
	\theta(b)&\simeq& -\frac{2}{\sqrt{u_2-u_3)}}\Big(F(z,k)-K(k)\Big),
\end{eqnarray}
where
\begin{eqnarray}
	k^2 =\frac{u_c-u_3}{u_2-u_3}, \qquad z^2 =\frac{u_3}{u_3-u_c}.
\end{eqnarray}
In order to obtain the logarithmic deflection formula, we employ an approximation such that $k\simeq 1$ \cite{crispino2009scattering},
\begin{eqnarray}
	K(k)\simeq \frac{1}{2}\ln\Big(\frac{16}{1-k^2}\Big),\quad F(z,k) = \frac{1}{2}\ln\Big(\frac{1+z}{1-z}\Big)\nonumber\\
\end{eqnarray}
With the help of the above expression, the deflection angle (\ref{nfs}) can be recast as
\begin{eqnarray}
	\theta(b)&\simeq& -\frac{1}{\sqrt{u_2-u_3}}\Big[\ln\Big(\frac{\sqrt{u_3-u_c}+\sqrt{u_3}}{\sqrt{u_3-u_c}-\sqrt{u_3}}\Big)\nonumber\\&-&\ln\Big(\frac{16(u_2-u_3)}{u_2-u_c}\Big)\Big].
\end{eqnarray}
Table(\ref{tab3}) shows the values of the critical impact parameter using the logarithmic approximation.   We  obtain  a small difference (nearly  $1\%$) in relation with the numerical integration results displayed in Tab.(\ref{tab:tab2}).
\begin{table}
	\centering
	\begin{tabular}{c|c|c}
		$q$ & $b_g/M$ & $\frac{b_g^2}{M^2}\Big|\frac{db}{d\theta}\Big|_{\theta\simeq\pi}$\\
		\hline  \hline  
		0&  5.35 & 4.85 \\
		\hline  
		0.2 & 5.28 & 4.81 \\
		\hline  
		0.4 & 4.95 & 4.55\\
		\hline  
		0.6 & 4.66 & 4.23\\
		\hline  
		0.8 & 3.96 & 3.54\\
		\hline  
		1 & 2.04 & 2.18\\
	\end{tabular}
	\caption{Impact parameter for the glory effect using a logarithmic approximation for a dilatonic black hole.}
	\label{tab3}
\end{table}
In Fig. (\ref{fig:scompa1}), we compare the complete numerical differential scattering cross-section, the differential cross-section for the backward glory effect, and the logarithmic approximation for the differential cross-section of the glory effect. Generally speaking, the logarithmic approximation is qualitatively good enough for describing the differential cross near $\pi$. The same results are obtained by varying between $q=0.4$ and $q=0.6$.

\end{document}